\documentclass[iop]{emulateapj}

\usepackage{natbib}
\usepackage{amsmath}
\usepackage{amsfonts}
\usepackage{graphicx}
\usepackage{comment}

\shorttitle{Radio Sources in the Nearby Universe}
\shortauthors{Condon, Matthews, \& Broderick}

\begin{document}

\title{Radio Sources in the Nearby Universe}


\author{J.~J.~Condon\altaffilmark{1}}
\affiliation{National Radio Astronomy Observatory,
520 Edgemont Road, Charlottesville, VA 22903, USA}
\email{jcondon@nrao.edu}
\author{A.~M.~Matthews}
\affiliation{Department of Astronomy, University of Virginia,
  Charlottesville, VA 22904, USA}
\email{amm4ws@virginia.edu}
\and
\author{J.~J.~Broderick}
\affiliation{Department of Physics, Virginia Polytechnic Institute and
State University, Blacksburg, VA 24061, USA}


\altaffiltext{1}{The National Radio Astronomy Observatory is a facility
of the National Science Foundation operated under cooperative agreement
by Associated Universities, Inc.}


\begin{abstract}  
We identified 15,658 NVSS radio sources among the 55,288 2MASX
galaxies brighter than $k_\mathrm{20fe} = 12.25$ at $\lambda =
2.16\,\mu\mathrm{m}$ and covering the $\Omega =7.016$~sr of sky
defined by J2000 $\delta > -40^\circ$ and $\vert b \vert > 20^\circ$.
The complete sample of 15,043 galaxies with 1.4~GHz flux densities $S
\geq 2.45 \mathrm{~mJy}$ contains a 99.9\% spectroscopically complete
subsample of 9,517 galaxies with $k_\mathrm{20fe} \leq 11.75$.  We
used only radio and infrared data to quantitatively
distinguish radio sources powered primarily by recent star formation
from those powered by active galactic nuclei.  The radio sources with
$\log[L(\mathrm{W~Hz}^{-1})] > 19.3$ that we used to derive the local
spectral luminosity and power-density functions account for $>99$\% of
the total 1.4~GHz spectral power densities $U_\mathrm{SF} = (1.54 \pm
0.20) \times 10^{19} \mathrm{~W~Hz}^{-1} \mathrm{~Mpc}^{-3}$ and
$U_\mathrm{AGN} = (4.23 \pm 0.78) \times 10^{19} \mathrm{~W~Hz}^{-1}
\mathrm{~Mpc}^{-3}$ in the universe today, and the spectroscopic
subsample is large enough that the quoted errors are dominated cosmic
variance. The recent comoving star-formation rate density indicated by
$U_\mathrm{SF}$ is $\psi \approx 0.015~ M_\odot \mathrm{~yr}^{-1}
\mathrm{~Mpc}^{-3}$.

\end{abstract}





\section{Introduction}

The 1.4~GHz continuum emission from galaxies is powered by a
combination of recent star formation in star-forming galaxies (SFGs)
and supermassive black holes (SMBHs) in active galactic nuclei (AGNs).
The tight and nearly linear far-infrared (FIR)/radio correlation
observed among low-redshift galaxies makes 1.4~GHz spectral luminosity
a good dust-unbiased tracer proportional to the recent star-formation
rate (SFR) \citep{con92}, while sources that are radio-loud relative
to the FIR/radio correlation reveal the presence of
radio-dominant AGNs, even those deeply embedded in dust.

This paper presents separate local radio luminosity functions
  for both source types.  When used in conjunction with sensitive
  radio surveys made by the JVLA, MeerKAT, the SKA, or the ngVLA,
  local luminosity functions anchor models for the cosmological
  co-evolution of star formation and SMBH growth.  Our large ($N =
  9,517$) spectroscopically complete sample of the brightest
  ($k_\mathrm{20fe} \leq 11.75$ and $S_\mathrm{1.4~GHz} \geq 2.45
  \mathrm{~mJy}$) galaxies covers most of the extragalactic sky
  ($\Omega = 7.016 \mathrm{~sr}$) in order to (1) reach the low radio
  spectral luminosities $\log[L_\mathrm{1.4~GHz}\mathrm{(W~Hz)^{-1}}]
  \geq 19.3$ needed constrain the full range of sources accounting for
  nearly all ($>99$\%) recent star formation and SMBH growth and (2)
  minimize cosmic variance.
  
{Bright galaxies are also more likely to have the
  multiwavelength data needed to distinguish between radio sources
  powered by star formation and by AGNs. The total radio emission from
  any galaxy is actually the sum of both types, so quantitatively
  accurate criteria are needed to determine which is dominant.  We
  used only quantitative FIR, MIR (mid-infrared), and radio data to
  determine which type is energetically dominant. We did not use BPT
  diagrams \citep{bal81} or other optical emission-line diagnostics
  because they are not good \emph{quantitative} measures of
  AGN-powered radio emission.  It turns out, however, that the
  \citet{mau07} AGN/SFG classifications based on optical spectra agree
  surprisingly well with ours.

The cosmological evolution of radio sources is so strong that nearby
sources comprise only a small fraction of all sources in flux-limited
samples. Radio continuum emission alone cannot separate the nearby
needles from the haystack of distant sources, so statistically
complete and reliable samples of nearby radio sources are usually
selected by position-coincidence cross-identifications with optical or
infrared samples of bright galaxies.  For example, of all NRAO VLA Sky
Survey (NVSS) \citep{con98} sources stronger than $S \approx 2.5
\mathrm{~mJy}$ at $\nu = 1.4 \mathrm{~GHz}$, $<1$\% can be identified
with galaxies brighter than $m_\mathrm{p} = 14.5$ \citep{con02}.
About 85\% of those sources are in relatively low-luminosity
star-forming galaxies (SFGs) whose median face-on surface brightness
is just $\langle T_\mathrm{b} \rangle \sim 1 \mathrm{~K}$ at $\nu =
1.4 \mathrm{~GHz}$ \citep{hum81}, so reasonably complete samples of
nearby radio sources can be constructed only from radio surveys having
lower surface-brightness detection limits.  The NVSS is suitable
because its sensitivity limit is $T_\mathrm{b} = 5 \sigma_\mathrm{T}
\approx 0.7 \mathrm{~K}$.

This paper presents and analyzes a large catalog of NVSS sources
identified with 2 Micron All-Sky Survey eXtended (2MASX) galaxies
\citep{jar00} brighter than $k_\mathrm{20fe} = 12.25$ at $\lambda =
2.16\,\mu \mathrm{m}$,
where $k_\mathrm{20fe}$ is the magnitude
measured inside the $20 \mathrm{~mag~arcsec}^{-2}$ isophote.  The 2MASX
  galaxy sample is described in Section~\ref{sec:2masx}, and the NVSS
  radio identification procedure is explained in
  Section~\ref{sec:nvssids}.  The resulting 2MASX/NVSS catalog
  (Section~\ref{sec:catalog}) contains a statistically complete sample
  of 15,043 galaxies brighter than $k_\mathrm{20fe} = 12.25$ and
  1.4~GHz flux densities $S \geq 2.45 \mathrm{~mJy}$.
  Most of the analysis in this paper is based on the spectroscopically
  complete
   subsample of 9,517 galaxies with
  $k_\mathrm{20fe} \leq 11.75$ and $S \geq 2.45 \mathrm{~mJy}$. All
  but 19 had published spectroscopic redshifts, and we measured new
  spectroscopic redshifts for 12 of the 19
  (Appendix~\ref{sec:redshiftappendix}).  Only FIR, MIR
  (mid-infrared), and radio data were used to distinguish 2MASX/NVSS
  radio sources primarily powered by recent star formation from those
  dominated by AGNs (Section~\ref{sec:energytype}).  The counts of
  2MASX/NVSS sources powered by star formation and AGNs as functions
  of 1.4~GHz flux density are plotted and discussed in
  Section~\ref{sec:counts}.  Separate 1.4~GHz local luminosity
  functions for SFGs and AGNs are reported in Section~\ref{sec:lumf},
  and Section~\ref{sec:udex} presents the corresponding spectral power
  density functions.  Cosmic variance exceeds the Poisson variance for
  the large 2MASX/NVSS spectroscopic sample
  (Section~\ref{sec:cosmicvar}).  The total 1.4~GHz spectral energy
  density produced by SFGs today, $U_\mathrm{SF} = (1.54 \pm 0.20)
  \times 10^{19} \mathrm{~W~Hz}^{-1} \mathrm{~Mpc}^{-3}$, indicates
  that the recent SFRD is $\psi \approx 0.015 ~M_\odot
  \mathrm{~yr}^{-1} \mathrm{~Mpc}^{-3}$.

All calculations of absolute quantities (comoving distance, spectral
luminosity,\dots) from the observables (redshift, flux density,\dots)
are based on the relativistically correct equations for a $\Lambda$CDM
universe from \citet{con18} with $\Omega_\mathrm{m} = 0.3$,
$\Omega_\Lambda = 0.7$, and $H_0 =
70 \mathrm{~km~s}^{-1} \mathrm{~Mpc}^{-1}$ ($h = 0.70$).

\section{The 2MASX Galaxy Sample}\label{sec:2masx}

Large samples of bright galaxies necessarily cover a significant
fraction of the sky.  The Two Micron All Sky Survey (2MASS)
\citep{skr06} Extended Source Catalog (2MASX) \citep{jar00} is ideal
because:
\newline\noindent (1) It is complete and reliable over the whole
extragalactic sky for galaxies brighter than $k_\mathrm{s} \approx
k_\mathrm{20fe} + 0.2 \approx 13.5$ ($S_{2.16\,\mu\mathrm{m}} \approx
2.9 \mathrm{~mJy}$) at the longest infrared wavelength ($\lambda
\approx 2.16\,\mu\mathrm{m}$) yielding good atmospheric
transparency. Dust extinction in our Galaxy and dust absorption in
nearby galaxies are both small at this wavelength, and confusion by
stars is negligible at galactic latitudes $\vert b \vert \geq
20^\circ$.
\newline\noindent (2) The $\lambda = 2.16 ~\mu\mathrm{m}$ luminosity
of a normal galaxy is nearly proportional to its total stellar mass
\citep{bel03} because dust absorption is low and late-type stars
dominate the near-infrared (NIR) luminosity.  Thus the 2MASX sample
most directly samples the stellar masses in galaxies; it is less
biased than optical or FIR samples by recently formed massive stars. The
NVSS/2MASX flux-density ratio is a good measure of the recent star
formation rate per unit stellar mass, or the specific star-formation
rate (SSFR), which is a constraint on the star-formation history of
the universe.
\newline\noindent (3) The 2MASX photometric errors for galaxies
brighter than $k_\mathrm{s} \sim 12.5$ are $\lesssim 4$\%.
\newline\noindent (4) Nearly all 2MASX galaxies have such small absolute
position errors ($< 1\arcsec$ rms) that complete and reliable radio
identifications of 2MASX galaxies can be made by position coincidence
alone.
\newline\noindent (5) Spectroscopic redshifts are now available for
nearly all galaxies brighter than $k_\mathrm{20fe} = 11.75$
\citep{huc12}.

Our 2MASX galaxy sample includes all galaxies with:
\newline\noindent (1) 2MASX catalog fiducial magnitudes
$k_\mathrm{20fe} \leq 12.25$\allowbreak mag measured within the $20
\mathrm{~mag~arcsec}^{-2}$ ($\approx 3\sigma$) fiducial elliptical
aperture. According to the 2MASS Explanatory Supplement
(\url{https://www.ipac.caltech.edu/\allowbreak
 2mass/releases/allsky/\allowbreak doc/explsup.html}), the measured
fiducial flux density typically contains about 85\% of the
extrapolated total flux density.
\newline\noindent (2) 2MASX  fiducial semi-major
axes $r_\mathrm{20fe} \geq 5''$, above which the 2MASX catalog is
nominally complete, and
\newline\noindent (3) J2000 $\delta > -40^\circ$, the NVSS southern
declination limit, and absolute galactic latitude $\vert b \vert >
20^\circ$, the limit of the \emph{InfraRed Astronomical Satellite
  (IRAS)} Faint Source Catalog (FSC) \citep{mos92}.  This $\Omega
\approx 7.016~\mathrm{sr}$ (Appendix~\ref{sec:skyareaappendix}) sky
area is shown  in Figure~\ref{fig:figure1}.

The 2MASX all-sky data release catalog
(\url{https://www.ipac.caltech.edu/2mass/releases/allsky/}) contains
$N_\mathrm{IR} =55,288$ infrared galaxies satisfying all three
requirements.  Their mean sky density is $\rho_\mathrm{IR} \approx
2.40~\mathrm{deg}^{-2}$.

\begin{figure}
\includegraphics[trim=75 130 0 120,scale=0.4]{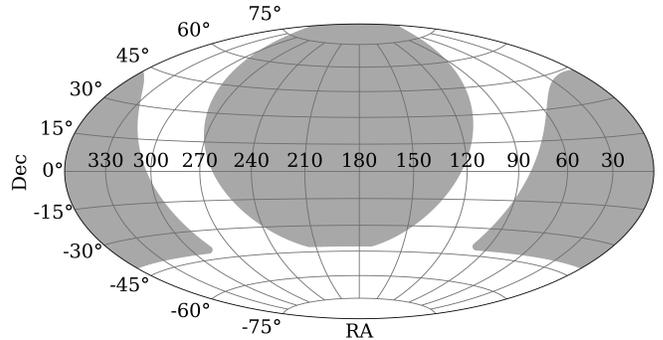}
\caption{The shaded 2MASX/NVSS area in this equal-area Hammer
projection covers the $\Omega \approx 7.016 \mathrm{~sr}$ (56\% of
the sky) with J2000 $\delta > -40^\circ$ and absolute galactic
latitude $\vert b \vert >20^\circ$.
\label{fig:figure1}}
\end{figure}

\section{NVSS Identifications of 2MASX Galaxies}\label{sec:nvssids}

To find all plausible NVSS identification candidates for the 2MASX
galaxies, we used the NVSS catalog browser
(\url{http://www.cv.nrao.edu/nvss/NVSSlist.shtml}) to select the
18,360 2MASX sample galaxies having (1) at least one NVSS radio
component within a search radius $r_\mathrm{s} = 60\arcsec$ or (2) at
least 2 NVSS components within $r_\mathrm{s} = 120\arcsec$.  These
search radii are compromises large enough to ensure high completeness
but small enough to avoid including too many unrelated background
sources.  Note that the NVSS catalog lists elliptical Gaussian radio
\emph{components} fitted to peaks on NVSS images, so the extended radio
\emph{source} produced by one galaxy may be represented by more than
one radio component.  If the radio emission from a galaxy is confused,
asymmetric, or significantly larger than the $\theta = 45\arcsec$ FWHM
Gaussian NVSS beam, the radio positions may be significantly offset
from the host galaxy position.  This is the case for $\sim5$\% of
2MASX/NVSS galaxies, so initial search radii much larger than the
combined 2MASX and NVSS position errors are needed to capture all
radio identifications and include all of their radio emission.

Our large search areas contain an unacceptable number of unrelated
background sources because the mean sky density of NVSS components is
$\rho \approx 53 \mathrm{~deg}^{-2}$.  The rms statistical sampling
error in a catalog of $N \sim 10^4$ identifications is $N^{1/2} \sim
10^2$, so exploiting the statistical power of such a large catalog
requires identification reliability $\gtrsim 99$\%.  Most background
radio sources are so distant (mean $\langle z \rangle \sim 1$) that
they are quite randomly distributed on the sky.  Thus the Poisson
probability $P$ that one or more unrelated NVSS components will lie
within $r_\mathrm{s} = 60\arcsec$ of any 2MASX galaxy is
\begin{equation}
P(\geq 1) = 1 - P(0) = 1 - \exp ( - \pi \rho
r_\mathrm{s}^2) \approx 0.045~,
\end{equation}
where the mean number of unrelated components in a search circle is
$\mu = \pi \rho r_\mathrm{s}^2$ and the probability of finding none is
$P(0) = \exp (-\mu)$.  In addition, some 2MASX galaxies are members of
physical groups and clusters, so the radio emission from close
companion galaxies must be excluded.  Thus \emph{at least}
$N_\mathrm{IR} P(\geq 1) = 55,288 \times 0.045 \gtrsim 2500$ of the
18360 candidate fields with $r_\mathrm{s} = 60\arcsec$ are likely to
contain unrelated NVSS components, leaving $\lesssim 16,000$ genuine
2MASX/NVSS identifications.  The probability of finding two or more
background sources within $r_\mathrm{s} = 120\arcsec$ is
\begin{eqnarray}
P(\geq 2) = & 1 - P(0) -
P(1) ~~~~~~~~~~~~~~~~~~~~~~~~ \\
=&1 - (1 - \pi \rho r_\mathrm{s}^2)\exp(-\pi \rho r_\mathrm{s}^2)
\approx 0.015~. \nonumber
\end{eqnarray}
Genuine 2MASX/NVSS identifications with neither a single component
within $60\arcsec$ nor two or more components within $120\arcsec$ are
 rare but may have been missed.

Recognizing and weeding out the background sources required extensive
and time-consuming human intervention, as described below.

Most of the radio sources produced by bright 2MASX galaxies are fairly
compact or at least symmetric.  Nearly all 2MASX galaxies have rms
position errors $\sigma_\alpha \approx \sigma_\delta \ll 1\arcsec$,
and NVSS position errors for unresolved components decline with
catalog flux density from $\sigma_\alpha \approx \sigma_\delta \approx
5\arcsec$ at $S = 2.45 \mathrm{~mJy}$ to $\lesssim 1\arcsec$ for $S >
15 \mathrm{~mJy}$.  Such candidates can be reliably accepted or
rejected on the basis position coincidence alone.  We define $\sigma$
as the quadratic sum of the 2MASX and NVSS rms position errors in each
coordinate,
\begin{equation}
  m \equiv r_\mathrm{s} / \sigma
\end{equation}
as the identification search radius in units of $\sigma$, and
\begin{equation}
  k \equiv 1 + 2 \pi \rho \sigma^2~.
\end{equation}
For  $\rho = 53
\mathrm{~deg}^{-2} \approx 4.09 \times 10^{-6} \mathrm{~arcsec}^{-2}$
and the worst case $\sigma \approx 5\arcsec$,
$k \approx 1.000642$.  In terms of $m$ and $k$ the completeness of the
identifications is \citep{con75}
\begin{equation}
  C = \frac{1 - \exp (-m^2 k / 2)} {k}~.
\end{equation}
Even when $\sigma = 5\arcsec$, $m = 3$ ($r_\mathrm{s} = 15 \arcsec$)
ensures $C \approx 0.99$.  The fraction of 2MASS galaxies actually
having NVSS counterparts is $f \approx 15,658 / 55,288 \approx 0.3$. 
The identification reliability \citep{con75}
\begin{equation}
R = C \Biggl[ \frac{1}{f} + \Biggl( 1 - \frac{1}{f} \Biggr)
\exp[ m^2 (1 - k) / 2] - \exp \Biggl( - \frac {m^2 k} {2} \Biggr) \Biggr]^{-1}
\end{equation}
is also $\gtrsim 99$\% because the probability that an unrelated NVSS
source lies within $3 \sigma \approx 15 \arcsec$ of any position is $
< 0.003$.

Figures~\ref{fig:figure2} and \ref{fig:figure3}
present examples illustrating both typical and difficult 2MASX/NVSS
cross-identifications.  The upper left panel of
Figure~\ref{fig:figure2} shows the Digitized Sky Survey (DSS)
gray-scale optical image, the NVSS 1.4~GHz brightness contours, and the
\emph{IRAS} $2\sigma$ position error ellipse for the typical spiral
galaxy IC 1526. Its 1.4~GHz flux density $S = 5.4 \pm 0.5
\mathrm{~mJy}$ and its 2MASX/NVSS position offset $r = 3\,\farcs8$ ($m
= 1.3$) are close to the sample medians. The radio sources in nearly
all spiral galaxies are fairly symmetric and roughly coextensive with
their optical host galaxies of stars.

Nonetheless, some radio sources in spiral galaxies could not be found
by position coincidence alone.  In the upper right panel of
Figure~\ref{fig:figure2}, the confused 2MASX position of NGC
5668 is marked by the cross on a bright spot $\sim 22\arcsec$ north of
the galaxy nucleus.  A few large face-on and edge-on spiral galaxies
have significantly offset or even multiple 2MASX positions that can be
recognized most easily by visual inspection of finding charts like
this one.

The NVSS contours and accurate 2MASX position for the very extended
low-brightness galaxy M74 are shown in the middle row, left panel of
Figure~\ref{fig:figure2}.  The closest NVSS catalog component is
$93\arcsec$ from the 2MASX position, so M74 is not in the list of
candidates within the $r_\mathrm{s} = 60 \arcsec$ search radius.  To
find similar cases, we searched for identifications among all galaxies
in the 1.49~GHz atlas of spiral galaxies with $B_\mathrm{T} \leq 12$
\citep{con87}.  M74 emphasizes the importance of high
surface-brightness sensitivity for identifying reasonably complete
radio samples of nearby galaxies.  Its total 1.4~GHz flux density is
$S \approx 180 \mathrm{~mJy}$, but its surface brightness is barely
above the NVSS $5 \sigma \approx 5 \times 0.45 \mathrm{~mJy~beam}^{-1}
\approx 2.3 \mathrm{~mJy~beam}^{-1}$ detection limit.

The price of high brightness sensitivity is low angular resolution.
The $\theta = 45''$ NVSS beam only marginally resolves the pair of
galaxies UGC 00644 and UGC 00644 NOTES01
(Figure~\ref{fig:figure2} middle row, right panel), and the NVSS
catalog lists only a single extended Gaussian component whose radio
centroid position is midway between the galaxies.  Finding charts make
it easy for humans to recognize such blends and decompose the radio
sources into unresolved components on the galaxy positions.

The majority of AGN-powered radio galaxies are also sufficiently
compact and/or symmetric to permit simple position-coincidence
identifications.  The lower left panel of Figure~\ref{fig:figure2}
shows the radio emission from an anonymous $S = 15.0 \mathrm{~mJy}$
(about the median flux density of AGNs in the sample) galaxy.
However, a significant minority of low-luminosity radio galaxies are
distinctly asymmetric.  In the lower right panel in
Figure~\ref{fig:figure2} are the 2MASX position cross and NVSS
contours of a head-tail radio galaxy whose centroid is significantly
offset to the north.  The head-tail morphology of this source is
confirmed by the high-resolution VLA image of \citet{owe93}.  Slightly
bent radio jets are common, but truly one-sided radio jets are rare in
low-luminosity radio sources.  The radio galaxy IC 1695 in the cluster
Abell 0193 appears in the upper left panel of
Figure~\ref{fig:figure3}.  About half of its flux density arises from
a compact component in the galaxy, and half originates in a slightly
curved one-sided jet extending $\sim 1\arcmin$ to the northeast
\citep{owe97}.

Radio galaxies powered by AGNs may emit most or even all of their
power in jets and lobes lying well outside the host galaxies of
stars. Thus it is necessary to search for radio components quite far
from each 2MASX position, \emph{whether or not} there is a radio
component close to the 2MASX position.  The right panel in the top row
of Figure~\ref{fig:figure3} is centered on an anonymous elliptical
galaxy at redshift $z \approx 0.0885$, so this $6\arcmin \times
6\arcmin$ finding chart is $ \approx 640 \mathrm{~kpc}$ on a side and
the triple radio source is even larger. \citet{mau07} identified only
the $S = 17.5 \mathrm{~mJy}$ central NVSS component with the 2MASX
galaxy, even though other NVSS components lie within their
$r_\mathrm{s} = 3 \arcmin$ candidate search radius and yield a total
flux density $S \approx 680 \mathrm{~mJy}$.  Sources like this are
difficult to recognize from component lists alone; there is no
substitute for visual inspection of finding charts that extend at
least $\pm 3 \arcmin$ in both directions.  The much larger $\pm 8
\arcmin$ finding chart in the left panel, middle row of
Figure~\ref{fig:figure3} is centered on the $S \approx 430
\mathrm{~mJy}$ triple radio galaxy 2MASX J15280499+0544278.  Only the
$S = 45.1 \mathrm{~mJy}$ central component was identified by
\citet{bes12}, and the lobes are only partially visible and not easily
recognized on our usual $6\arcmin \times 6\arcmin$ finding chart, so
we might have missed other sources with even more widely separated
lobes.  Although large triple sources like these are rare among bright
galaxies, they are usually so luminous that capturing their total flux
densities is important for deriving accurate radio luminosity
functions.

\begin{figure*}[ht]
  \plottwo{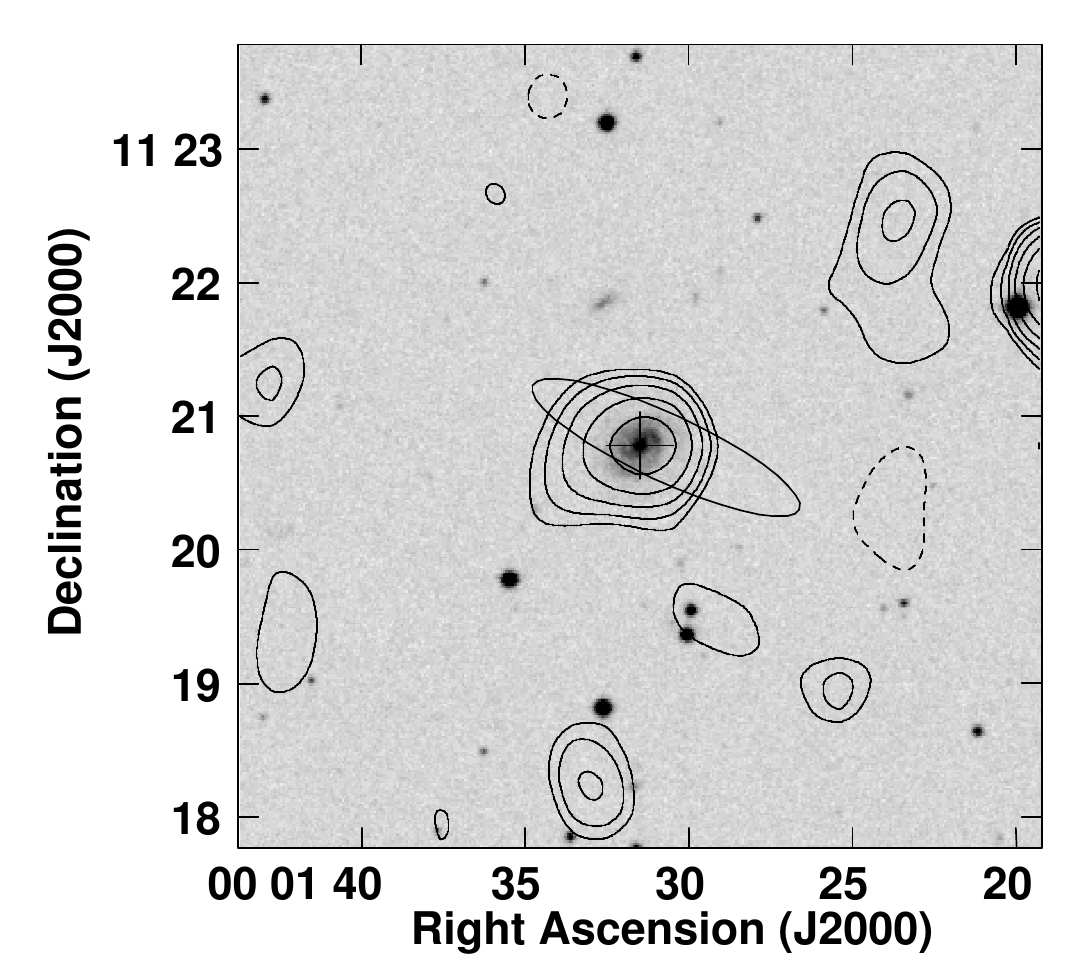}{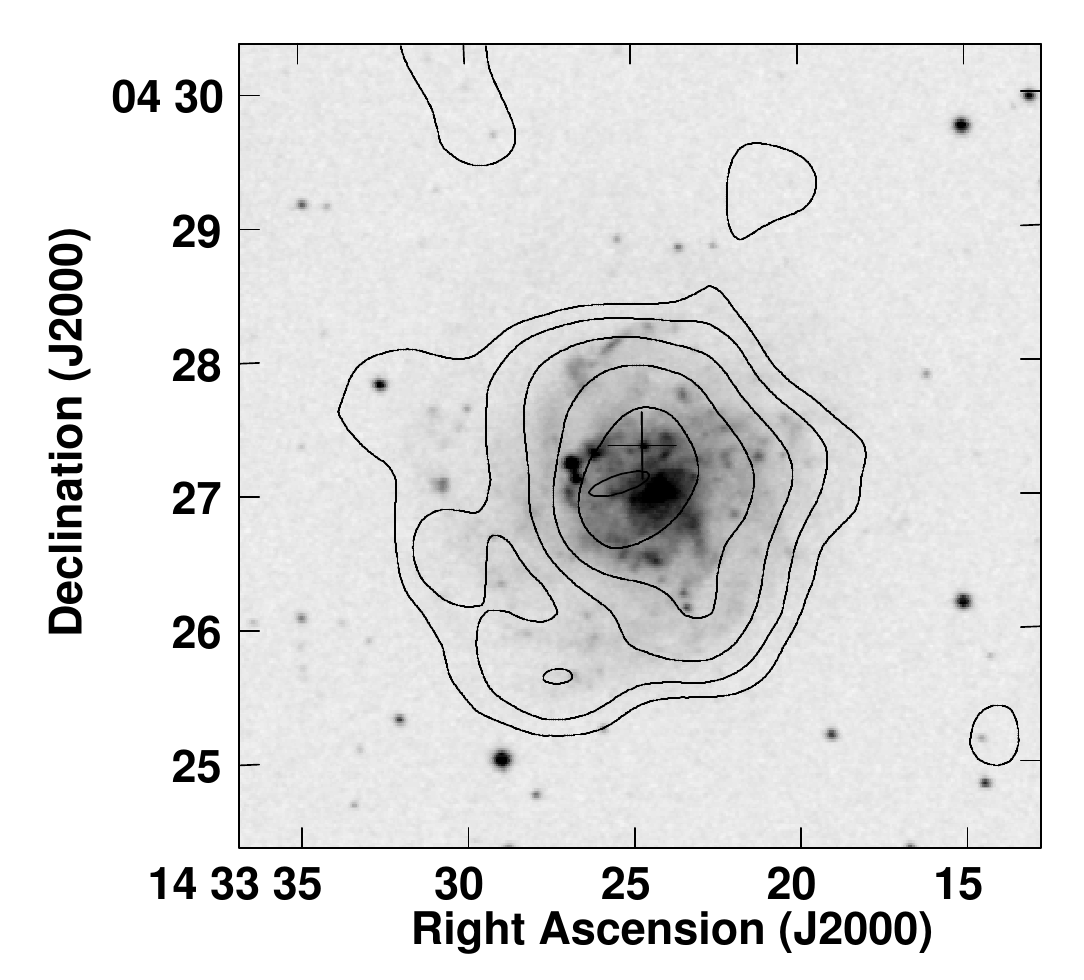}
  \plottwo{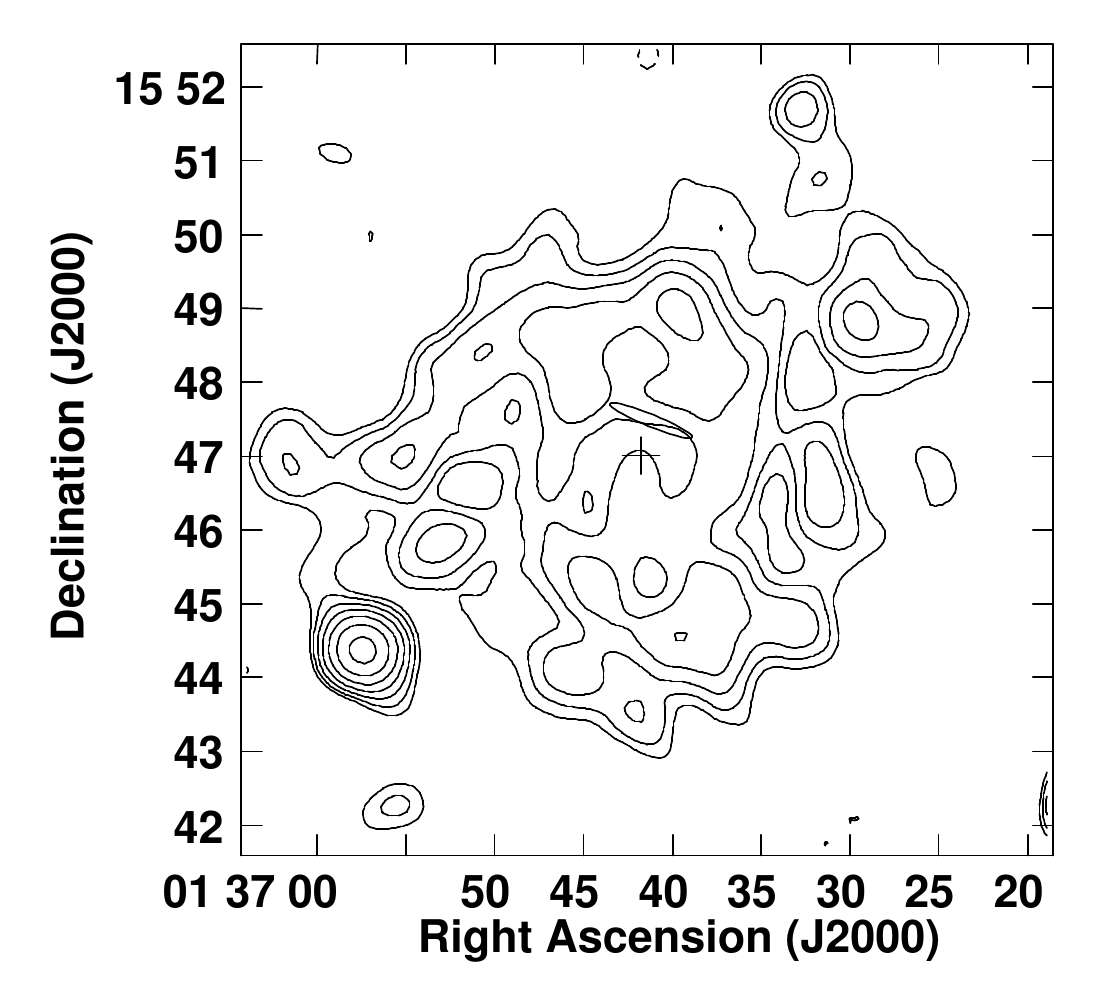}{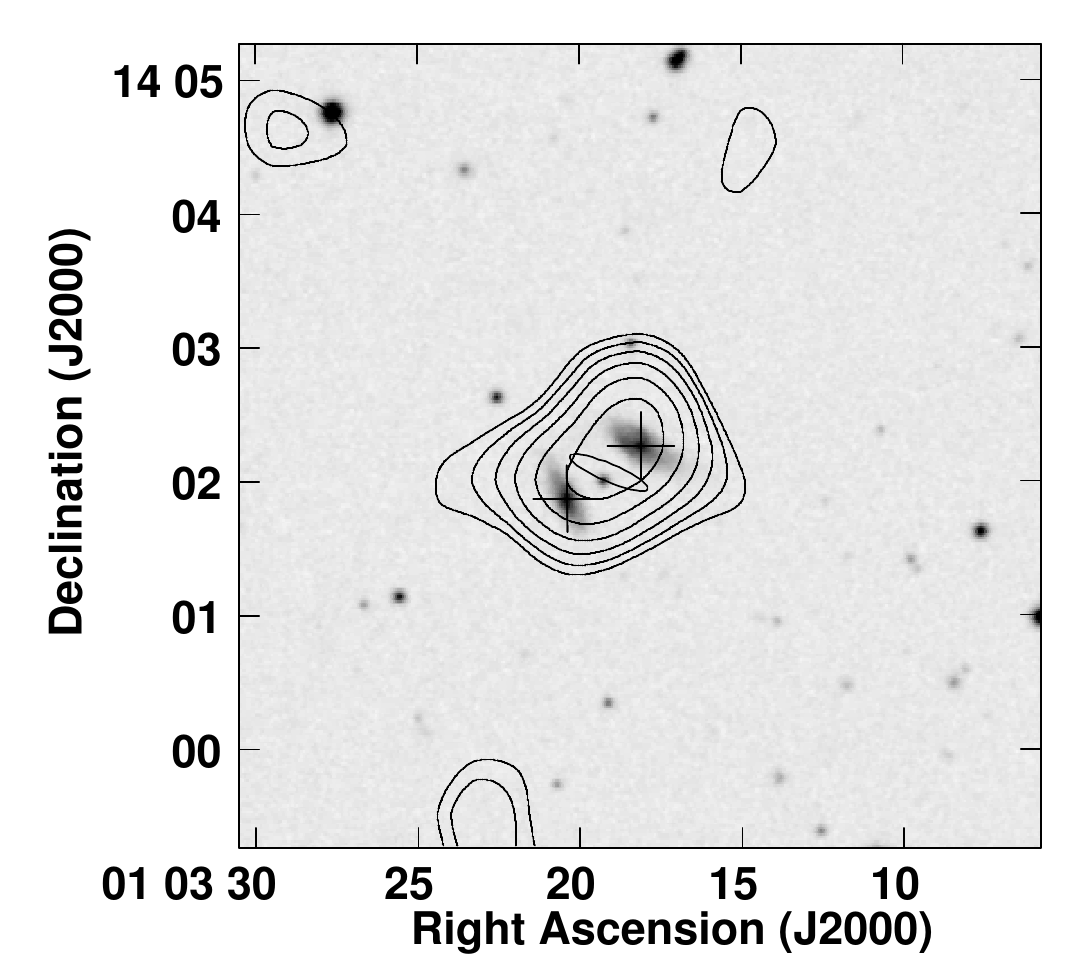}
  \plottwo{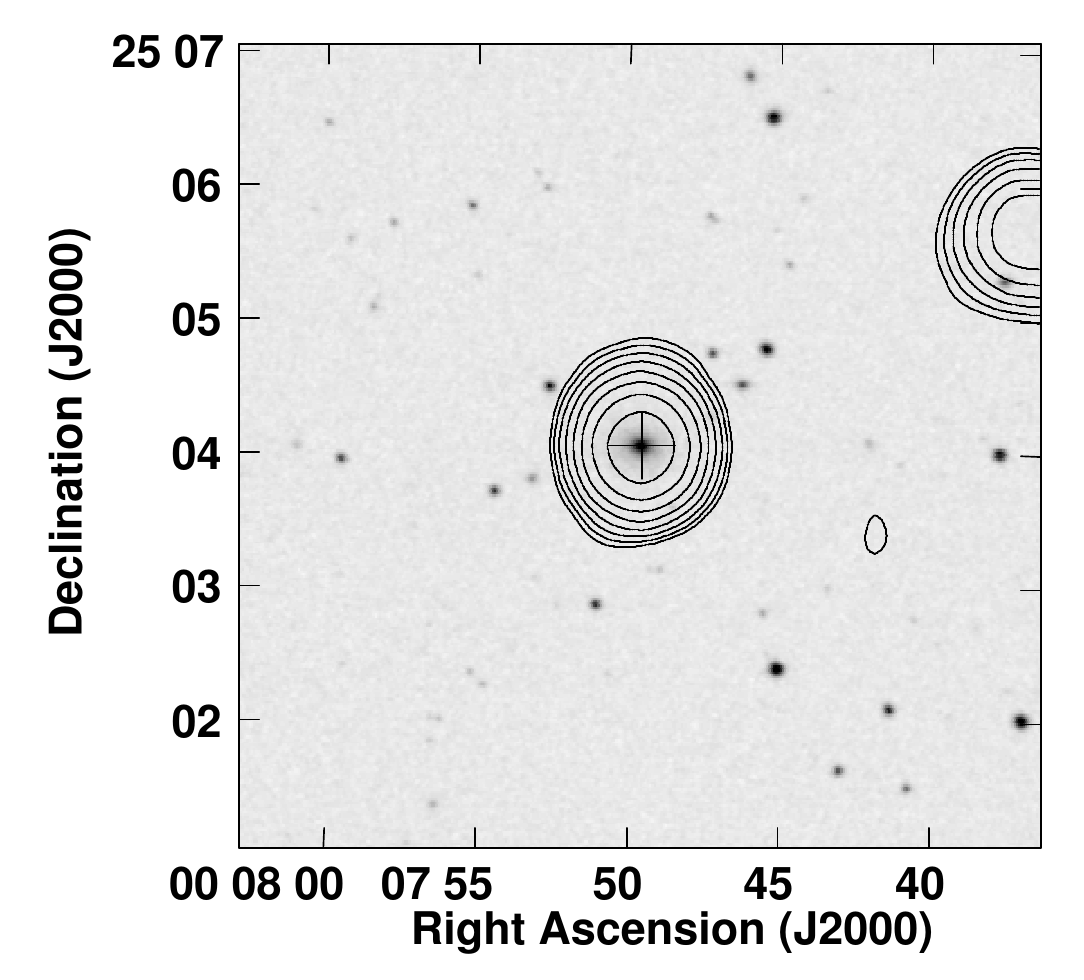}{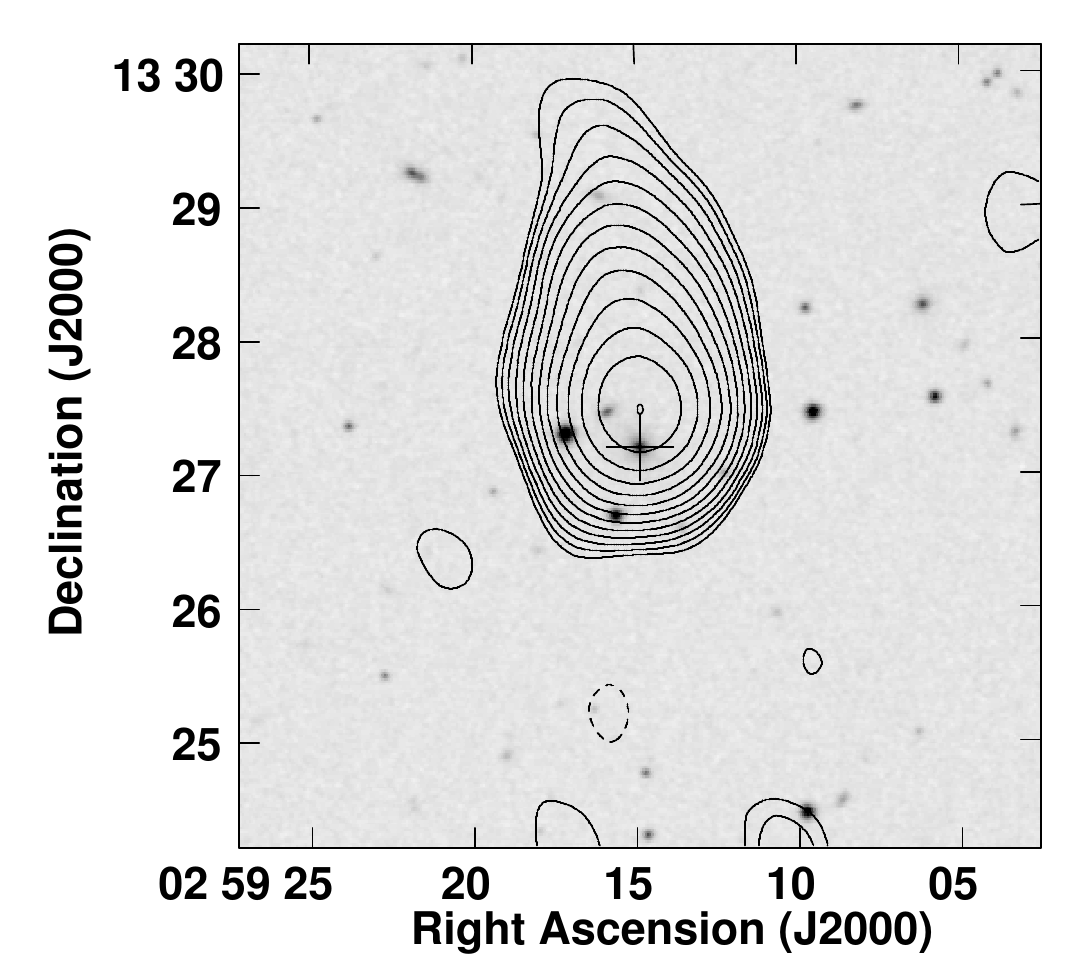}
  \caption{Selected finding charts. DSS gray-scale
    images are shown under NVSS contours plotted at $S_\mathrm{p} = \pm 1
    \mathrm{~mJy~beam}^{-1} \times 2^0,~2^{1/2},~2^1, \dots$~~ 2MASX
    source positions are marked by
    crosses, ellipses outline \emph{IRAS} position errors.
\label{fig:figure2}}
\end{figure*}

\begin{figure*}[ht!]
  \plottwo{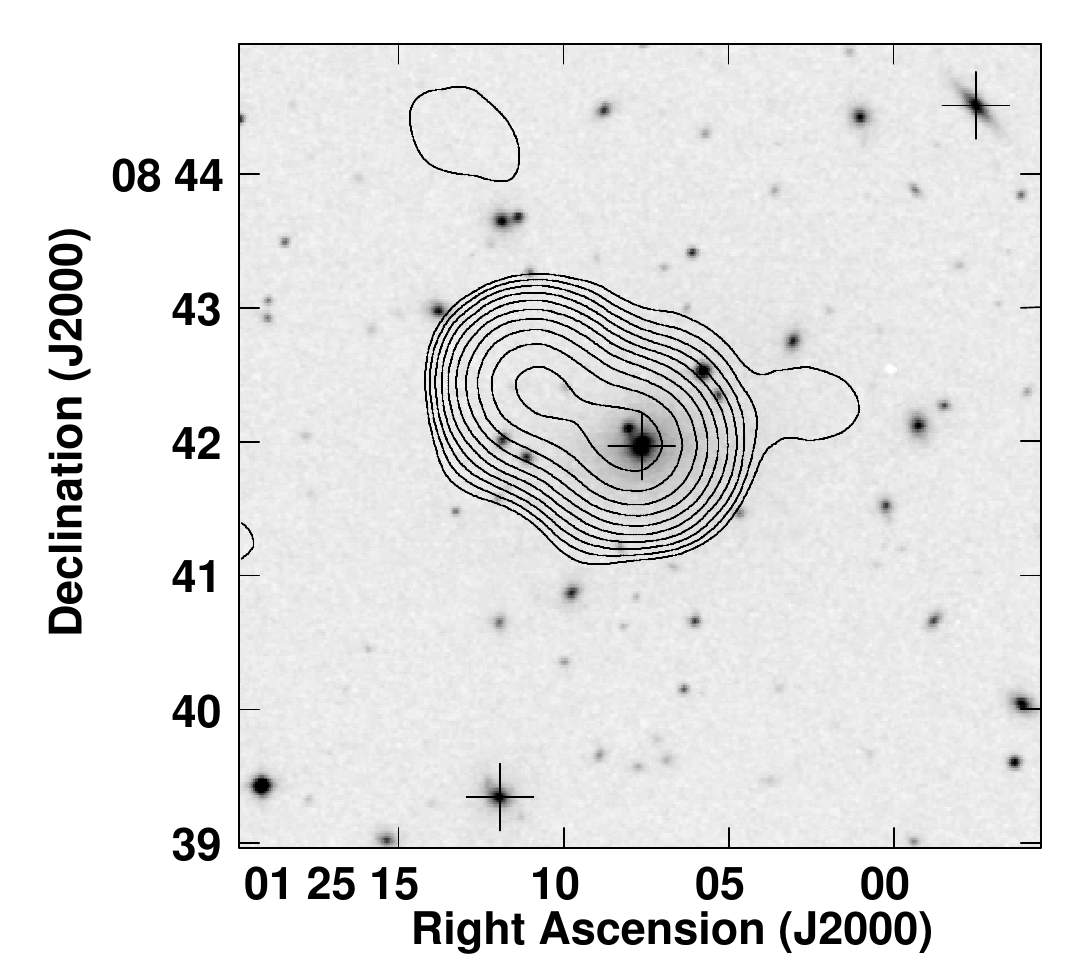}{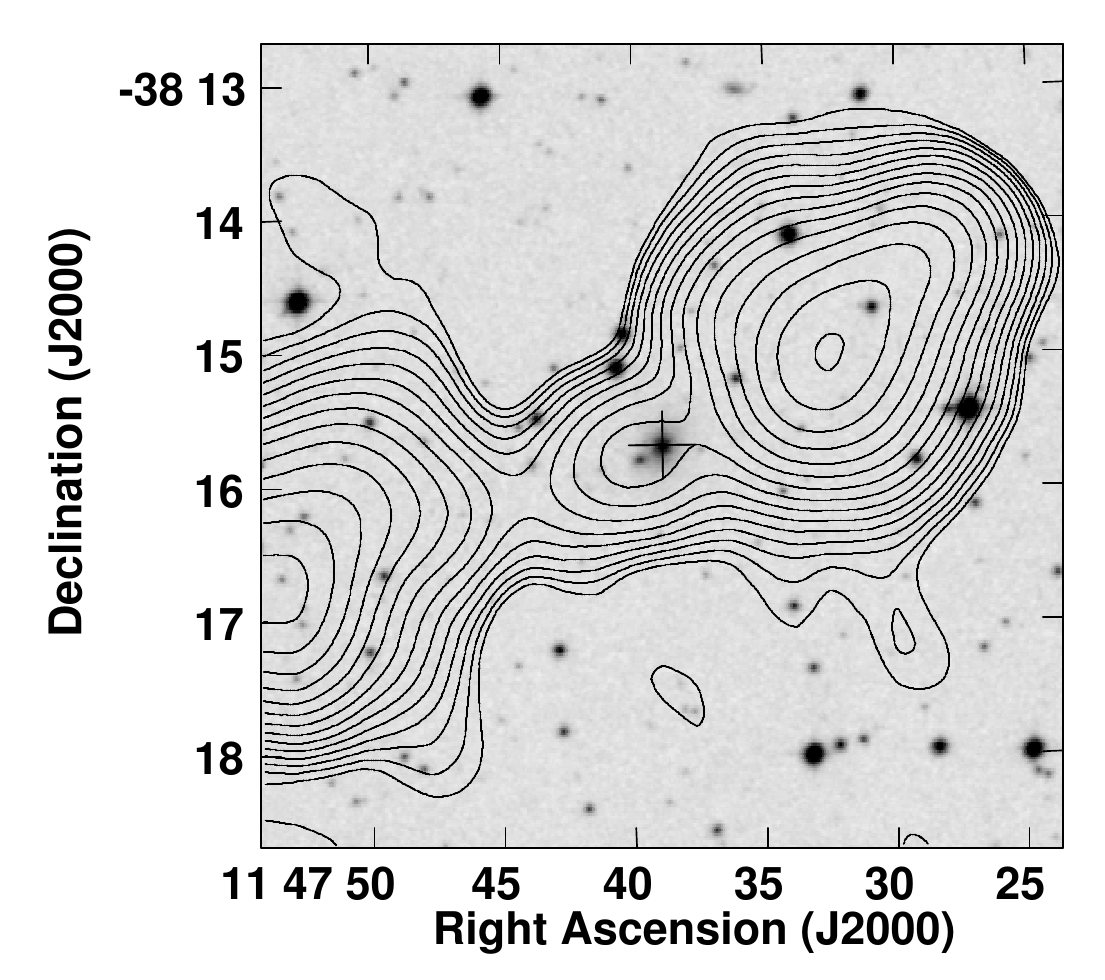}
  \plottwo{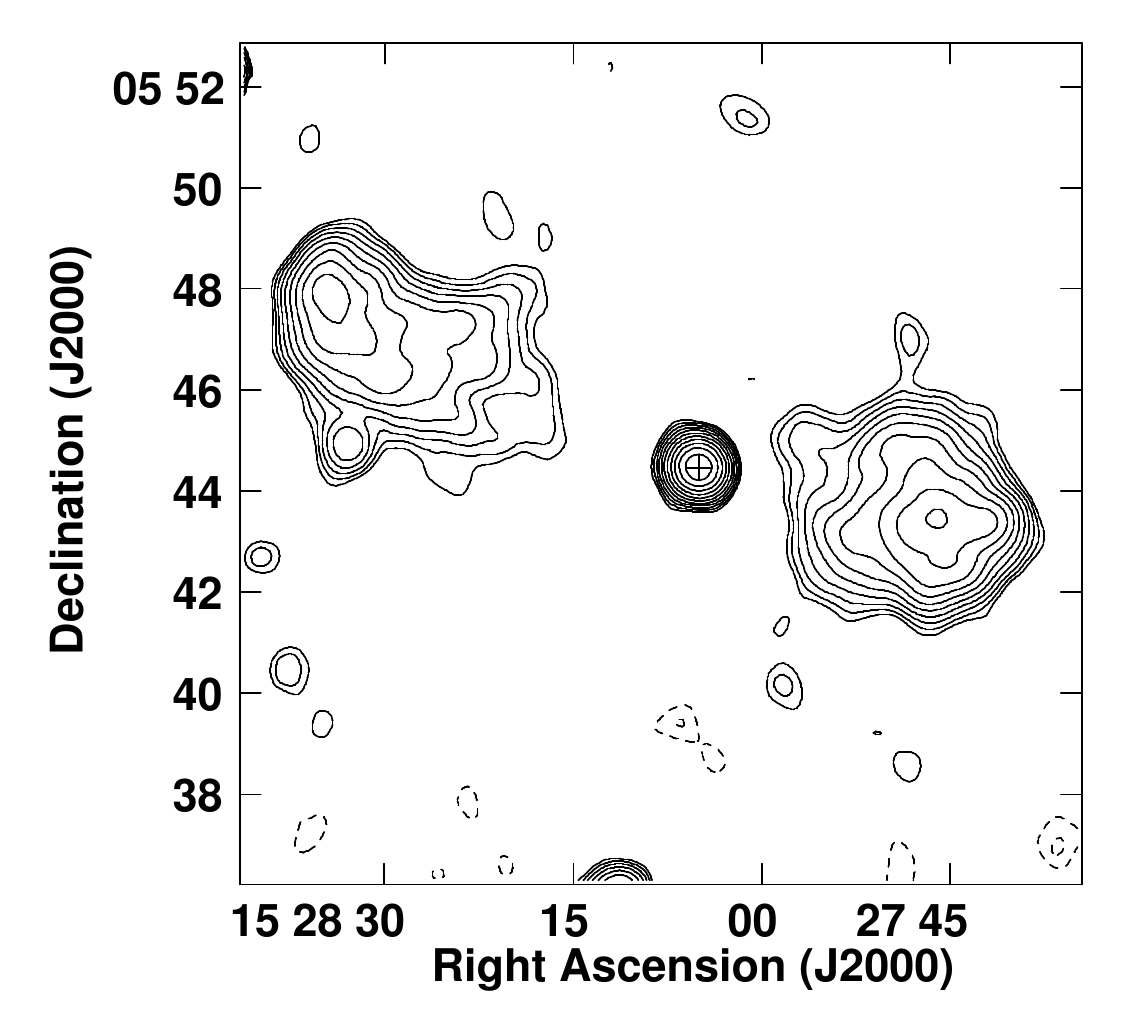}{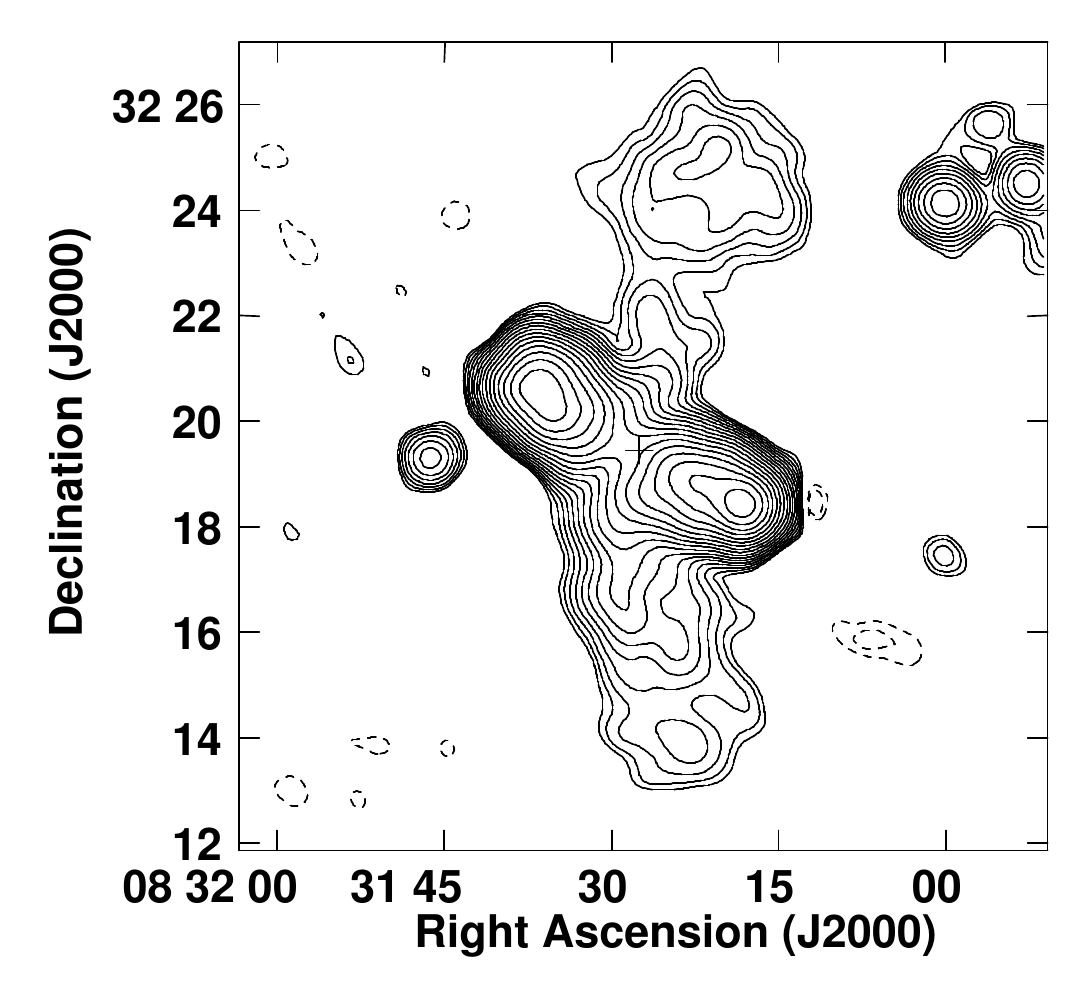}
  \plottwo{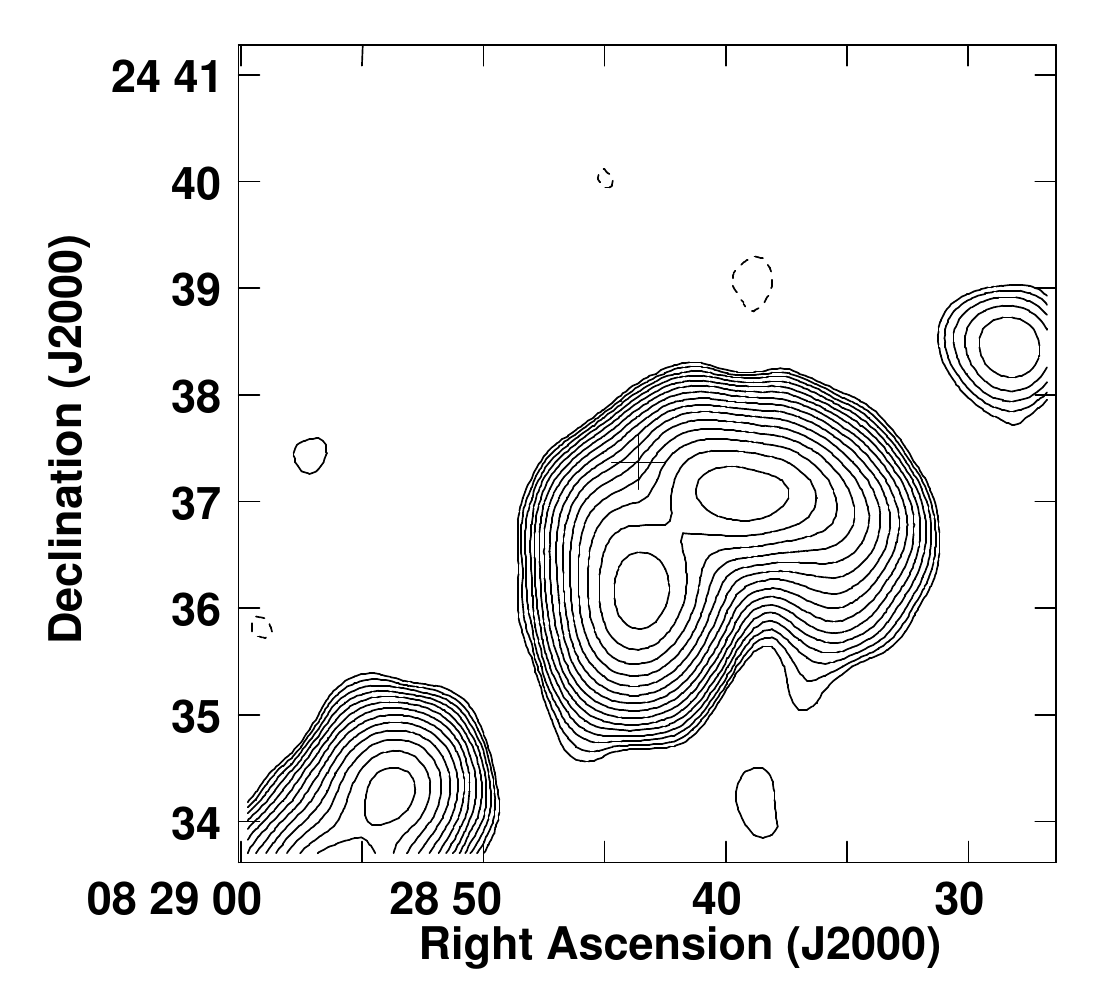}{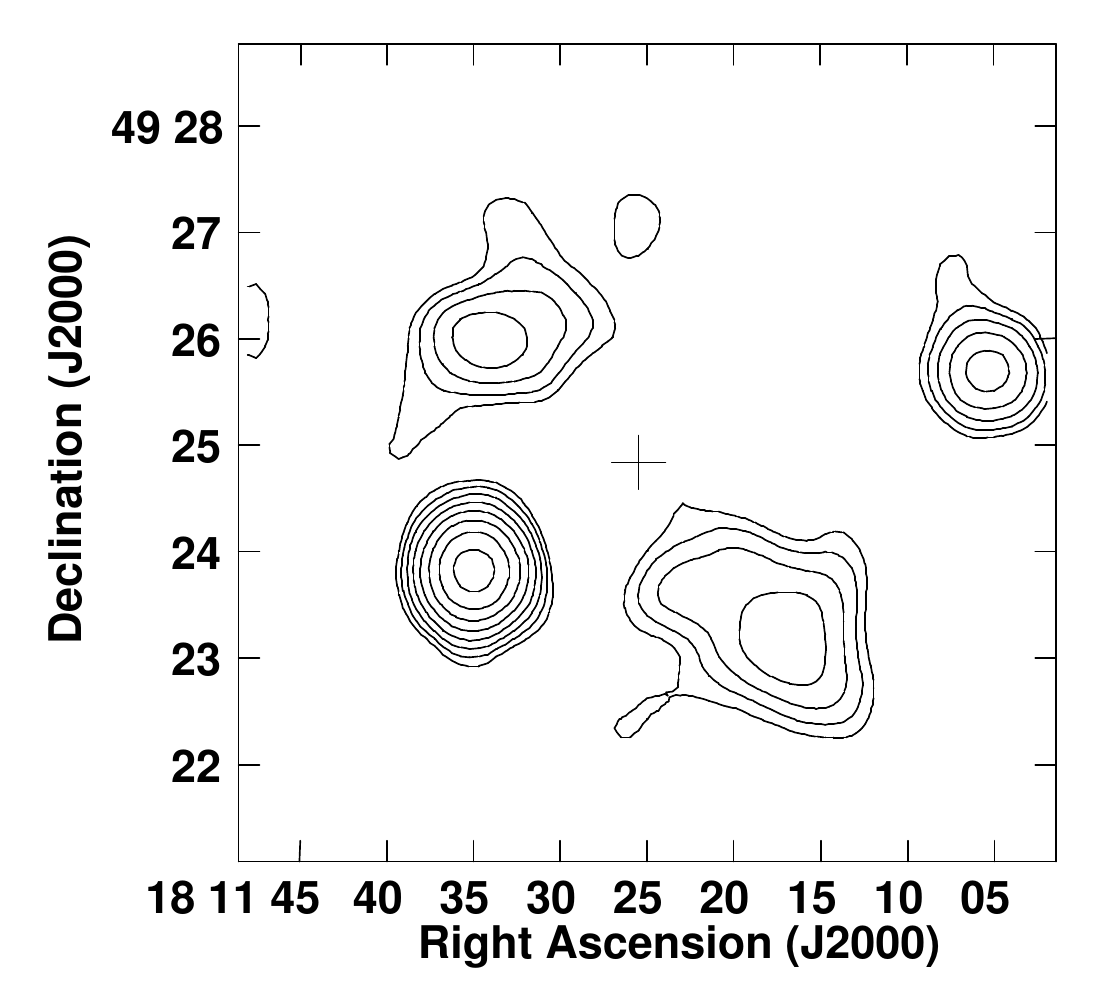}
  \caption{Additional selected finding charts. DSS gray-scale
    images are shown under NVSS contours plotted at $S_\mathrm{p} = \pm 1
    \mathrm{~mJy~beam}^{-1} \times 2^0,~2^{1/2},~2^1, \dots$~~ 2MASX
    source positions are marked by
    crosses. \label{fig:figure3}}
\end{figure*}

Some ``empty double'' radio sources have no NVSS components within
$60\arcsec$ of their 2MASX host galaxies.  To find them, we searched
for pairs or multiple components offset by up to $120\arcsec$.  The
nearest NVSS components in the X-shaped radio source 4C~+32.25 =
B2~0828+32 (right panel, middle row of
Figure~\ref{fig:figure3}) are the bright FR~II lobes
symmetrically offset from 2MASX J08312752+3219270 by $104\arcsec$ and
$119\arcsec$.  The larger but fainter north-south extension has a
steep radio spectrum and may be the relic of an earlier outburst in a
precessing system \citep{par85}.  Most coreless double sources can be
recognized because their lobes have roughly equal flux densities, are
about equally distant from their host galaxies, and are at position angles
differing by $\sim 180^\circ$. Somewhat more difficult to recognize
are bent coreless doubles.  The left panel, bottom row of
Figure~\ref{fig:figure3} shows the luminous ($S \approx 650
\mathrm{~mJy}$ at $z \approx 0.0830$) bent double source having no
NVSS components within $60\arcsec$ of the cross on 2MASX
J08284360+2437220.  Finally, the lower right panel of
Figure~\ref{fig:figure3} shows a large but faint double source
that illustrates the limit of reliable identifications.  Secondary
evidence supporting this identification as a double source includes
(1) the two components are roughly equidistant from the 2MASX galaxy,
(2) the two components have comparable brightness, (3) the line
between them passes close to the galaxy, and (4) the southwest
component has a tail pointing back toward the galaxy.

\section{The 2MASX/NVSS Catalog and Samples}\label{sec:catalog}

Following the procedures described in Section~\ref{sec:nvssids}, we
identified NVSS sources with 15,658 of the 55,288 2MASX galaxies
having $k_\mathrm{20fe} \leq 12.25$ and semi-major axes
$r_\mathrm{20fe} \geq 5\arcsec$ in the $\Omega \approx 7.016$~sr solid
angle defined by J2000 declination $\delta > -40^\circ$ and absolute
galactic latitude $\vert b \vert \geq 20^\circ$.  The resulting
2MASX/NVSS galaxy catalog is displayed in part as
Table~\ref{tab:table1}, which lists for each galaxy its 2MASX J2000
coordinate name, 2MASX fiducial $\lambda = 2.16\,\mu\mathrm{m}$
magnitude $k_\mathrm{20fe}$, 2MASX fiducial major-axis diameter
$d_\mathrm{20fe} = 2\,r_\mathrm{20fe}$ in arcsec, 1.4~GHz NVSS total
flux density $S$ in mJy, dominant radio energy source type (either
recent star formation S or active galactic nucleus A) derived from
FIR data, from MIR data, and the final type derived from both as
explained in Section~\ref{sec:energytype}, heliocentric radial
velocity $cz$ in km~s$^{-1}$ usually from the NASA/IPAC Extragalactic
Database (NED), and the most common alternative galaxy name (e.g., UGC
12890) from NED.

All NVSS catalog flux densities are rounded to the nearest $0.1
\mathrm{~mJy}$, so the 15,043 galaxies with 1.4~GHz catalog flux
densities $S \geq 2.5 \mathrm{~mJy}$ comprise a flux-limited sample
complete to $S = 2.45 \mathrm{~mJy}$.  The spectroscopically complete
subsample of the 9,517 galaxies with $k_\mathrm{20fe} \leq 11.75$ and $S
\geq 2.45 \mathrm{~mJy}$ now has redshifts for all but 7 (99.9\%
redshift completeness).

\begin{deluxetable*}{c r r r c c c c l}
\tablecaption{2MASX/NVSS catalog\label{tab:table1}}
\tablewidth{0pt}
\tablehead{
  \colhead{2MASX} & \colhead{$k_\mathrm{20fe}$} & \colhead{$d_\mathrm{20fe}$} &
  \colhead{$S_\mathrm{1.4}$} & \multicolumn{3}{c} {\llap{E}nergy Source Typ\rlap{e}} &
  \colhead{$cz$} &  \\
  \colhead{J2000 name} & \colhead{(mag)} & \colhead{\llap{(}$\arcsec$\rlap{)}} &
  \colhead{(mJy)} & \colhead{~FIR} & \colhead{MIR} & \colhead{Final~~} &
  \colhead{\llap{(}km~s$^{-1}$\rlap{)}} & \colhead{\llap{NE}D name}
  }

\startdata
00000701$+$0816448 &  10.779 &  23.6 &    82.7 & A & A &  A  &    11602 &   UGC 12890           \\
00001278$+$0107123 &  11.839 &  15.3 &     2.1 & S & S &  S  & \phn7390 &   CGCG 382-016        \\
00002880$+$3246563 &  11.108 &  13.4 &     5.2 & S & S &  S  & \phn9803 &   IC 5373             \\
00003138$+$2619318 &  11.967 &  13.2 &     7.5 & S & S &  S  & \phn7653 &   UGC 12896           \\
00003564$-$0145472 &  11.488 &  16.1 &     2.8 & ? & S & (S) & \phn7274 &   CGCG 382-017        \\
00005234$-$3550370 &  11.548 &  15.2 &    48.4 & A & A &  A  &    15581 &                       \\
00010444$+$0430001 &  12.013 &  15.4 &     2.7 & S & S &  S  & \phn8932 &   IC 5374             \\
00011996$+$1306406 &   9.920 &  27.4 &    12.3 & S & S &  S  & \phn5366 &   NGC 7803            \\
00013148$+$1120465 &  11.341 &  16.0 &     5.4 & S & S &  S  & \phn9099 &   IC 1526             \\
00013359$+$0900445 &  11.234 &  15.0 &     2.7 & ? & S & (S) & \phn9252 &   UGC 12912           
\enddata


\tablecomments{Table \ref{tab:table1} is published in its entirety in
the electronic edition of the {\it Astrophysical Journal}. A portion
is shown here for guidance regarding its form and content.}
\end{deluxetable*}


\subsection{Radio Energy Sources}\label{sec:energytype}

The ultimate energy sources powering the radio continuum continuum
emission from galaxies are recently formed massive short-lived stars
and SMBHs in AGNs.  In order to use radio continuum luminosity as a
quantitative tracer of the SFR, we classified the radio emission of
each galaxy in Table~\ref{tab:table1} as being powered \emph{primarily} by
recent star formation ``S'' or by an AGN ``A''.  Labels ``(S)'' and
``(A)'' indicate uncertain classifications.  Note that these
  are \emph{quantitative} classifications because both star formation and
  an AGN may contribute to the total radio luminosity a single galaxy.

Optical emission- and absorption-line spectra have often been used
  to classify galaxies as SFGs or AGNs. For example, \citet{sad02} and
  \citet{mau07} classified as AGNs all galaxies with absorption-line spectra
  like those of giant elliptical galaxies, absorption-line spectra
  with weak LINER-like emission lines, or stellar continua dominated
  by nebular emission lines stronger than Balmer emission lines; and
  they classified as SFGs all galaxies with spectra dominated by strong
  narrow H$\alpha$ and H$\beta$ emission lines.

We decided not to use any optical indicators to determine the
  dominant radio energy sources in our galaxies.  AGN signposts such
as [\ion{O}{3}] luminosity are often not correlated with radio
luminosity \citep{bes05}.  We did not assume that star formation
powers the radio sources in spiral galaxies or that AGNs drive radio
emission from E and S0 galaxies. We did not use optical colors and
fluxes, which may be biased by dust absorption.  We  did not use BPT
\citep{bal81} diagrams which plot the [\ion{O}{3}]/H$\beta$ ratio as a
function of the [\ion{N}{2}]/H$\alpha$ ratio because $\sim 40$\% of
nearby radio-loud AGN are too gas poor and optically inactive to be
detected this way \citep{ger15}.  Thus our energy-source classification method
is independent of the \citep{sad02} and \citep{mau07} classification method
based on optical spectra.

Instead, we used only a combination of radio and infrared data to
classify our radio sources.  Radio sources powered by stars can be
recognized because (1) $>99$\% obey the tight and nearly linear
FIR/radio flux correlation \citep{con91}, (2) they have the steep FIR
spectral indices $\alpha(25\,\mu\mathrm{m}, 60\,\mu\mathrm{m}) < -1.5$
characteristic of cold dust emission \citep{deg85}, (3) they usually
reside in galaxies having ``dusty'' MIR colors, and (4) they are
roughly coextensive with their optical host galaxies.  Radio sources
powered by AGNs (1) are usually much stronger than expected from the
FIR/radio correlation, (2) may be associated with warmer FIR sources,
(3) usually reside in galaxies having the nearly blackbody MIR colors
of ``naked'' stars, and (4) may contain jets and lobes extending well
outside their host galaxies.

We used a combination of these four indicators as described in detail below to
assign a primary energy source type to each 2MASX/NVSS galaxy in
Table~\ref{tab:table1}:
\newline

(1) The \emph{IRAS} FIR/NVSS 1.4 GHz flux-density ratio was
parameterized by the quantity
\begin{equation}\label{eqn:q}
  q \equiv \log \Biggl[ \frac{\mathit{FIR} / (3.75 \times 10^{12} \mathrm{~Hz})}
    {S_\mathrm{1.4~GHz}\, \mathrm{(W~m}^{-2}\mathrm{~Hz}^{-1}\mathrm{)}} \Biggr] ~,
\end{equation}
where
\begin{eqnarray}
  \mathit{FIR} \mathrm{~(W~m}^{-2}) \equiv ~~~~~~~~~~~~~~~~~~~~~~~~~~~~~~~~~~~~\\
  1.26 \times 10^{-14}
  [2.58S_{60\,\mu\mathrm{m}}\, \mathrm{(Jy)} + S_{100\,\mu\mathrm{m}}\, \mathrm{(Jy)}]
  \nonumber
\end{eqnarray}
\citep{hel88}.

If a galaxy was detected by \emph{IRAS} (\emph{IRAS} flux quality code
2 or 3) at both $60\,\mu\mathrm{m}$ and $100\,\mu\mathrm{m}$, the
value of $q$ was calculated directly from Equation~\ref{eqn:q}.  If a
galaxy was detected at $60\,\mu\mathrm{m}$ but not detected
(\emph{IRAS} flux quality code 1) at $100\,\mu\mathrm{m}$, an
approximate $q$ was estimated using the median observed
$S_{100\,\mu\mathrm{m}} \sim 2 S_{60\,\mu\mathrm{m}}$ \citep{yun01}.
Conversely, if a galaxy was detected at $100\,\mu\mathrm{m}$ but not
at $60\,\mu\mathrm{m}$, $q$ was estimated assuming
$S_{60\,\mu\mathrm{m}} \sim S_{100\,\mu\mathrm{m}} / 2$.  If a galaxy
was observed by \emph{IRAS} but not detected at either
$60\,\mu\mathrm{m}$ or $100 \,\mu\mathrm{m}$, an upper limit to $q$
was calculated from the \emph{IRAS} FSC 90\%-completeness upper limits
$S_{60\,\mu\mathrm{m}} < 0.36 \mathrm{~Jy}$ and
$S_{100\,\mu\mathrm{m}} < 1.2 \mathrm{~Jy}$.  Finally, if a galaxy was
in an area not adequately covered by \emph{IRAS}, we set $q =\,?$
and used only other classification methods.

The normalized probability distribution $P(q)$ of all galaxies in the
complete
1.4~GHz
flux-limited sample that were observed by \emph{IRAS} is
plotted as a histogram in Figure~\ref{fig:figure4}. Within that
histogram the unshaded area indicates upper limits to $q$ for galaxies
observed but not detected by \emph{IRAS} at either
$60\,\mu\mathrm{m}$ or $100\,\mu\mathrm{m}$, and the shaded
area shows measured or estimated $q$ values.  Star-forming galaxies
obeying the FIR/radio correlation are clustered in the narrow peak
with mean $\langle q \rangle \approx 2.30$ and rms scatter $\sigma_q
\approx 0.17$.  The intrinsic scatter in $q$ \citep{con91} is nearly
equal to our measured $\sigma_q$, so the peak in
Figure~\ref{fig:figure4} has not been broadened significantly by
flux-density measurement errors.  Adding a dominant AGN to a $q = 2.3$
SFG would result in $q < 2.0$.  To allow for the observed scatter in
$q$, we classified galaxies with measured $q > 1.8$ as SFGs and
galaxies with upper limits or measured values of $q < 1.8$ as primarily
AGN-powered.

\begin{figure}[ht]
\includegraphics[trim=65 280 0 185,scale=0.6]{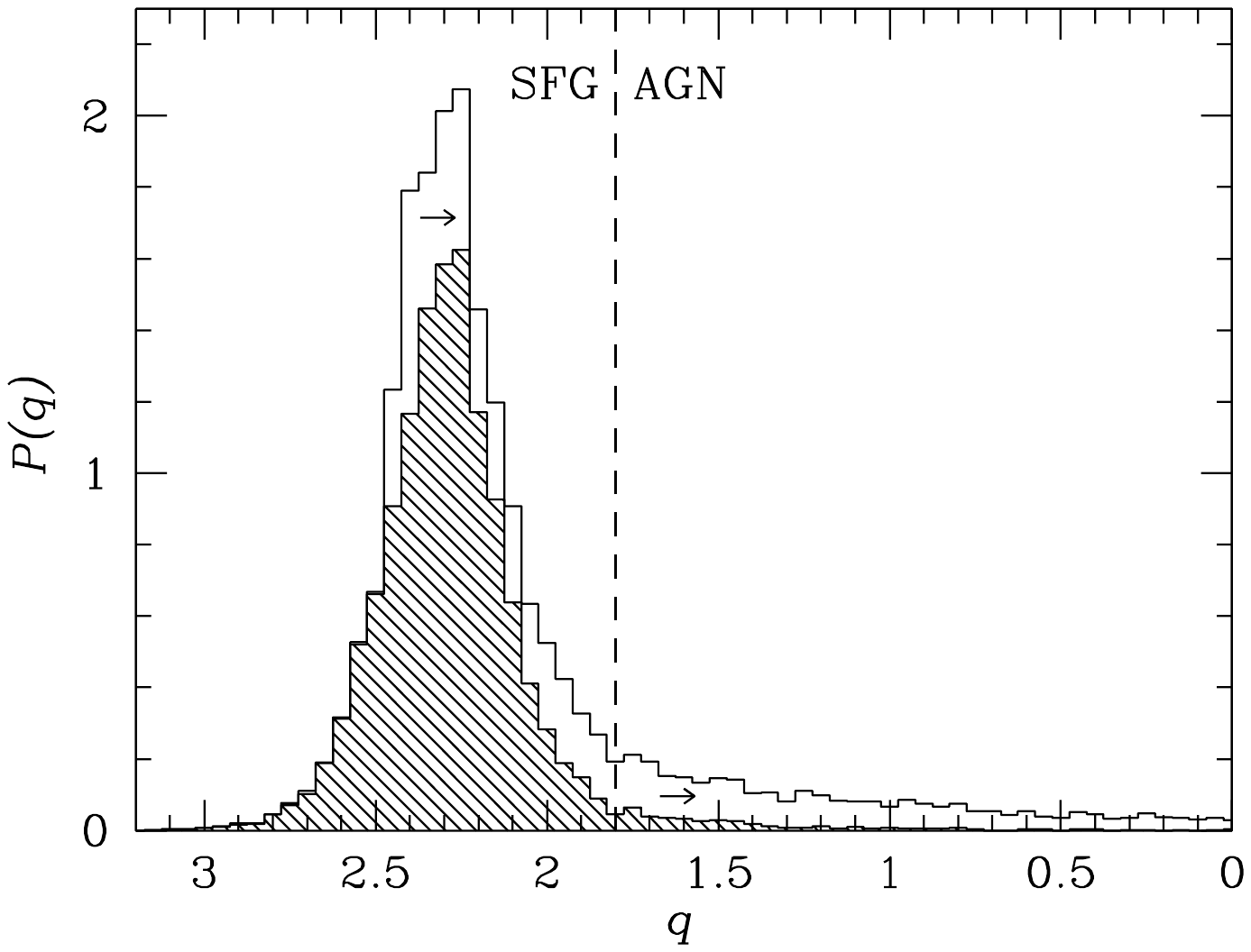}
\caption{For 2MASX/NVSS galaxies with $S_\mathrm{1.4~GHz} \geq 2.45
  \mathrm{~mJy}$, the normalized probability distribution of the
  measured FIR/radio ratio $q$ is shown by the shaded histogram and
  upper limits for those observed but not detected by \emph{IRAS} are
  indicated by the unshaded part of the histogram.
  The vertical dashed line separates SFGs
   with measured $q > 1.8$ from AGNs with
  upper limits or measured $q < 1.8$. Abscissa:
  FIR/radio flux ratio $q$ (Equation~\ref{eqn:q}).  Ordinate:
  Probability density $P(q)$.
  \label{fig:figure4}}
\end{figure}

Galaxies with upper limits to $q$ larger than 1.8 and galaxies not
observed by \emph{IRAS} could not be classified by this method.  The
NVSS and \emph{IRAS} are comparably sensitive to SFGs: the value of
$q$ corresponding to the sensitivity limits $S_\mathrm{1.4~GHz} = 2.45
\mathrm{~mJy}$, $S_{60\,\mu\mathrm{m}} = 0.36 \mathrm{~Jy}$, and
$S_{100\,\mu\mathrm{m}} = 1.2 \mathrm{~Jy}$ is $q \approx 2.4$.  Thus
many galaxies not detected by \emph{IRAS} do have upper limits to $q$
larger than 1.8 (Figure~\ref{fig:figure4}).
\newline

(2) A FIR source warm enough to have
\begin{equation}
  \alpha(25\,\mu\mathrm{m},\,60\,\mu\mathrm{m}) > -1.5
\end{equation}
indicates concentrated dust heating by a single AGN, rather than by a
comparably luminous but more extended cluster of stars \citep{deg85}.
The spectral-index error resulting from a 20\% error in the
$25\,\mu\mathrm{m}$ flux density is $\Delta\alpha(25\,\mu\mathrm{m},
\,60\,\mu\mathrm{m}) \sim \pm 0.25$.  To allow for spectral-index errors
of sources with $\lambda = 25\,\mu\mathrm{m}$ signal-to-noise ratios
as low as 5, we conservatively classified only galaxies with
$\alpha(25\,\mu\mathrm{m},\,60\,\mu\mathrm{m}) > -1.25$ as primarily
AGN-powered.  There are 247 such ``warm'' galaxies, of which 77 also have
$q < 1.8$ and the remaining 170 were newly classified as AGN-powered by their
warm FIR spectra.
\newline

(3) \emph{Wide-field Infrared Survey Explorer (WISE)} \citep{wri10}
MIR magnitudes in bands W1 ($\lambda = 3.4\,\mu\mathrm{m}$), W2
($\lambda = 4.6 \,\mu\mathrm{m}$), and W3 ($\lambda =
12\,\mu\mathrm{m}$) determine the colors $(W1 - W2)$ and $(W2 - W3)$
that help to distinguish AGNs residing in elliptical galaxies and
Seyfert galaxies from dusty spiral galaxies dominated by ongoing star
formation, as illustrated in Figure~\ref{fig:figure5}.  Stars alone and
dustless elliptical galaxies (lower left circle in
Figure~\ref{fig:figure5}) have low values of $(W1 - W2)$ and $(W2 - W3)$
because the W1, W2, and W3 wavelengths are on the Rayleigh-Jeans side
of the blackbody peak of most stars, and the limiting Rayleigh-Jeans
spectral index $\alpha = +2$ corresponds to $(W1 - W2) \approx -0.05$
and $(W2 - W3) \approx -0.07$ for the \emph{WISE} flux-density scales
and frequencies listed in Table 1 of \citet{jar11}.  The sublimation
temperature of large interstellar dust grains is too low for them to
affect $(W1 - W2)$ significantly, but dust in SFGs
increases $(W2 - W3)$.  Nuclear emission from Seyfert galaxies (upper
right circle in Figure~\ref{fig:figure5}) can increase $(W1 - W2)$
enough to separate AGNs from SFGs.  Thus radio
sources in galaxies above the broken line specified by
\begin{eqnarray}\label{eqn:wisecolors}
W1 - W2  = & +0.8 ~~~~~~~~~~~~~~~~ \quad (W2 - W3 \geq 3.1)\quad\quad \\
W1 - W2  = &
  (W2 - W3 - 1.82) / 1.6 \quad (W2 - W3 < 3.1)  \nonumber
\end{eqnarray}
are probably AGN-powered, and those below the line are likely powered
by ongoing star formation.  Although the \emph{WISE} MIR colors are
less reliable indicators than the \emph{IRAS} FIR/radio correlation,
they are available for nearly all 2MASX/NVSS galaxies, so we used them
to classify cases that have neither measured $q$ values nor upper
limits $q < 1.8$.

\begin{figure}
\includegraphics[trim=20 120 0 20,scale=0.47]{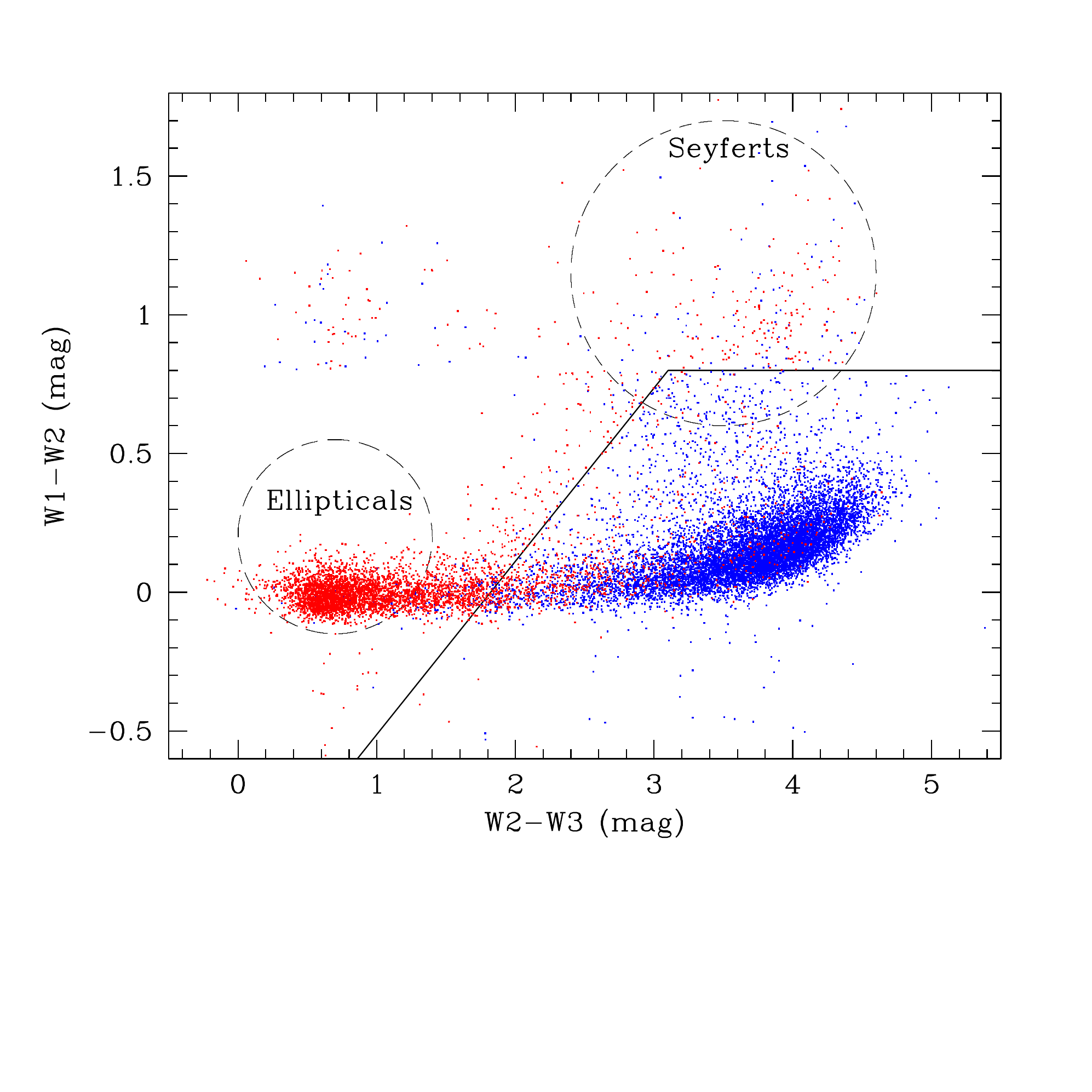}
\caption{The mid-infrared colors $(W1-W2)$ ($[3.4\,\mu\mathrm{m}] -
  [4.6\,\mathrm{m}]$) and $(W2-W3)$ ($[4.6\, \mu\mathrm{m}] -
  [12\,\mu\mathrm{m}]$) can be used to separate galactic stars and
  elliptical galaxies (lower left dashed circle) from Seyfert galaxies
  (upper right dashed circle) and from dusty spiral galaxies with
  ongoing star formation (below the broken line), as shown in
  \citet{wri10} Figure~12.  Galaxies listed as A or (A) in
  Table~\ref{tab:table1} are shown as red points; S or (S) galaxies
  are blue.  Abscissa: $(W2 - W3)$~(mag).  Ordinate: $(W1 -
  W2)$~(mag).
\label{fig:figure5}}
\end{figure}

\emph{WISE} does not have the far-infrared coverage needed to yield
$q$ (Equation~\ref{eqn:q}), but the $\lambda = 22\,\mu\mathrm{m}$
\emph{WISE} magnitude $W4$ can be used to define a similar quantity
\begin{equation}\label{eqnq22}
  q_{22} \equiv \log[S(22\,\mu\mathrm{m}) / S(1.4\,\mathrm{GHz})]~,
\end{equation}
where
\begin{equation}
  \log \biggl[ \frac {S(22\,\mu\mathrm{m})} {\mathrm{Jy}} \biggr] =
    0.918  - 0.4 \, W4
\end{equation}
\citep{jar11}.  The normalized probability distribution $P(q_{22})$ is
shown in Figure~\ref{fig:figure6} for all 2MASX/NVSS galaxies with
$k_\mathrm{20fe} \leq 12.25$ and $S(\mathrm{1.4~GHz}) \geq 2.45
\mathrm{~mJy}$ (black histogram).
Galaxies with \emph{WISE} MIR colors below the broken line in
Figure~\ref{fig:figure5} (\emph{WISE} energy source S) are represented by the
blue histogram, and galaxies above the broken line in Figure~\ref{fig:figure5}
(\emph{WISE} energy source A) by the red
histogram.  Like the distribution of $q$ in Figure~\ref{fig:figure4},
the distribution of $q_{22}$ in Figure~\ref{fig:figure6} has a narrow
peak dominated by SFGs and a long tail of galaxies containing
radio-loud AGNs.  Thus \emph{WISE} MIR colors and \emph{WISE}
MIR/radio flux ratio parameters $q_{22}$ provide independent energy-type
classifications that largely agree.

\begin{figure}[ht]
\includegraphics[trim=50 270 0 185,scale=0.58]{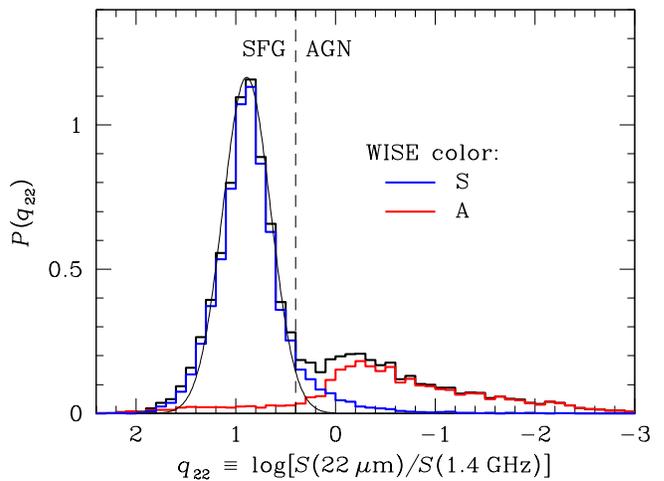}
\caption{The normalized probability distribution $P(q_{22})$ of
  $q_{22} \equiv \log[S(22\,\mu\mathrm{m}) / S(\mathrm{1.4\,GHz})]$
  for all 2MASX/NVSS galaxies with $S(\mathrm{1.4\,GHz}) \geq 2.45
  \mathrm{~mJy}$ is indicated by the black histogram.  The blue and
  red histograms show the galaxies classified as S or A by their
  \emph{WISE} MIR colors (Equation~\ref{eqn:wisecolors} and
  Figure~\ref{fig:figure5}).
The vertical dashed line at $q_{22} = +0.4$ separates
  most SFGs ($q_{22} > +0.4$) from most AGNs ($q_{22} < +0.4$).
  The Gaussian fit to the narrow peak has
  mean $\langle q_{22} \rangle = 0.89$ and rms width $\sigma = 0.24$.
  Abscissa: \emph{WISE} MIR/radio flux-ratio parameter $q_{22}$.
  Ordinate: Probability density $P(q_{22})$.
\label{fig:figure6}}
\end{figure}

The main advantage of the \emph{WISE} $q_{22}$ distribution over the
\emph{IRAS} $q$ distribution is that all but a handful of 2MASX/NVSS
galaxies were detected by \emph{WISE} at $\lambda = 22
\,\mu\mathrm{m}$.  The drawback of $q_{22}$ is contamination of
$S(22\,\mu\mathrm{m})$ by emission from warm dust heated by AGNs,
making it a somewhat less reliable parameter than $q$ for
distinguishing SFGs from AGNs.  For our final MIR classifications, we
used the \emph{WISE} color criterion (Equation~\ref{eqn:wisecolors}).
We used $q_{22}$ only for a few galaxies having no \emph{IRAS} data,
as described in items 3. through 5. in the list at the end of
Subsection~\ref{sec:energytype}.
\newline

(4) Radio morphology complements the three photometric indicators
above.  Radio sources powered by star formation are
roughly coextensive with the star-forming regions, their synchrotron
emission broadened only slightly by diffusion of cosmic-ray electrons
\citep{mur08}.  Coextensive synchrotron emission and free-free
absorption by ionized hydrogen at electron temperature $T_\mathrm{e}
\sim 10^4 \mathrm{~K}$ limits the 1.4~GHz brightness temperature of
SFGs to $T_\mathrm{b} \lesssim 10^5 \mathrm{~K}$
\citep{con92}.  AGNs can produce radio jets and lobes that extend
well outside their host galaxies, and they can produce compact radio
cores with brightness temperatures $T_\mathrm{b} \gg 10^5
\mathrm{~K}$.

To identify very extended radio jets and lobes, we inspected the
finding charts of all galaxies having two or more NVSS components.
Most are either elliptical galaxies or spiral galaxies larger than the
radio sources and much larger than the $\theta = 45\arcsec$ FWHM NVSS
beam.

The only spiral galaxy with radio emission outside the galaxy of stars
is the Seyfert NGC 4258 (2MASX J12185761+4718133) with unique
``anomalous radio arms'' \citep{vdk72}.  NGC 4258 was not covered by
the \emph{IRAS} FSC, but we classified its radio source as primarily
AGN-powered on the basis of \emph{WISE} photometry ($q_{22} = -0.10$).

All but one of the multicomponent NVSS sources not identified with
large spiral galaxies are so radio-loud (either $q < 1.8$ or $q_{22} <
0.4$) that they had already been photometrically classified as
AGN-powered.  The sole exception is luminous triple radio source in
2MASX J23415138$-$3729306, which has neither \emph{IRAS} nor
\emph{WISE} photometry and was classified as AGN-powered on the basis
of radio morphology alone.

Sub-arcsecond resolution is needed to resolve
sources brighter than $T_\mathrm{b} \sim 10^5 \mathrm{~K}$, so the
NVSS alone is unable to distinguish AGN cores from compact
SFGs.
\newline

The four indicators above do not always agree, so the final energy
source types A, (A), S, and (S) listed in Table~\ref{tab:table1} were
derived by reconciling the various \emph{IRAS} and \emph{WISE}
classifications as follows:
\begin{enumerate}
  \item If \emph{IRAS} and \emph{WISE} agree on A or S, the final
    classification is A or S.
  \item if \emph{IRAS} and \emph{WISE} disagree on A or S, the
    \emph{IRAS} result was kept but qualified as uncertain (A) or (S).
  \item If \emph{IRAS} = ? (no \emph{IRAS} data) and $q_{22} < 0.4$
    (radio loud), then the final classification is M if the
    \emph{WISE} MIR color classification = ? and (M) if it = S.
  \item If \emph{IRAS} = ?, $0.7 > q_{22} \geq 0.4$, and \emph{WISE} =
    S, then the final classification is (S).
  \item If \emph{IRAS} = ?, $q_{22} \geq 0.7$, and \emph{WISE} = S,
    then the final classification is S.
\end{enumerate}    

Among our 15,043 galaxies classified by radio and infrared
criteria are 3,466 that had been classified by \citet{mau07} on the
basis of optical line spectra.  Their star-forming galaxies were
labeled SF, and their AGNs were divided into three subtypes: Aa (pure
absorption-line spectra like those of giant elliptical galaxies), Aae
(spectra with absorption lines and weak narrow LINER-like emission
lines), or Ae (conventional Type II AGN spectra with nebular emission
lines such as} [O\textsc{ii}], [O\textsc{iii}], or [N\textsc{ii}]
that are stronger than any hydrogen Balmer emission lines, or
conventional Type I AGN spectra with strong and broad hydrogen Balmer
emission lines).  Uncertain optical classifications were indicated by
\lq?\rq.

Table~\ref{tab:table2} compares our independent galaxy
classification methods for these 3,466 galaxies, and the agreement is
better than we had expected.  Of the 888 galaxies we classified as A,
\citet{mau07} classified 867, 847 (97.7\%) as various AGN types and
only 20 (2.3\%) as SF or SF?.  We classified 2218 galaxies as S and
\citet{mau07} classified 2185 of them, 2082 (95.3\%) as SF or SF? and
103 (4.7\%) as various AGN types.  Of their 2186 SF galaxies, we
classified 2121 (97.0\%) as S or (S), 17 (0.8\%) as A, and 48 (2.2\%)
as (A).  They classified 690 galaxies as Aa; we classified 676
(98.0\%) as A or (A) and 14 (2.0\%) as S or (S).  The agreement is
lower for the 161 Aae galaxies (74\%) and the 104 Ae galaxies (66\%).
Most of these are star-forming LINERs or Seyfert II galaxies whose AGN
radio luminosities appear to be less than half their total radio
luminosities.

\begin{deluxetable}{r r r r r}
\tablecaption{Source classification matrix\label{tab:table2}}
\tablewidth{0pt}
\tablehead{
  \colhead{\hphantom{x}} & \colhead{~~~~A} & \colhead{~(A)~} &
  \colhead{~~~~~~S~} & \colhead{~~(S)} 
  }

\startdata
  Aa & 644 & 32 & 6 & 8 \\
  Aa\rlap{?} & 27  &  5 & 1 & 5 \\
  Aae & 101 & 18 & 27 & 15 \\
  Aae\rlap{?} & 24 & 10 & 27 & 10 \\
  Ae & 41 & 28 & 19 & 16 \\
  Ae\rlap{?} & 10 & 17 & 23 & 5 \\
  SF & 17 & 48 & 2018 & 103 \\
  SF\rlap{?} & 3 & 12 & 64 & 18 \\
  ?~ & 21 & 5 & 33 & 5 
\enddata

\end{deluxetable}

\subsection{1.4 GHz Nearby Galaxy Counts}\label{sec:counts}

\begin{figure}[ht]
\includegraphics[trim=70 130 0 120,scale=0.64]{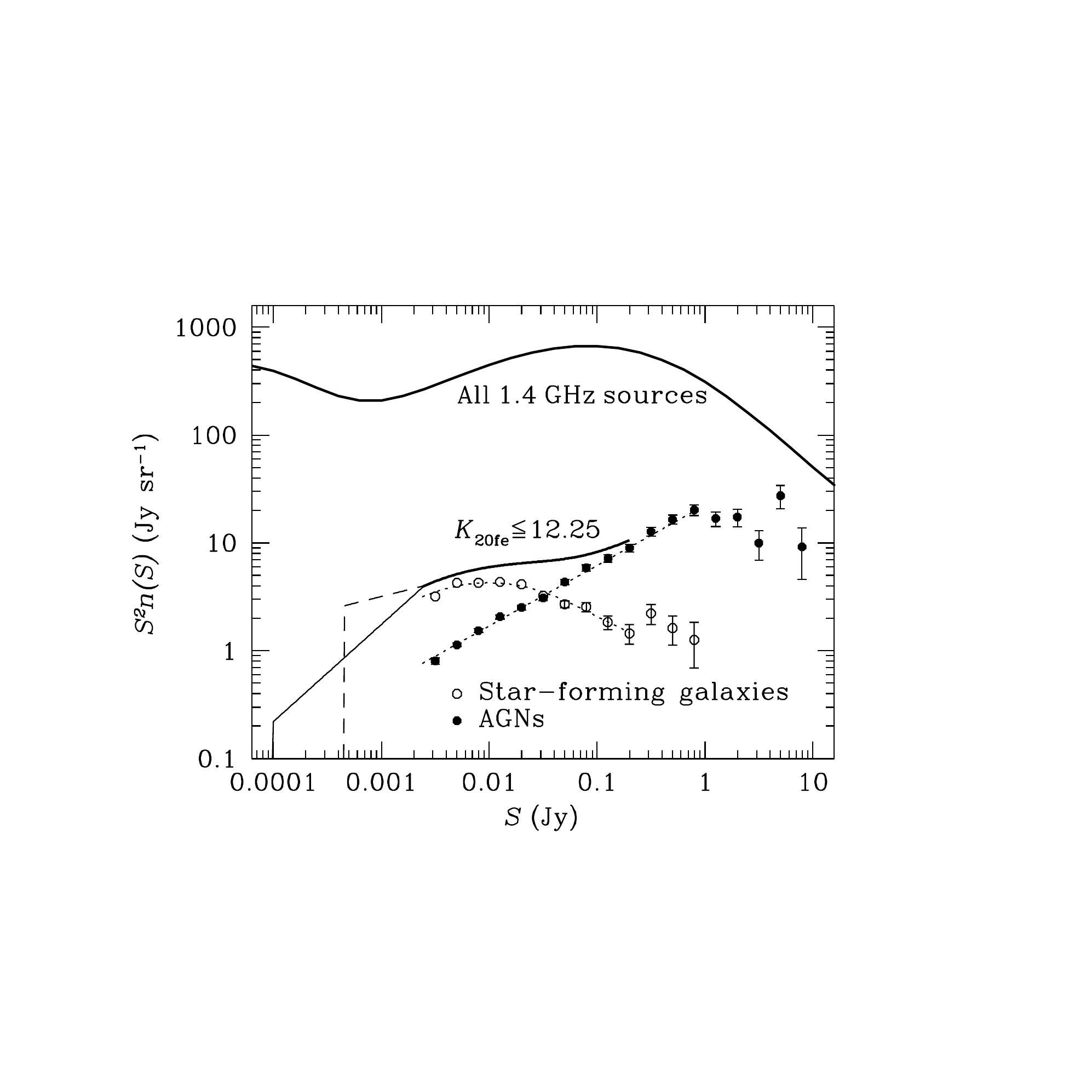}
\caption{The brightness-weighted 1.4~GHz differential counts $S^2
  n(S)$ are plotted separately for all radio sources
\citep{con84} (solid curve) and
  for 2MASX/NVSS galaxies brighter than $k_\mathrm{20fe} = 12.25$
  whose radio sources are powered by stars (open circles) or AGNs
  (filled circles).  $S^2 n(S)$ is proportional to the contribution
  per decade of flux density to the 1.4~GHz sky brightness
  temperature.  The light straight line matching the $P(S_\mathrm{p})$
  distribution (Figure~\ref{fig:figure8}) suggests that most galaxies
  with $k_\mathrm{20fe} \leq 12.25$ have 1.4 GHz flux densities $S
  \gtrsim 0.1 \mathrm{~mJy}$.  Abscissa: 1.4~GHz flux density $S$
  (Jy).  Ordinate: Brightness-weighted differential count $S^2 n(S)$
  (Jy~sr$^{-1}$).
\label{fig:figure7}}
\end{figure}

The differential source count $n(S) dS$ is the number of sources per
steradian with flux densities between $S$ and $S+dS$.  The
differential contribution $d\,T_\mathrm{B}$ of radio sources between $S$
and $S + d\,\log(S)$ to the Rayleigh-Jeans sky brightness temperature
$T_\mathrm{b}$ is
\begin{equation}
 \frac {d\,T_\mathrm{b}}{d\,\log(S)} =
  S^2 n(S) \biggl[ \frac {\ln (10)c^2}{2 k_\mathrm{B} \nu^2} \biggr]~,
\end{equation}
where $k_\mathrm{B} \approx 1.38 \times 10^{-23} \mathrm{~J~K}^{-1}$
is the Boltzmann constant.
Figure~\ref{fig:figure7} is a logarithmic plot comparing the
brightness-weighted 1.4~GHz counts $S^2 n(S)$ for all extragalactic
sources
\citep{con84} (upper solid curve), all 2MASX/NVSS sources with
$k_\mathrm{20fe} \leq 12.25$ and $S_\mathrm{1.4~GHz} \geq 2.45
\mathrm{~mJy}$ (lower solid curve), 2MASX/NVSS sources powered
primarily by star formation (open circles), and 2MASX/NVSS sources
powered by AGNs (filled circles). Below $S \approx 0.1 \mathrm{~Jy}$
the nearby ($z \lesssim 0.1$) 2MASX/NVSS sources contribute $\lesssim
1$\% of the total radio-source background.

The 2MASX catalog is quite complete for $k_\mathrm{20fe} \leq 12.25$
and bright 2MASX galaxies have a nearly static Euclidean source count
$n(S) \propto S^{-5/2}$ \citep{jar04}.  Radio sources powered by star
formation typically have relatively low absolute spectral luminosities
$L \sim 10^{22} \mathrm{~W~Hz}^{-1}$ \citep{con02} yielding flux
densities $S \sim 0.01 \mathrm{~Jy}$ at the distances $d \sim 100
\mathrm{~Mpc}$ ($cz \sim 7,000 \mathrm{~km~s}^{-1}$) typical of
2MASX/NVSS SFGs.  At flux densities $S > 0.01 \mathrm{~Jy}$ the
volume accessible to the 2MASX/NVSS sample of SFGs is almost
completely IR-limited, so their radio source count should also be
nearly Euclidean: $S^2 n(S) \propto S^{-1/2}$ as suggested by the
dotted line connecting the open points in Figure~\ref{fig:figure7}.
At lower flux densities the 2MASX/NVSS sample depth becomes more
radio-limited, so the contribution by SFGs to $S^2 n(S)$ flattens out
and eventually turns over.  Nearby 2MASX/NVSS radio AGNs are typically
more luminous than SFGs by factors of $\sim 10^{2.5}$ \citep{con02},
so their source count is IR-limited and nearly Euclidean only for the
small number of AGNs stronger than $S \sim 1 \mathrm{~Jy}$.  Nearly
all 2MASX/NVSS AGNs are much weaker and the volume sampled is strongly
limited by the radio sensitivity limit.  The dotted straight line
fitted to the AGN count (filled circles) in Figure~\ref{fig:figure7}
has a slope $d \,\log [S^2 n(S)] / d \log(S) \approx +0.6$ for $S \ll
1 \mathrm{~Jy}$.  This slope is close to the slope $\alpha \approx
+0.5$ of the 1.4~GHz spectral power density function
$U_\mathrm{dex}(L)$ at spectral luminosities $L \ll 10^{24}
\mathrm{~W~Hz}^{-1}$ (see Section~\ref{subsec:agns}).

\begin{figure}[ht]
  \includegraphics[trim=25 120 0 110,scale=0.58]{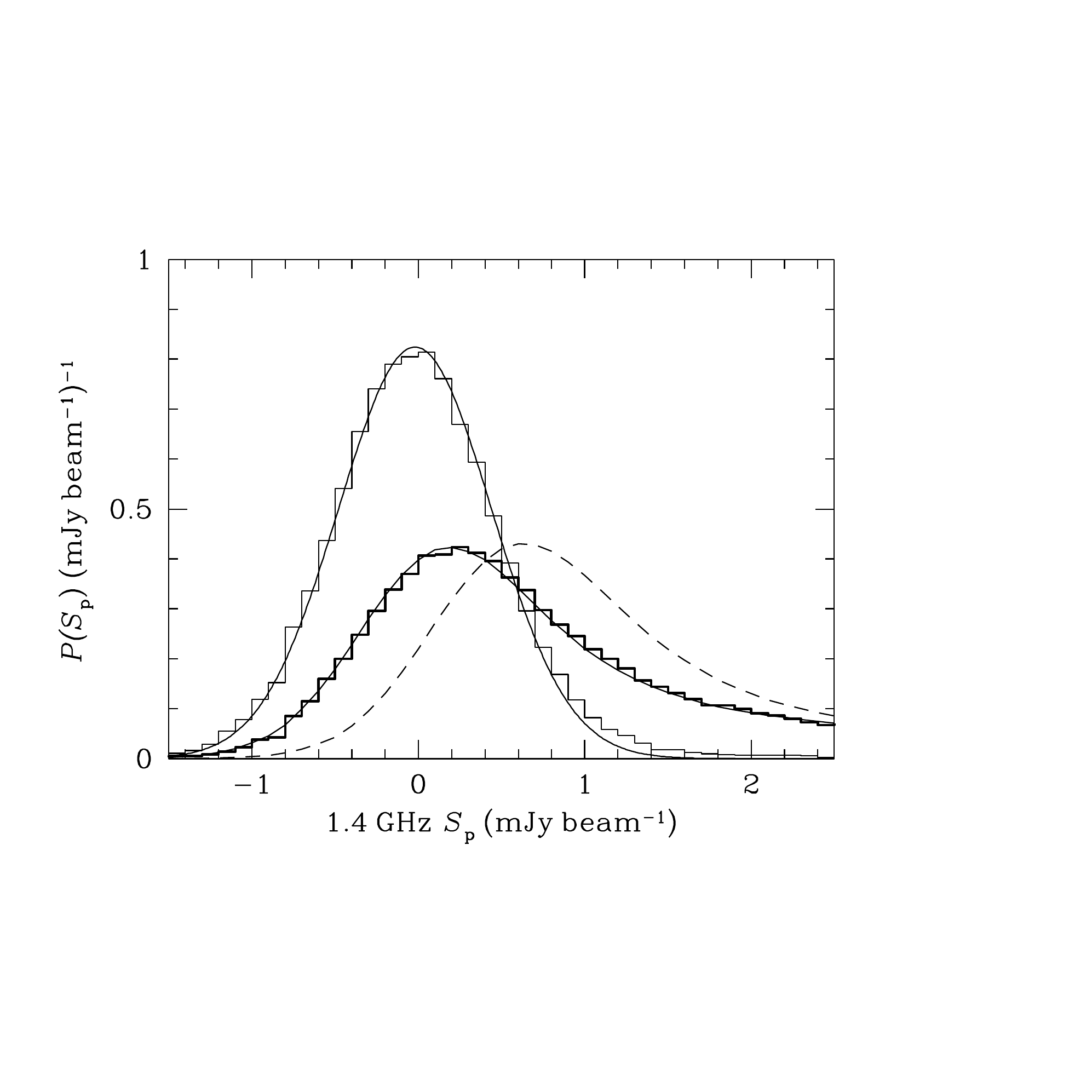}
  \caption{The distribution of NVSS peak flux densities $S_\mathrm{p}$
    at the positions of all 2MASX sample galaxcies is plotted as the
    heavy histogram, and the light histogram shows the corresponding
    peak flux densities at ``blank'' positions offset $1^\circ$ to the
    north.  The source count indicated by the light solid line in
    Figure~\ref{fig:figure7} yields the best fits (solid curves) to
    these histogrrams.  The dashed curve that doesn't fit the data is
    the $P(S_\mathrm{p})$ distribution corresponding to the dashed
    source count in Figure~\ref{fig:figure7}.
    \label{fig:figure8}}
\end{figure}

Only 15,043 of the 55,288 (27\%) 2MASX galaxies brighter than
$k_\mathrm{20fe} = 12.25$ contain NVSS sources stronger than 2.45~mJy
at 1.4~GHz.  However, the normalized probability distribution
$P(S_\mathrm{p})$ of NVSS peak flux densities $S_\mathrm{p}$ at the
positions of all 55,288 2MASX sample galaxies (heavy histogram in
Figure~\ref{fig:figure8}) constrains the radio source count well below
2.45~mJy.  The light histogram shows the matching distribution of peak
flux densities at ``blank'' positions offset $1^\circ$ to the north.
It is well fit by a noise Gaussian with rms width $\sigma = 0.47
\mathrm{~mJy~beam}^{-1}$ (light curve) plus a thin positive-going
tail produced by background radio sources.  The NVSS on-source
distribution is the convolution of the off-source distribution and the
peak flux density distribution of 2MASX galaxies.  To the extent that
most weak 2MASX radio sources are unresolved in the $\theta =
45\arcsec$ FWHM NVSS beam, $S_\mathrm{p} \approx S$ and
\begin{equation}
  \Omega \, n(S)\, dS = N\,P(S_\mathrm{p})\,d S_\mathrm{p}~.
\end{equation}
For any $S^2 n(S)$ (Figure~\ref{fig:figure7}) we can solve for
\begin{equation}
  P(S_\mathrm{p}) = \biggl( \frac{\Omega}{N S^2}\biggr) S^2 n(S)~.
\end{equation}
To the degree that the weighted differential count can be approximated
by a power law $S^2n(S) \approx kS^\gamma$ over the flux-density range
$S_1$ to $S_2$, the number of sources $\Delta N$ with flux densities
between $S_1$ and $S_2$ is
\begin{equation}\label{eqn:integral}
  \Delta N \approx \Omega \int_{S_1}^{S_2} k S^{\gamma - 2}\,dS =
  \biggl( \frac {\Omega k}{1 - \gamma} \biggr)
  (S_1^{\gamma-1} - S_2^{\gamma - 1})~.
\end{equation}
Among the $N = 55,288$ 2MASX galaxies brighter than $k_\mathrm{20fe} =
12.25$ in $\Omega = 7.016 \, \mathrm{sr}$, there are $\Delta N =
40,245$ with peak flux densities  $S_2 < 2.45 \mathrm{~mJy}$.

Equation~\ref{eqn:integral} is an integral constraint on $S_1$ (there
can't be more 2MASX/NVSS radio sources than 2MASX galaxies) as a
function of the other two source-count variables $k$ and $\gamma$.
Continuity of the direct source count $S^2 n(S) \approx 4
\mathrm{~sr}^{-1}$ at $S \approx 0.003 \mathrm{~Jy}$
(Figure~\ref{fig:figure7}) fixes $k$ for any $\gamma$.  The best value
of the remaining unknown $\gamma$ is the one that yields the best fit
to the heavy $P(S_\mathrm{p}$ histogram in Figure~\ref{fig:figure8}.
For example, the power-law extrapolation $S^2n(S) = 17.9 S^{0.25}$ of
the direct count of 2MASX galaxies with $k_\mathrm{20fe} \leq 12.25$
above $S = 2.45 \mathrm{~mJy}$ yields the dashed line in
Figure~\ref{fig:figure7} that must break at $S_1 \approx
0.46 \mathrm{~mJy}$ lest the number of radio sources exceed the number
of galaxies.  However, this solution is unsatisfactory because it
predicts the dashed $P(S_\mathrm{p})$ distribution in
Figure~\ref{fig:figure8} that is shifted far to the right of the
observed distribution (heavy histogram).

The best power-law fit is $S^2 n(S) \approx 880 S^{-0.90}$ cutting off
at $S_1 \approx 0.0001 \mathrm{~Jy}$, as shown by the light straight
line in Figure~\ref{fig:figure7} and the continuous curve that is a
good match to the heavy histogram in Figure~\ref{fig:figure8}.  We
conclude that (1) the brightness-weighted count of nearby sources
fainter than 2.45~mJy must converge rapidly, and (2) the NVSS is
sufficiently sensitive to have detected individually those sources
that contribute most of the low redshift ($z \lesssim 0.05$) sky
brightness.  It also appears that most 2MASX galaxies brighter than
$k_\mathrm{20fe} = 12.25$ are radio sources stronger than $S \sim
0.1$~mJy and should be detectable above the planned $S_\mathrm{p}
\approx 0.05 \mathrm{~mJy}$ sensitivity limit of the upcoming EMU
survey \citep{nor11}.

\section{The Spectroscopically Complete Subsample}\label{sec:zdist}

All but 7 of the 9,517 2MASX/NVSS galaxies with $k_\mathrm{20fe} \leq
11.75$ have published spectroscopic velocities $cz$ or
new velocities reported in Appendix~\ref{sec:redshiftappendix}.


To estimate accurate distances from the observed heliocentric
velocities, we first converted the heliocentric velocities $v \equiv
cz$ to velocities $v_\mathrm{CMB}$ in the frame of the cosmic
microwave background (CMB) using
\begin{equation}
  \begin{split}
    {v}_{\rm CMB} &= v + {v}_{\rm apex}[\sin(b)\sin(b_{\rm apex}) \\
      &+\cos(b)\cos(b_{\rm apex})\cos(l-l_{\rm apex})]~,
  \end{split}
\end{equation}
where
 $(l_{\rm apex},b_{\rm
  apex})=(264.14^{\circ},48.26^{\circ})$ and 
$v_\mathrm{apex} = 371.0 \mathrm{~km~s}^{-1}$ \citep{fix96}.

Large-scale structures (e.g., galaxy clusters) cause additional
deviations from the local Hubble flow that depend on position and
redshift. We adopted the local bulk flow models of \cite{car15} to
correct  for this effect.

The \cite{car15} model of the peculiar velocity field is given as a
$257^3$ voxel cube in right-handed galactic Cartesian coordinates with
$i, j,$ and $k$ indices corresponding to galactic $X, Y, Z$ in
$\mathrm{Mpc~h}^{-1}$, with the $i$ index running fastest. The voxel
centers run from $-200h^{-1}$ to $200 h^{-1} \mathrm{~Mpc}$ so the
voxel spacing is $1.5625 h^{-1} \mathrm{~Mpc}$. The $i,j,k$ indices
can be converted to Cartesian galactic coordinates using
\begin{eqnarray}
  X &= (i-128)\times400./256.\label{eq:x}\\
  Y &= (j-128)\times400./256.\label{eq:y}\\
  Z &= (k-128)\times400./256.\label{eq:z}
\end{eqnarray}
The center of the cube [128,128,128] represents the Local
Group. All peculiar velocities $v_\mathrm{pec}$ in the cube are
relative to the CMB and are generated by the galaxy density models of
\cite{car15}, which depend upon the cosmological density of matter
$\Omega_\mathrm{m}$ (taken to be 0.3 in this study) and the bias $b^*$
of an $L^*$ galaxy.  Along the radial line to each galaxy, we solved
\begin{equation}
  H_0\,r+v_{\rm pec}(r)=v_\mathrm{CMB}\label{eq:vel}
\end{equation}
to obtain the corrected galaxy velocity $v_\mathrm{c} = c
z_\mathrm{c}$.

\begin{figure}[!ht]
\includegraphics[trim=45 255 0 160,scale=0.56]{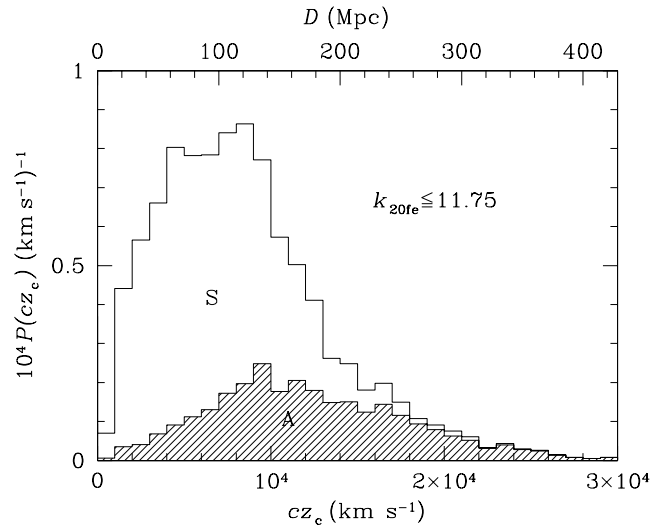}
\caption{All but 7 of the 9,517 2MASX/NVSS galaxies with
  $k_\mathrm{20fe} \leq 11.75$ have spectroscopic redshifts.
  Histograms of their corrected velocities $cz_\mathrm{c}$ are shown
  separately for galaxies whose radio sources are powered by AGNs (A)
  or by stars (S). The corresponding Hubble distances for $H_0 = 70
  \mathrm{~km~s}^{-1} \mathrm{~Mpc}^{-1}$ are also shown. Lower
  Abscissa: $cz_\mathrm{c}\,\mathrm{(km~s}^{-1}$).  Upper Abscissa:
  Hubble distance $D \mathrm{~(Mpc)}$ Ordinate:
  $10^4\,P(cz_\mathrm{c})\, \mathrm{(km~s}^{-1}\mathrm{)}^{-1}$.
\label{fig:figure9}}
\end{figure}

The histograms in Figure~\ref{fig:figure9} show the normalized
probability distributions $P(cz)$ of corrected velocities $cz_\mathrm{c}$ for
galaxies whose radio sources are powered by stars (unshaded area) or
by AGNs (shaded).  Star-forming galaxies outnumber AGNs by a ratio of
$>$2:1 in this sample of bright galaxies, especially at lower
redshifts.  The median velocity of star-forming galaxies is only
$\langle cz_\mathrm{c} \rangle \approx 0.6 \times 10^4 \mathrm{~km~s}^{-1}$,
about half the median velocity $\langle cz_\mathrm{c} \rangle \approx 1.2 \times
10^4 \mathrm{~km~s}^{-1}$ of galaxies with AGN-powered radio sources.

\section{Local 1.4 GHz Luminosity Functions}\label{sec:lumf}

The local luminosity function specifies the mean space density of
galaxies in the nearby universe as a function of spectral luminosity.
The universe is evolving and homogeneous on large scales, so the local
luminosity function more usefully represents the universal average
space density during the present epoch rather than our particular
location in space.  We derived separate 1.4\,GHz luminosity functions
for radio sources powered by star formation and by AGNs.  They are
today's benchmarks for comparing with higher-redshift samples to
constrain models for the cosmological evolution of star formation and
AGN activity.

The 2MASX/NVSS spectroscopic subsample should yield reliable 1.4~GHz
luminosity functions because redshifts are available for nearly all
galaxies and it is complete for galaxies that are brighter than
$k_{\rm 20fe} = 11.75$ at $\lambda = 2.16\,\mu\mathrm{m}$, stronger
than $S = 2.45$\,mJy at 1.4~GHz, and lie in the solid angle $\Omega =
7.016$~sr defined by J2000 $\delta > -40^\circ$ and $\vert b \vert >
20^\circ$. The 2MASX catalog itself is actually complete and reliable
for galaxies much fainter than $k_{\rm 20fe} = 11.75$; our
magnitude limit reflects the availability of spectroscopic redshifts.
The NVSS sample includes all sources with catalog flux densities $S
\geq 2.5 \mathrm{~mJy}$.  However, the NVSS catalog flux densities are
rounded to the nearest 0.1\,mJy, so sources as faint as $S =
2.45 \mathrm{~mJy}$ are listed as having $ S = 2.5 \mathrm{~mJy}$ in
the catalog.  The spectroscopic redshifts are from \citet{huc12} or
from new optical and NIR spectra obtained with the Apache Point
Observatory (APO) 3.5~m telescope, as described in
Appendix~\ref{sec:redshiftappendix}.

The 1.4~GHz spectral luminosity function $\rho(L)\,dL$ is defined as
the space density of sources with 1.4~GHz spectral luminosities
between $L$ and $L+dL$. The range of spectral luminosities spanned by
galaxies is so large that it is convenient to define a logarithmic
spectral luminosity function
\begin{equation}
\rho_\mathrm{dex}(L) \equiv \rho(L) \frac{dL}{d\log(L)} = \ln(10)L\rho(L)
\end{equation}
specifying the space density of sources per decade of spectral
luminosity.

The 1.4\,GHz spectral luminosity of each source is
\begin{equation}
L = 4 \pi D_L^2 S_{1.4} (1+z)^{-(1+\alpha)}~,
\end{equation}
where $D_\mathrm{L} = (1+z)D_\mathrm{C}$ is the luminosity distance to
the radio source, $D_\mathrm{C}$ is the comoving distance, $S_{1.4}$
is the 1.4\,GHz NVSS flux density, and $\alpha = -0.7$ is the mean
spectral index ($S \propto \nu^{\alpha}$) of sources selected at
1.4\,GHz \citep{con84}. The absolute magnitude $K_\mathrm{20fe}$ was
calculated using
\begin{equation}
  K_{\rm 20fe} = k_{\rm 20fe} -
  5 \log \biggl( \frac {D_L} {10\mathrm{~pc}} \biggr) - k(z)~,
\end{equation}
where $k(z) = -6.0\,\log(1+z)$ is the $k$-correction that is
independent of galaxy type and valid for all $z \leq 0.25$
\citep{koc01}.  We therefore used $z_\mathrm{max} = 0.25$ as the
maximum possible redshift when calculating $V_\mathrm{max}$ values for
our galaxy sample.


\subsection{Maximum Redshifts}

\begin{figure}[h]
  \centering
  \includegraphics[trim={0.5cm 1.5cm 0.5cm 4.5cm},clip,width=0.49\textwidth]
                  {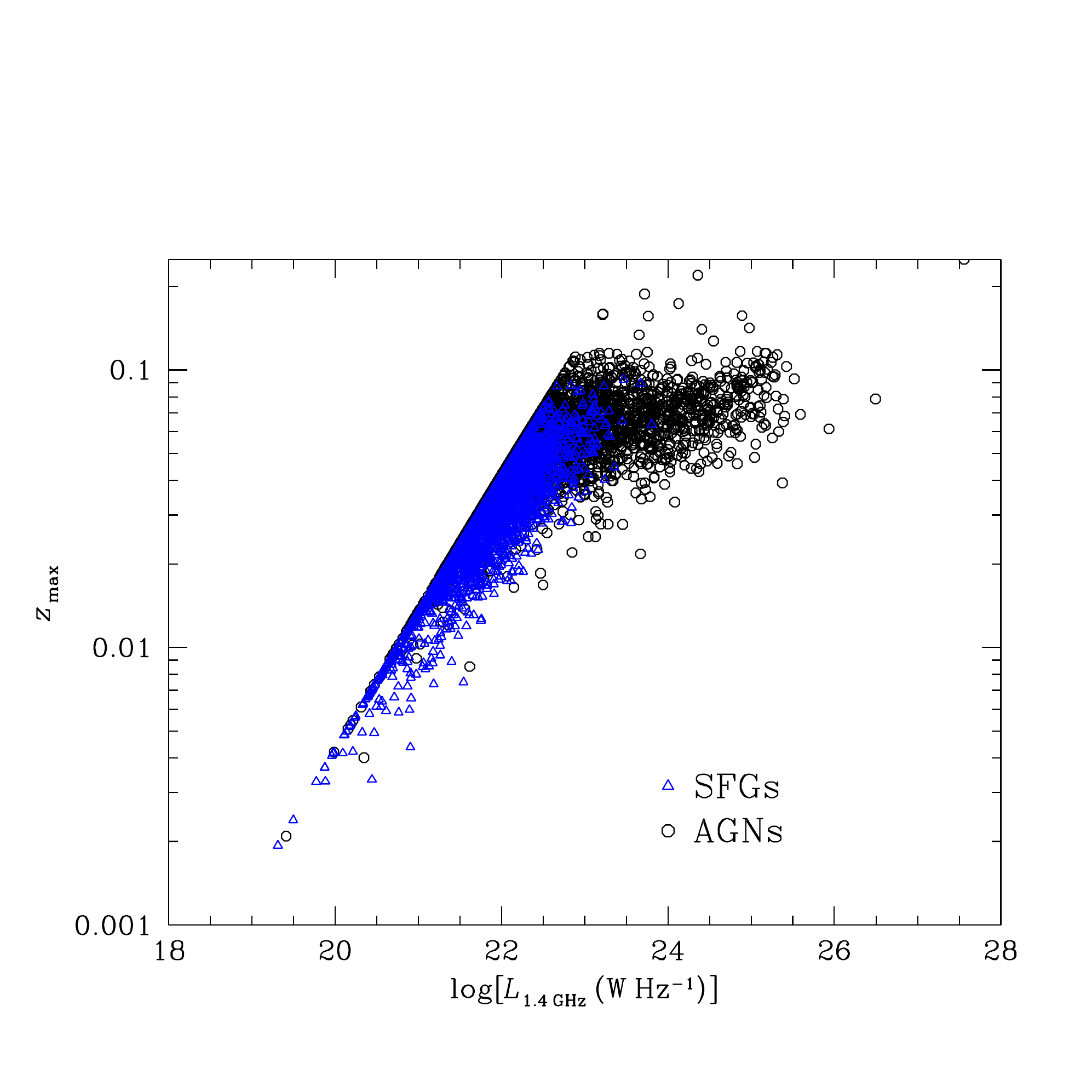}
  \caption{Maximum redshifts out to which a 2MASX/NVSS source could be
    moved and remain in the spectroscopic subsample, as a function of
    1.4~GHz luminosity. The maximum redshifts for star-forming
    galaxies (blue triangles) and AGNs (black circles) are shown as a
    function of 1.4\,GHz radio luminosity. \label{fig:figure10}}
\end{figure}

\begin{figure}[h]
  \centering
  \includegraphics[trim={0.5cm 1.5cm 0.5cm 4.5cm},clip,width=0.49\textwidth]
                  {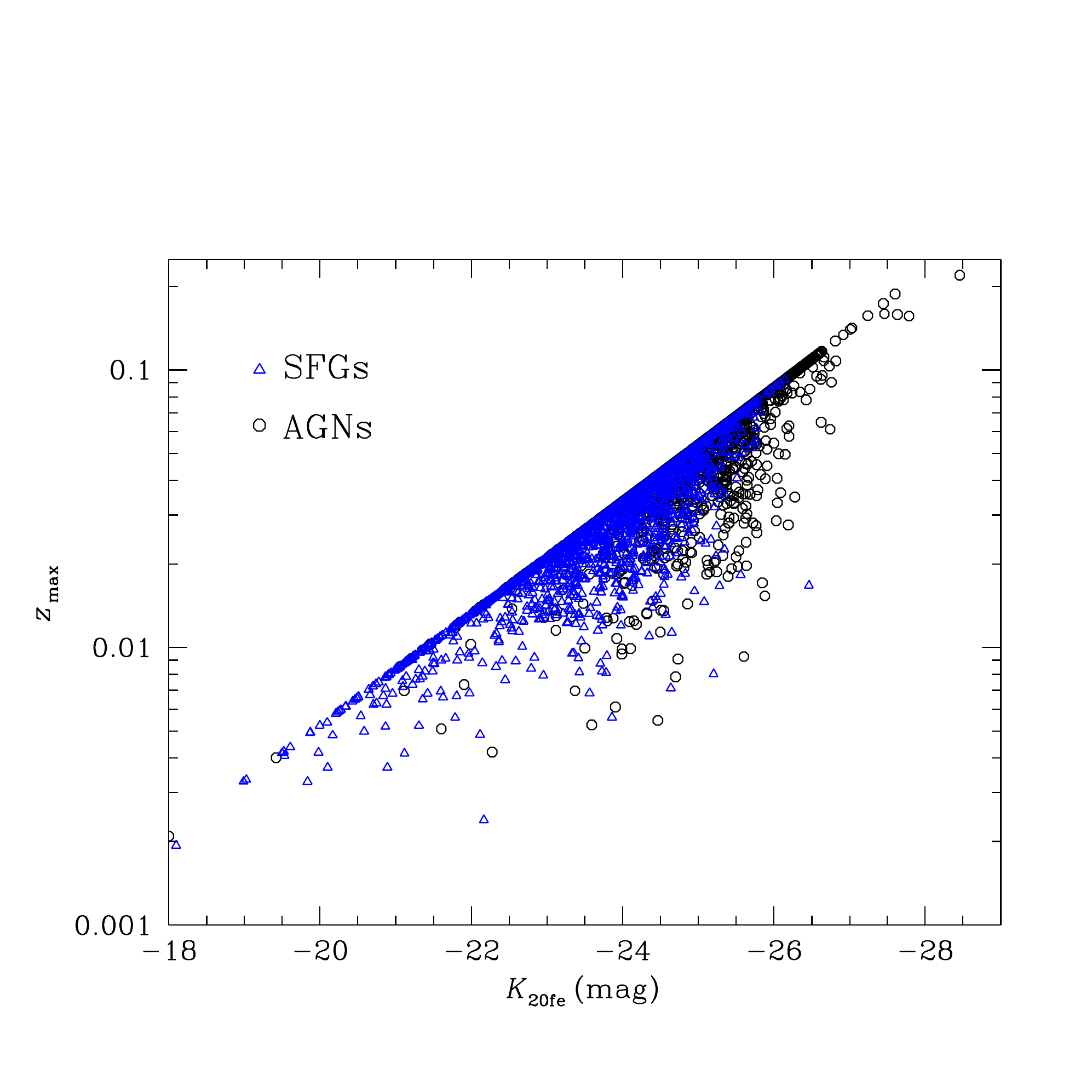}
  \caption{Maximum redshifts out to which a 2MASX/NVSS source could be
    moved and remain in the spectroscopic subsample, as a function of
    $K_\mathrm{20fe}$. The maximum redshifts for star-forming galaxies
    (blue triangles) and AGNs (black circles) are shown as a function
    of $K_\mathrm{20fe}$. \label{fig:figure11}}
\end{figure}

We plotted the maximum redshifts $z_\mathrm{max}$ at which galaxies
could remain in our spectroscopic subsample as functions of both 1.4
GHz spectral luminosity (Figure~\ref{fig:figure10}) and $\lambda =
2.16~\mu\mathrm{m}$ absolute magnitude $K_\mathrm{20fe}$
(Figure~\ref{fig:figure11}).  Star-forming galaxies (blue triangles)
span the majority of the redshift range ($0.0017 \lesssim z \lesssim
0.12$) and absolute magnitudes ($ -18 \gtrsim K_\mathrm{20\rm fe}
\gtrsim -27$), but are mainly limited to 1.4~GHz luminosities
$\log[L(\mathrm{W~Hz}^{-1})] \lesssim 23$. AGNs (black circles)
dominate both the high radio luminosities and absolute magnitudes
$K_\mathrm{20fe}$, but are fewer in number at the lowest redshifts ($z
\lesssim 0.007$).

\subsection{Correction for Local Overdensity}

Galaxies cluster and we are located in a galaxy, so the space density
$\rho_\mathrm{P}$ of the nearest galaxies is somewhat greater than the mean
density $\rho$ of all galaxies. We corrected our local luminosity
function for the local overdensity within a distance $r$ using the
equation \citep{fis94}
\begin{equation}
  \frac{\rho_\mathrm{P}}{\rho} =
  1 + \frac{3}{3-\gamma}\biggl(\frac{r_0}{r}\biggr)^{\gamma}~.
\end{equation}
The correlation function parameters of {\it IRAS} galaxies are
appropriate for describing the clustering of the 2MASX/NVSS galaxies;
they are $r_0 = 3.76 \, h^{-1}$ and $\gamma = 1.66$ for $r < 20 \,
h^{-1}\,{\rm Mpc}$ \citep{fis94}. For $r < 20\,h^{-1}\,{\rm Mpc} $,
the volume within $r$ was multiplied by $\rho_\mathrm{P}/\rho$ in our
calculation of $V_{\rm max}$; otherwise the volume was left
unchanged. To minimize uncertainties introduced by large values of
this correction for local-group galaxies, we excluded 48 galaxies in
the volume with $r < 5 \mathrm{~Mpc}$ (corrected $cz_\mathrm{c} < 350
\mathrm{~km~s}^{-1}$, or $z \lesssim 0.0017$) when calculating
luminosity functions. Only about 5\% of our sample galaxies have $r <
20\,h^{-1} \mathrm{~Mpc} \approx 29 \mathrm{~Mpc}$, so correcting for
the local overdensity has only a small effect on our radio luminosity
functions.

\subsection{The Distribution of $V/V_\mathrm{max}$}

If the radio sources are randomly distributed throughout the corrected
volume, the distribution of $V/V_{\rm max}$ should be uniform in the
interval [0,1] and have a mean $\langle V/V_\mathrm{max} \rangle
\approx 0.5$. The standard deviation of a uniform distribution on the
interval [0,1] is $12^{-1/2}$, so the rms uncertainty in $\langle
V/V_{\rm max}\rangle$ of $N \gg 1$ radio sources is $\sigma \approx
(12N)^{-1/2}$. A statistically significant departure from a uniform
distribution with mean $0.5$ may indicate one or more of the
following: poor corrections for the local overdensity, incorrect
sample limits, strong clustering, or monotonic evolution of sources
during the lookback times spanned by the sample volume. For the
2MASX/NVSS galaxies used to determine the local luminosity function,
the 6699 star-forming galaxies have $\langle V/V_{\rm max} \rangle=
0.500 \pm 0.004$, the 2763 AGNs have $\langle V/V_{\rm max} \rangle =
0.494 \pm 0.005$, and all 9462 galaxies have $\langle V/V_{\rm max}
\rangle = 0.497 \pm 0.003$.  Thus our $\langle
V/V_\mathrm{max}\rangle$ test detects no monotonic evolution during
the sample-limited lookback time $\tau \sim 1-2$~Gyr.

\begin{figure}[ht]
  \centering
  \includegraphics[trim={1cm 6.cm 0.5cm 1.5cm},clip,width=0.49\textwidth]
    {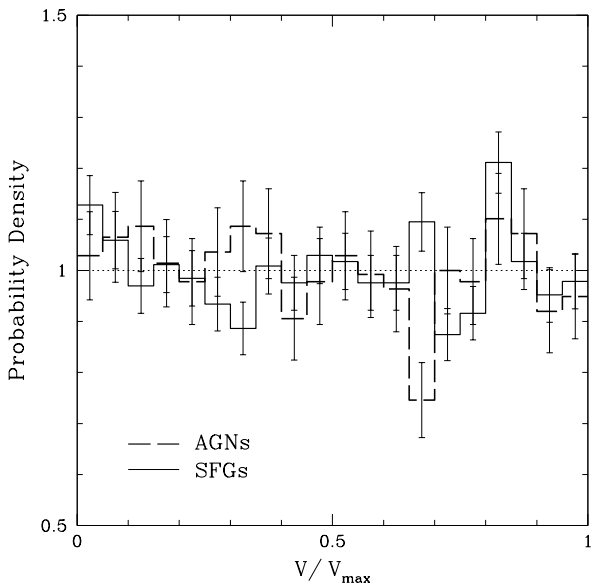}
  \caption{Binned distributions of $V/V_\mathrm{max}$ for star-forming
    galaxies (solid line) and AGNs (dashed line) with the rms
    uncertainties expected for randomly distributed galaxies.
    Abscissa: $V/V\mathrm{max}$.  Ordinate: Binned probability
    density. \label{fig:figure12}}
\end{figure}

The normalized probability densities of $V/V_{\rm max}$ in 20 bins of
width $\Delta (V/V_\mathrm{max}) = 0.05$ are plotted separately for
star-forming galaxies and AGNs in Figure \ref{fig:figure12}. The
$V/V_{\rm max}$ distribution for AGNs closely follows a uniform
distribution with a $\chi^2_{\nu} \approx 1.08$. In contrast, the
distribution for star-forming galaxies appears to deviate slightly,
with a $\chi^2_{\nu} \approx 2.04$, marginally significant at the
$\sim 0.01$ level. This slight deviation from a uniform distribution
can be mostly attributed to the peak in the bin of $V/V_{\rm max}$
from $0.80 - 0.85$. This is caused mainly by galaxies whose $z_{\rm
  max}$ is limited by radio luminosity rather than $K$ band magnitude.
This peak would be a marginally statistically significant $3.6\sigma$
bump for the star-forming galaxies if galaxies were distributed
randomly in space. However, our $V/V_{\rm max}$ fluctuations are
consistent with the statistical fluctuations expected in clustered
galaxy samples mildly exacerbated by the NVSS catalog flux-density
quantization.

\subsection{Luminosity Function Results}

\begin{figure}[ht]
  \centering
  \includegraphics[trim={1cm 1.5cm 0.5cm 3cm},clip,width=0.49\textwidth]
                  {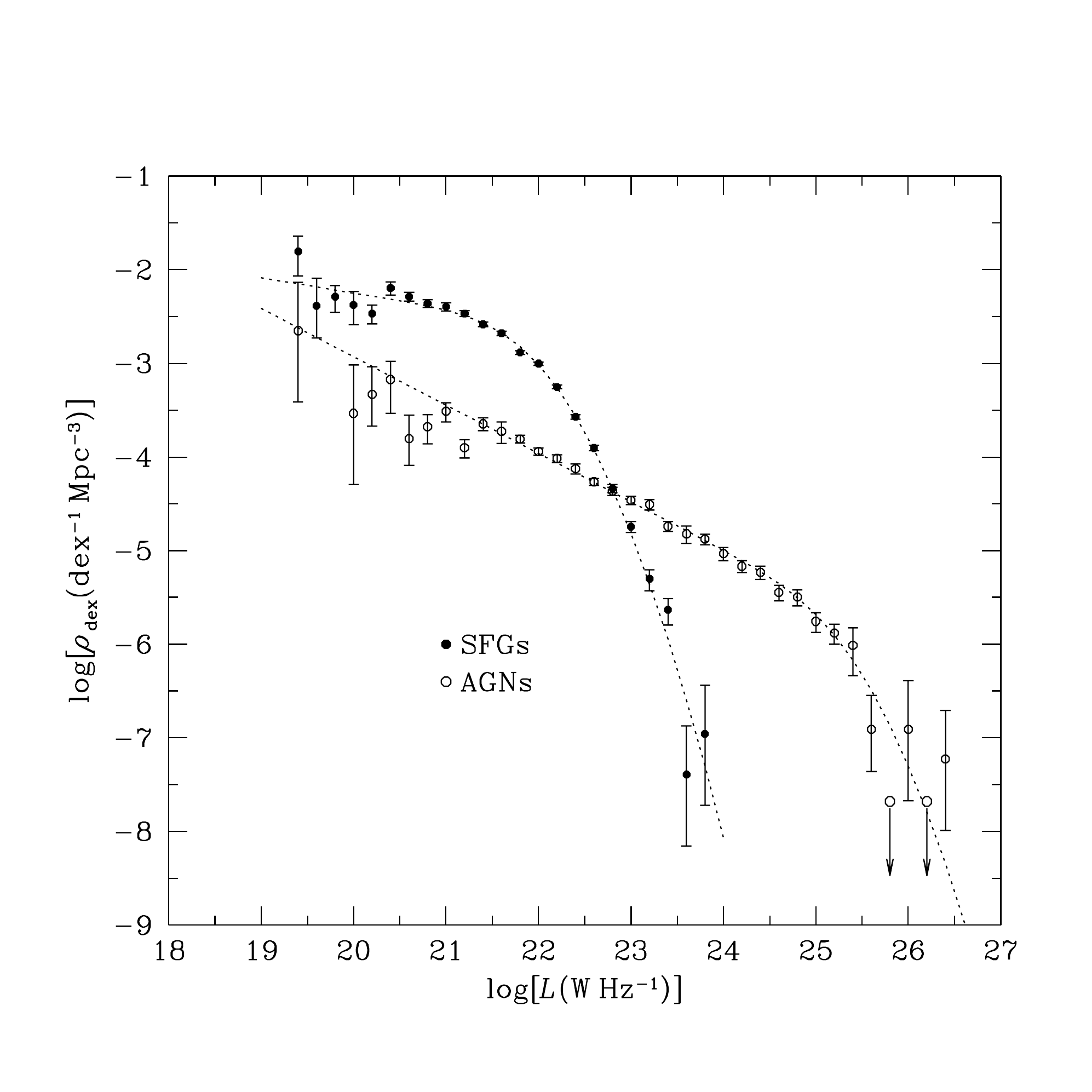}
  \caption{The 2MASX/NVSS 1.4\,GHz logarithmic luminosity functions
    for sources whose radio emission is dominated by star formation
    (filled circles) and AGNs (unfilled circles). The two dotted
    curves represent the \cite{sau90} and double power-law fits to the
    local luminosity functions of star-forming galaxies and AGNs,
    respectively. \label{fig:figure13}}
\end{figure}

We sorted our galaxies into luminosity bins of width $\Delta \log(L) =
0.2$ (5 bins per decade) centered on $\log[L(\mathrm{W~Hz}^{-1})]
= 19.4$ to 27.6 and calculated separate local luminosity functions of
star-forming galaxies and AGNs using
\begin{equation}
  \rho_\mathrm{dex} = 5 \sum_{i=1}^N \Biggl( \frac{1}{V_\mathrm{max}}\Biggr)_i~.
\end{equation}
Our 1.4\,GHz local luminosity functions
$\rho_\mathrm{dex}$ for star-forming galaxies and AGNs are listed in
Table \ref{tab:table3} and plotted in Figure \ref{fig:figure13}. The
listed errors are the rms Poisson counting errors for
independent galaxies
\begin{equation}
  \sigma = 5 \Biggl[ \sum_{i=1}^N \biggl( \frac {1} {V_\mathrm{max}} \Biggr)^2_i \Biggr]^{1/2}
\end{equation}
quadratically summed with a 3\% flux-scale
uncertainty.  If the number $N$ of galaxies in a luminosity bin is
small ($N < 5$), the quoted errors are the 84\% confidence limits
tabulated in \citet{geh86}.  Clustering and cosmic variance are
addressed in Section~\ref{sec:cosmicvar}.

The luminosity functions of SFGs and AGNs intersect at $\log[L(1.4
  \mathrm {~GHz})] \approx 22.7$ in agreement with the earlier result
of \cite{con02} and close to the $\log[L(\mathrm{W~Hz}^{-1})] \approx
22.9$ found by \citet{mau07} despite the different samples and
classification methods used. This crossover marks the 1.4~GHz spectral
luminosity below which star-forming galaxies outnumber AGNs within the
local universe.

\begin{deluxetable}{crcrc}
  \tabletypesize{\footnotesize}
  \tablewidth{0pt}
  \tablecolumns{5}
  \tablecaption{1.4 GHz Local Luminosity Functions ($h=0.70$)\label{tab:table3}}
  \tablehead{\colhead{} & \multicolumn{2}{c}{Star-forming Galaxies} & \multicolumn{2}{c}{\quad AGNs} \\
    \colhead{$\log\,L$} & \colhead{} & \colhead{$\log\,\rho_{\rm dex}$} & \colhead{} & \colhead{$\log\,\rho_{\rm dex}$} \\
    \colhead{$\rm (W\,Hz^{-1})$} & \colhead{$N$} & \colhead{(dex$^{-1}\,\rm Mpc^{-3}$)} & \colhead{$N$} & \colhead{(dex$^{-1}\,\rm Mpc^{-3}$)}}
  \startdata
    19.4 & 6 & $-1.81_{-0.26}^{+0.16}$ & 1 & $-2.65_{-0.76}^{+0.52}$ \\
    19.6 & 3 & $-2.39_{-0.34}^{+0.30}$ & 0 & \nodata \\
    19.8 & 10 &$ -2.29_{-0.17}^{+0.12}$ & 0 & \nodata \\
    20.0 & 11 & $-2.37_{-0.21}^{+0.14}$ & 1 & $-3.53_{-0.76}^{+0.52}$ \\
    20.2 & 21 & $-2.46_{-0.11}^{+0.09}$ & 3 & $-3.33_{-0.34}^{+0.30}$ \\
    20.4 & 59 & $-2.20_{-0.08}^{+0.07}$ & 5 & $-3.17_{-0.36}^{+0.20}$ \\
    20.6 & 103 & $-2.29_{-0.05}^{+0.05}$ & 4 & $-3.80_{-0.28}^{+0.25}$ \\
    20.8 & 147 & $-2.36_{-0.04}^{+0.04}$ & 9 & $-3.68_{-0.18}^{+0.13}$ \\
    21.0 & 244 & $-2.39_{-0.05}^{+0.04}$ & 22 & $-3.51_{-0.12}^{+0.09}$ \\
    21.2 & 411 & $-2.47_{-0.03}^{+0.03}$ & 22 & $-3.90_{-0.11}^{+0.09}$ \\
    21.4 & 584 & $-2.58_{-0.02}^{+0.02}$ & 65 & $-3.65_{-0.07}^{+0.06}$ \\
    21.6 & 823 & $-2.68_{-0.02}^{+0.02}$ & 85 & $-3.73_{-0.13}^{+0.10}$ \\
    21.8 & 975 & $-2.88_{-0.02}^{+0.02}$ & 171 & $-3.81_{-0.04}^{+0.04}$ \\ 
    22.0 & 1124 & $-3.00_{-0.02}^{+0.02}$ & 216 & $-3.94_{-0.04}^{+0.04}$ \\
    22.2 & 893 & $-3.25_{-0.02}^{+0.02}$ & 281 & $-4.02_{-0.04}^{+0.04}$ \\
    22.4 & 624 & $-3.57_{-0.02}^{+0.02}$ & 280 & $-4.12_{-0.06}^{+0.05}$ \\
    22.6 & 368 & $-3.90_{-0.03}^{+0.03}$ & 286 & $-4.26_{-0.04}^{+0.03}$ \\
    22.8 & 168 & $-4.34_{-0.05}^{+0.04}$ & 239 & $-4.37_{-0.05}^{+0.04}$ \\
    23.0 & 87 & $-4.74_{-0.06}^{+0.05}$ & 209 & $-4.46_{-0.05}^{+0.04}$ \\
    23.2 & 30 & $-5.30_{-0.13}^{+0.10}$ & 184 & $-4.51_{-0.06}^{+0.05}$ \\
    23.4 & 13 & $-5.63_{-0.17}^{+0.12}$ & 133 & $-4.74_{-0.06}^{+0.05}$ \\
    23.6 & 1 & $-7.39_{-0.76}^{+0.52}$ & 97 & $-4.82_{-0.10}^{+0.08}$ \\
    23.8 & 1 & $-6.96_{- 0.76}^{+0.52}$ & 103 & $-4.88_{-0.06}^{+0.05}$ \\
    24.0 & 0 & \nodata & 69 & $-5.03_{-0.08}^{+0.06}$ \\
    24.2 & 0 & \nodata & 70 & $-5.17_{-0.07}^{+0.06}$ \\
    24.4 & 0 & \nodata & 59 & $-5.23_{-0.08}^{+0.06}$ \\
    24.6 & 0 & \nodata & 41 & $-5.45_{-0.09}^{+0.07}$ \\
    24.8 & 0 & \nodata & 41 & $-5.50_{-0.10}^{+0.08}$ \\
    25.0 & 0 & \nodata & 30 & $-5.76_{-0.12}^{+0.09}$ \\
    25.2 & 0 & \nodata & 24 & $-5.88_{-0.12}^{+0.10}$ \\
    25.4 & 0 & \nodata & 8 & $-6.01_{-0.33}^{+0.18}$ \\
    25.6 & 0 & \nodata & 2 & $-6.91_{-0.45}^{+0.37}$ \\
    25.8 & 0 & \nodata & 0 & $\lesssim -7.68$ \\
    26.0 & 0 & \nodata & 1 & $-6.91_{-0.76}^{+0.52}$ \\
    26.2 & 0 & \nodata & 0 & $\lesssim -7.68$ \\
    26.4 & 0 & \nodata & 1 & $-7.23_{-0.76}^{+0.52}$ \\
    26.6 & 0 &  \nodata     &    0    &  \nodata \\
    26.8 & 0 &  \nodata     &    0    &  \nodata \\
    27.0 & 0 &  \nodata     &    0    &  \nodata \\
    27.2 & 0 &  \nodata     &    0    &  \nodata \\
    27.4 & 0 &  \nodata     &    0    &  \nodata \\
    27.6 & 0 &  \nodata     &    1    &  $-8.68_{-0.76}^{+0.52}$
    \enddata
  \end{deluxetable}

\subsubsection{Star-Forming Galaxies}

The FIR/radio correlation shows that the radio and FIR luminosities of
star-forming galaxies are nearly proportional, so their logarithmic
radio and FIR luminosity functions should be similar in
form. \cite{sau90} found that the FIR $(40\,\mu\rm m < \lambda <
120\,\mu\rm m)$ logarithmic luminosity function $\phi(L)$ derived from
seven large samples of IRAS sources is well fit by the parametric form
\begin{equation}\label{eqn:sau}
  \phi(L) = C \Biggl( \frac{L}{L_*} \Biggr)^{1-\alpha}
  \exp \Biggl[ - \frac {1} {2 \sigma^2} \log^2 \Biggl(1 +
    \frac {L}{L_*} \Biggr) \Biggr]
\end{equation}
that approaches a power law at with slope $(1-\alpha)$ when $L \ll L_*$
and falls like a Gaussian with $\log(L)$ when $L \gg L_*$.

Equation~\ref{eqn:sau} also fits the local 1.4 GHz logarithmic
luminosity function $\rho_\mathrm{dex}(L)$ of star-forming galaxies
very well, in congruence with the FIR/radio correlation, and it gives a
better fit than the \citet{sch76} luminosity function. The dotted
curve fitting the filled points in Figure \ref{fig:figure13} has the
best-fit parameters for the 2MASX/NVSS star-forming galaxies stronger
than $\log[L(\mathrm{W~Hz}^{-1})] = 19.3$: $C = 3.50 \times10^{-3}
  \,{\rm dex^{-1}\,Mpc^{-3}}, L_* = 1.9 \times 10^{21}\,{\rm
    W\,Hz^{-1}}, \alpha = 1.162$, and $\sigma = 0.558$.  Despite the
  good parametric fit to the data, we have not quoted errors on these
  four parameters because they are so highly correlated that they ``grossly
  overestimate the total acceptable volume of parameter space''
  \citep{sau90}. 

\citet{mau07} used a deeper ($k_\mathrm{s} < 12.75$) sample of 2MASX galaxies
identified with NVSS sources and having 6dF spectra in a smaller area
of sky ($\Omega \approx 2.16 \mathrm{~sr}$) to calculate
$\rho_\mathrm{dex}(L)$ in the luminosity range $19.8 \lesssim \log(L)
\lesssim 23.8$ for galaxies they classified as star-forming on the basis
of their optical spectra.  Their
fit to the \citet{sau90} form in Equation~\ref{eqn:sau} gave $C = 1.48
\pm 0.17 \times 10^{-3} \mathrm{~dex}^{-1} \mathrm{~Mpc}^{-3}$, $L_* =
1.5 \pm 0.5 \times 10^{21} \mathrm{~W~Hz}^{-1}$, $\alpha = 1.02 \pm
0.15$, and $\sigma = 0.60 \pm 0.04$.  Again, these four parameters are so
highly correlated that apparently significant differences between
their values and ours are not meaningful.  Direct comparisons of our binned
luminosity functions show that they agree within the expected errors after
cosmic variance (Section~\ref{sec:cosmicvar}) has been taken into account.

\subsubsection{AGNs}\label{subsubsec:agns}

For the high-luminosity bins in which no sources were detected, we are
able to place upper limits on the space density of AGNs. Given their
large radio luminosities, hypothetical sources in these empty bins
would likely be volume-limited by the $k_{\rm 20fe} = 11.75$
cutoff. The mean absolute magnitude of AGNs in our sample with
$\log[L_\mathrm{1.4~GHz}(\mathrm{W~Hz}^{-1})] > 24.5$ is $\langle
K_\mathrm{20fe}\rangle \approx -25.84$.  We used this value to
determine the maximum volume within which such a source would have
$k_\mathrm{20fe} \leq 11.75$. For luminosity bins with $N=0$, the
resulting 84\%-confidence $= 1 \sigma$ upper limit given by Poisson
statistics \citep{geh86} is
$\log[\rho_\mathrm{dex}(\mathrm{dex}^{-1}\mathrm{~Mpc}^{-3})] \lesssim
-7.68$. These limits are shown by downward pointing arrows in Figure
\ref{fig:figure13} and were used as additional constraints on the AGN
luminosity function. Above $\log[L(\mathrm{W~Hz}^{-1})] \sim 26.4$
those limits are well above the measured data points and provide no
useful constraints on the luminosity function.

A double power-law has traditionally been used to describe the local
logarithmic luminosity function of AGNs:
\begin{equation}\label{eqn:doublepowerlaw}
  \rho_{\rm dex}(L) =
  \frac{C}{\left(L/L_*\right)^{\alpha} + \left(L/L_*\right)^{\beta}}~.
\end{equation}
Here $\alpha$ is the power-law slope in the limit $L \ll L_*$ and
$\beta$ is the slope for $L \gg L_*$. Both $C$ and $\alpha$ are well
constrained by our data.  However, radio-luminous AGNs are so rare in
the local universe that we can only weakly constrain the local
turnover luminosity $L_*$ and high-luminosity slope $\beta$.  The
deeper ($k_\mathrm{s} < 12.75$) \citet{mau07} AGN luminosity function gives
a slightly better constraint on $\beta$.

The dotted curves matching the filled and unfilled points in Figure
\ref{fig:figure13} indicate the best-fitting \citet{sau90} parametric luminosity
functions for the 2MASX/NVSS star-forming galaxies and AGNs,
respectively.

\section{Local 1.4~GHz Spectral Power Density Functions}\label{sec:udex}

The spectral power density function $u(L)$ is defined as the spectral
power density generated by sources with 1.4 GHz spectral luminosities
in the range $L$ and $L+dL$:
\begin{equation}\label{eq:rhou}
u(L) \equiv L\rho(L).
\end{equation}
The symbol $u$ is a reminder that the dimensions of spectral power
density ($\mathrm{W~Hz}^{-1} \mathrm{~Mpc}^{-3}$) are the same as
those of energy density ($\mathrm{J~Mpc}^{-3}$).  The range of
spectral luminosities spanned by galaxies is so large that it is
convenient to define a logarithmic spectral power density function
\begin{equation}
u_\mathrm{dex}(L) \equiv \ln(10) L u(L) = L \rho_\mathrm{dex}(L)
\end{equation}
equal to the spectral power density (or energy density) per decade of
spectral luminosity.

To calculate $u_\mathrm{dex}(L)$ we separated our galaxies into bins of
logarithmic width $\Delta \log(L) = 0.2$ centered on
$\log[L(\mathrm{W~Hz}^{-1})] = 19.4,\, 19.6, \,\dots\,,27.6$ and
counted the number $N$ of galaxies in each bin.  There are 5 bins per
decade of luminosity, so each bin centered on luminosity $L$ yields
the estimate
\begin{equation}
u_{\rm dex}(L) = 5\sum_{i=1}^N\left(\frac{L}{V_{\rm max}}\right)_i~,
\end{equation}
with rms counting uncertainty
\begin{equation}
\sigma = 5 \left[\sum_{i=1}^N\left(\frac{L}{V_{\rm max}}\right)_i^2\right]^{1/2}~.
\end{equation}
Our 1.4\,GHz local power density functions for star-forming galaxies
and AGNs are listed  in Table \ref{tab:table4} with rms errors equal to the
quadratic sum of the rms counting uncertainty and 3\%.

\begin{deluxetable}{crcrc}
  \tabletypesize{\footnotesize}
  \tablewidth{0pt}
  \tablecolumns{5}
  \tablecaption{1.4\,GHz Spectral Power Density Functions ($h=0.70$)\label{tab:table4}}
  \tablehead{\colhead{} & \multicolumn{2}{c}{~~~Star-forming Galaxies~~} & \multicolumn{2}{c}{\quad AGNs} \\
    \colhead{$\log\,L$} & \colhead{\rlap{N}} & \colhead{$\log\,u_{\rm dex}$} & \colhead{\rlap{N\,}} &
    \colhead{$\log\,u_{\rm dex}$} \\
    \colhead{\llap{(W}\,Hz\rlap{$^{-1})$}} & \colhead{} &
    \colhead{\llap{(W}\,Hz$^{-1}$\,dex$^{-1}$\,Mpc\rlap{$^{-3})$}} &
    \colhead{} &
    \colhead{\llap{(W}\,Hz$^{-1}$\,dex$^{-1}$\,Mpc\rlap{$^{-3})$}} }
  \startdata
19.4 &  6    &  $17.56_{-0.25}^{+0.16}$  &    1    &  $16.76_{-0.76}^{+0.52}$ \\
19.6 &  3    &  $17.18_{-0.34}^{+0.30}$  &    0    &  \nodata \\
19.8 &  10   &  $17.53_{-0.17}^{+0.12}$  &    0    &  \nodata \\
20.0 &  11   &  $17.66_{-0.22}^{+0.14}$  &    1    &  $16.45_{-0.76}^{+0.52}$ \\ 
20.2 &  21   &  $17.72_{-0.11}^{+0.09}$  &    3    &  $16.85_{-0.34}^{+0.30}$ \\ 
20.4 &  59   &  $18.22_{-0.08}^{+0.07}$  &    5    &  $17.21_{-0.34}^{+0.19}$ \\ 
20.6 &  103  &  $18.32_{-0.05}^{+0.04}$  &    4    &  $16.83_{-0.28}^{+0.25}$ \\
20.8 &  147  &  $18.45_{-0.04}^{+0.04}$  &    9    &  $17.12_{-0.18}^{+0.13}$ \\ 
21.0 &  244  &  $18.60_{-0.04}^{+0.04}$  &    22   &  $17.47_{-0.12}^{+0.09}$ \\ 
21.2 &  411  &  $18.74_{-0.03}^{+0.03}$  &    22   &  $17.33_{-0.11}^{+0.09}$ \\ 
21.4 &  584  &  $18.81_{-0.02}^{+0.02}$  &    65   &  $17.74_{-0.07}^{+0.06}$ \\ 
21.6 &  823  &  $18.92_{-0.02}^{+0.02}$  &    85   &  $17.88_{-0.13}^{+0.10}$ \\
21.8 &  975  &  $18.91_{-0.02}^{+0.02}$  &    171  &  $18.00_{-0.04}^{+0.04}$ \\ 
22.0 &  1124 &  $19.00_{-0.02}^{+0.02}$  &    216  &  $18.06_{-0.04}^{+0.04}$ \\
22.2 &  893  &  $18.94_{-0.02}^{+0.02}$  &    281  &  $18.19_{-0.04}^{+0.04}$ \\ 
22.4 &  624  &  $18.82_{-0.02}^{+0.02}$  &    280  &  $18.29_{-0.06}^{+0.05}$ \\ 
22.6 &  368  &  $18.68_{-0.03}^{+0.03}$  &    286  &  $18.33_{-0.04}^{+0.03}$ \\ 
22.8 &  168  &  $18.45_{-0.05}^{+0.04}$  &    239  &  $18.43_{-0.05}^{+0.04}$ \\ 
23.0 &  87   &  $18.24_{-0.06}^{+0.05}$  &    209  &  $18.54_{-0.05}^{+0.04}$ \\ 
23.2 &  30   &  $17.86_{-0.12}^{+0.09}$  &    184  &  $18.69_{-0.06}^{+0.05}$ \\ 
23.4 &  13   &  $17.74_{-0.17}^{+0.12}$  &    133  &  $18.65_{-0.06}^{+0.05}$ \\ 
23.6 &  1    &  $16.28_{-0.76}^{+0.52}$  &    97   &  $18.80_{-0.11}^{+0.09}$ \\ 
23.8 &  1    &  $16.85_{-0.76}^{+0.52}$  &    103  &  $18.92_{-0.06}^{+0.05}$ \\ 
24.0 &  0    &  \nodata              &    69   &  $18.97_{-0.08}^{+0.07}$ \\ 
24.2 &  0    &  \nodata              &    70   &  $19.03_{-0.07}^{+0.06}$ \\ 
24.4 &  0    &  \nodata              &    59   &  $19.17_{-0.08}^{+0.06}$ \\ 
24.6 &  0    &  \nodata              &    41   &  $19.15_{-0.09}^{+0.07}$ \\ 
24.8 &  0    &  \nodata              &    41   &  $19.31_{-0.09}^{+0.08}$ \\
25.0 &  0    &  \nodata              &    30   &  $19.25_{-0.12}^{+0.10}$ \\ 
25.2 &  0    &  \nodata              &    24   &  $19.30_{-0.12}^{+0.10}$ \\ 
25.4 &  0    &  \nodata              &    8    &  $19.36_{-0.33}^{+0.18}$ \\ 
25.6 &  0    &  \nodata              &    2    &  $18.66_{-0.45}^{+0.37}$ \\ 
25.8 &  0    &  \nodata              &    0    &  $\lesssim 18.16$ \\
26.0 &  0    &  \nodata              &    1    &  $19.27_{-0.76}^{+0.52}$ \\
26.2 &  0    &  \nodata              &    0    &  $\lesssim 18.56$ \\
26.4 &  0    &  \nodata              &    1    &  $19.27_{-0.76}^{+0.52}$ \\
26.6 &  0    &  \nodata              &    0    &  \nodata \\
26.8 &  0    &  \nodata              &    0    &  \nodata \\
27.0 &  0    &  \nodata              &    0    &  \nodata \\
27.2 &  0    &  \nodata              &    0    &  \nodata \\
27.4 &  0    &  \nodata              &    0    &  \nodata \\
27.6 &  0    &  \nodata              &    1    &  $18.88_{-0.76}^{+0.52}$
\enddata
\end{deluxetable}

\subsection{Star-Forming Galaxies}

\begin{figure}[ht]
  \centering
  \includegraphics[trim={1.8cm 9.1cm 0.5cm 7cm},clip,width=0.52\textwidth]
                  {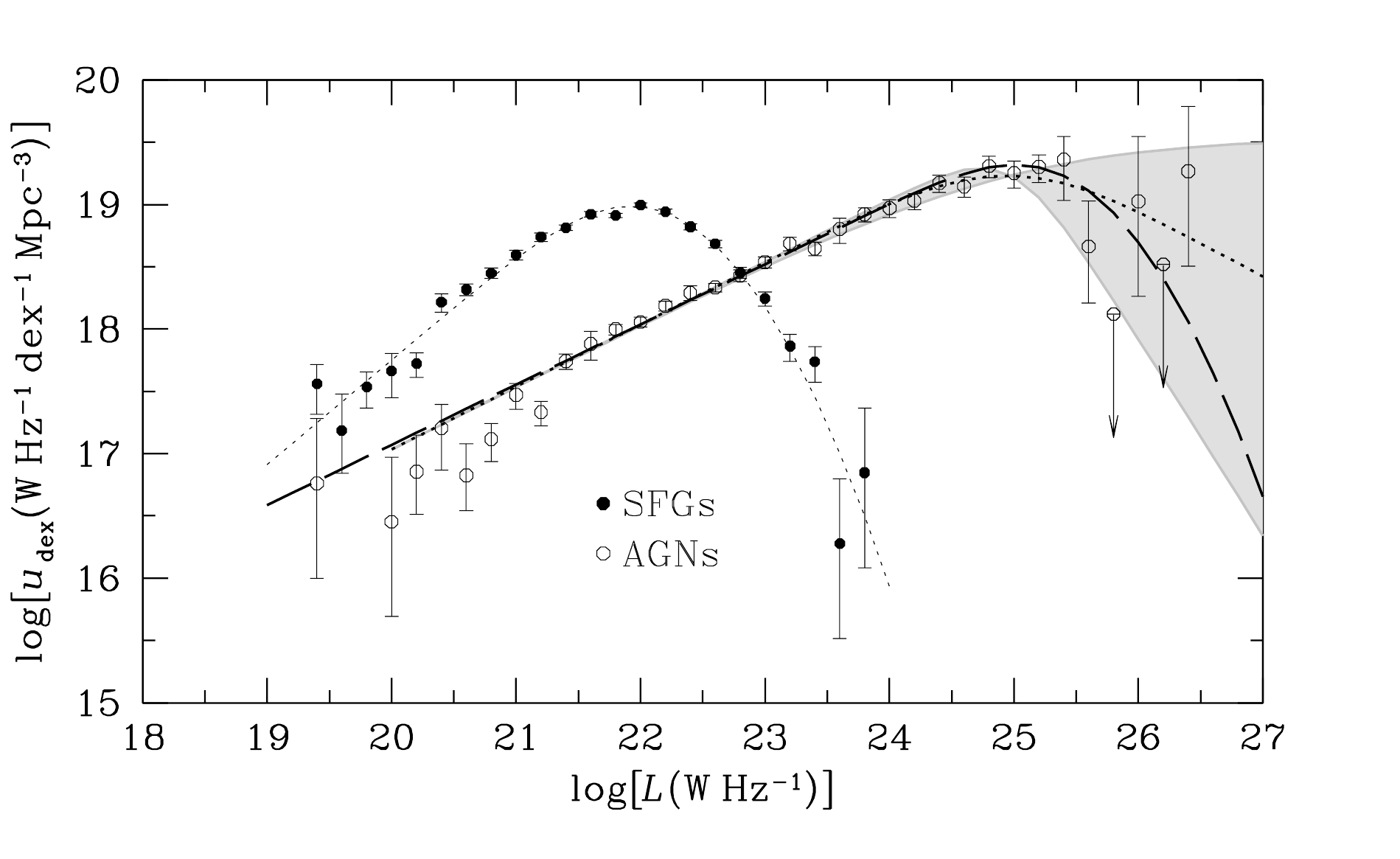}

 \caption{Local spectral power density functions for radio sources
 powered primarily by star formation (filled circles) and AGNs (open
 circles) derived from the 2MASX/NVSS spectroscopic sample shown as
 functions of radio luminosity $L_{\rm 1.4\,GHz}$. The SFGs were
 fitted by the \cite{sau90} parametric form (Equation~\ref{eqn:sau})
 multiplied by $L_{\rm 1.4\,GHz}$.  The AGNs were fitted by both the
 \citet{sau90} form (dashed curve) and by the
 Equation~\ref{eqn:doublepowerlaw} double power law (dotted curve).
 The shaded region shows the wide range of possible slopes $\beta$ in
 Equation~\ref{eqn:doublepowerlaw} such that $\chi^2<2$.
     \label{fig:figure14}}
\end{figure}

As expected, the local 1.4 GHz spectral power density function of
star-forming galaxies is well fit by
\begin{equation}\label{eqn:usffit}
  u_{\rm dex}(L) = C \left(\frac{L}{L_*}\right)^{2-\alpha} 
  \exp\left[-\frac{1}{2\sigma^2}\log^2\left(1+\frac{L}{L_*}\right)\right]
\end{equation}
with the same parameters $C = 3.50\times10^{-3} \mathrm{~
  dex}^{-1}\,\mathrm{Mpc}^{-3}$, $L_* = 1.9\times 10^{21}
\mathrm{\,W\,Hz}^{-1}$, $\alpha = 1.162$, and $\sigma = 0.558$ that
fit the local logarithmic luminosity function.  This fit is indicated by
the dotted curve matching the filled circles in Figure~\ref{fig:figure14}.

The total 1.4 GHz spectral power produced per unit volume by
star-forming galaxies $U_\mathrm{SF}$ is the integral the local power density
function of star-forming galaxies over spectral luminosity:
\begin{equation}
U_{\rm SF} = \int_{0}^{\infty} u_\mathrm{SF}(L) \,dL ~.
\end{equation}
$U_{\rm SF}$ is an extinction-free measurement proportional to the
SFRD $\psi_{\rm SF}~(\mathrm{M_\odot~yr}^{-1} \mathrm{~Mpc}^{-3})$.
We calculated $U_\mathrm{SF}$ directly by summing $L/V_{\rm max}$ over
the unbinned sample of all star-forming galaxies in the 1.4\,GHz
2MASX/NVSS spectroscopic subsample; it is
\begin{equation}\label{eqn:usf}
U_{\rm SF} = (1.54 \pm 0.05) \times 10^{19}\mathrm{~W~Hz}^{-1} \mathrm{~Mpc}^{-3}
\end{equation}
for $H_0 = 70 \mathrm{~km~s}^{-1} \mathrm{~Mpc}^{-1}$.  The rms error
in $U_\mathrm{SF}$ includes a 3\% flux-density calibration uncertainty.

\begin{figure}[!h]
  \centering
  \includegraphics[trim={1cm 1.2cm 0.5cm 7.5cm},clip,width=0.49\textwidth]
                  {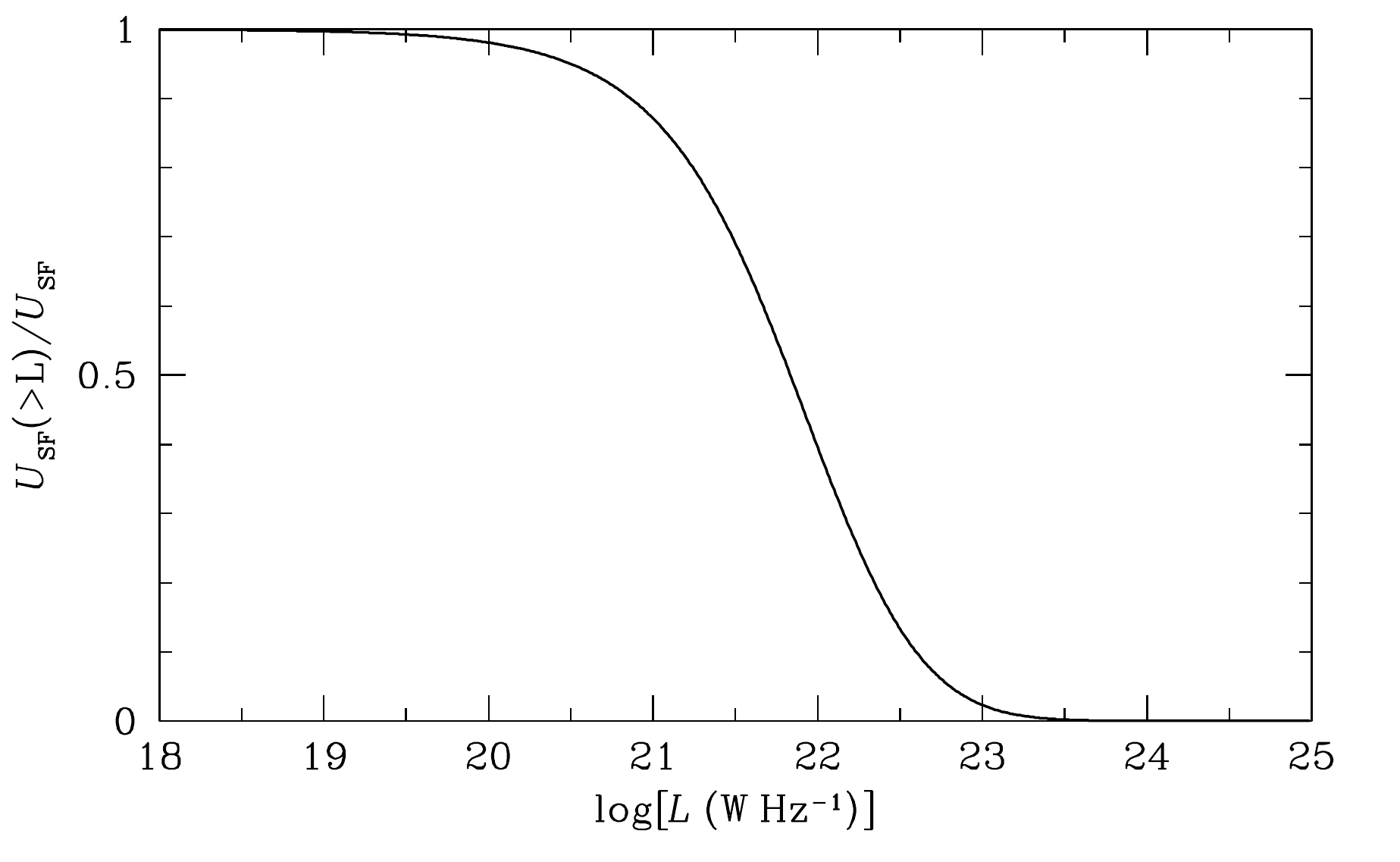}
  \caption{This curve shows the fraction $U_\mathrm{SF}(>L) /
    U_\mathrm{SF}$ of the 1.4~GHz spectral power density generated by
    star-forming galaxies with luminosities $>L$ predicted by
    extrapolating the fitting function in Equation~\ref{eqn:usffit}.
    \label{fig:figure15}}
\end{figure}

Let $U_\mathrm{SF}(>L)$ be the cumulative spectral power density
produced by star-forming galaxies with 1.4~GHz spectral luminosities
$>L$, so the ratio $U_\mathrm{SF}(>L)\, /\, U_\mathrm{SF}$ is the
fraction of $U_\mathrm{SF}$ produced by galaxies more luminous than
$L$.  The curve in Figure~\ref{fig:figure15} shows that ratio
calculated from our fit to Equation~\ref{eqn:usffit}.  It is $0.99$
for $\log[L(\mathrm{W~Hz}^{-1})] = 19.3$, the lowest luminosity in the
2MASX/NVSS spectroscopic subsample, suggesting that sources fainter
than our sample limit account for $<1\%$ of all nearby star formation.

\subsection{AGNs}\label{subsec:agns}

Excluding the anomalous quasar 3C~273 at
$\log[L(\mathrm{W~Hz}^{-1})] \sim 27.4$, the parameters of the double
power-law fit were determined by minimizing the reduced $\chi^2$
statistic of the fit to the measurements weighted by their
uncertainties. The dashed line in Figure \ref{fig:figure14} represents
this best fit to Equation~\ref{eqn:doublepowerlaw} with parameters $C
= 3.58\times10^{-6}$, $L_* = 9.55\times10^{24}\,{\rm W\,Hz^{-1}}$,
$\alpha = 0.498$, and $\beta = 1.55$. Because luminous AGNs are so
rare, the value for $\beta$ can range from $1$ to $2.58$ for $\chi^2 <
2$ (shaded region in Figure \ref{fig:figure14}).

Lacking the data needed to constrain the high-luminosity
power-law slope $\beta$ for the AGN luminosity function, we considered
an alternative approach. There is strong evidence supporting the
notion of co-evolution of star-forming host galaxies and AGNs
\citep[e.g.][]{geb00}. This co-evolution indicates that the luminosity
functions of these populations might be represented by the same
functional form, so we applied the \cite{sau90} form
(Equation~\ref{eqn:sau}) used for the SFGs to the AGNs. There remains
the issue of the poorly sampled high-$L$ end of the AGN luminosity
function, so we held our SFG value $\sigma = 0.558$ fixed while
fitting the AGN luminosity function. The dotted curve following the
unfilled points in Figure \ref{fig:figure13} represents the best-fitting
Equation~\ref{eqn:sau} parameters for the 2MASX/NVSS AGNs: $C =
4.59\times10^{-6}\, \rm dex^{-1}\,Mpc^{-3}$, $L_* =
4.65\times10^{24}\, \rm W\,Hz^{-1}$, $\alpha = 1.516$, and $\sigma =
0.558$.


We calculated the total 1.4~GHz spectral power density produced by
AGNs $U_\mathrm{AGN}$ directly by summing $L/V_\mathrm{max}$ over the
unbinned sample of all AGNs in 1.4~GHz 2MASX/NVSS spectroscopic
subsample; it is
\begin{equation}\label{eqn:uagn}
  U_\mathrm{AGN} = (4.23 \pm 0.55) \mathrm{~W~Hz}^{-1} \mathrm{~Mpc}^{-3}
\end{equation} 
for $H_0 = 70 \mathrm{~km~s}^{-1} \mathrm{~Mpc}^{-1}$.  The rms error
in $U_\mathrm{AGN}$ includes a 3\% flux-density calibration
uncertainty.

\section{Cosmic Variance}\label{sec:cosmicvar}

The small statistical errors quoted in Tables~\ref{tab:table3} and
\ref{tab:table4} and in Equations~\ref{eqn:usf} and \ref{eqn:uagn}
inlcude only the Poisson counting errors for unclustered galaxies
added in quadrature with the 3\% absolute flux-density calibration
uncertainty of the NVSS \citep{con98}.  The mean accessible redshifts
of galaxies used to estimate the local spectral luminosity and power
density functions, weighted by each source's contribution to the total
star formation density, are $\langle z \rangle= 0.026$ and $\langle z
\rangle = 0.070$ for the 2MASX/NVSS star-forming galaxies and AGNs,
respectively.  The corresponding distances $D \sim 100 - 300
\mathrm{~Mpc}$ are comparable with the size $D \sim 150 \mathrm{~Mpc}$
of baryon acoustic oscillations, so significant cosmic variance from
large-scale clustering is expected.  To extend our local results
(e.g., the local $U_\mathrm{SF}$) derived from observations made from
only one point in the universe to the whole universe (e.g, the recent
$U_\mathrm{SF}$ averaged over all space), it is necessary to add this
cosmic variance to the Poisson and calibration variances.

\begin{figure}[!h]
  \centering
  \includegraphics[trim={3.5cm 7cm 2.5cm 3.5cm},clip,width=0.49\textwidth]
                  {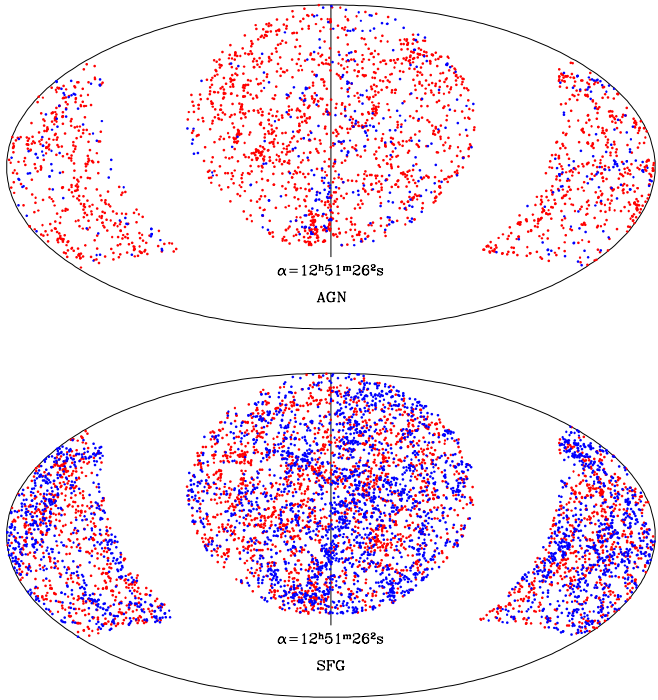}
  \caption{AGNs (upper panel) and SFGs (lower panel) in our sample are
    shown on Hammer equal-area projections of the sky centered on
    J2000 $\alpha = 12^\mathrm{h}51^\mathrm{m}26^\mathrm{s}$ and
    $\delta = 0$. Right ascension increases to the left, so the RA1
    hemisphere is to the right of the vertical dividing line and the
    RA2 hemisphere is to the left.  Blue indicates galaxies with $cz <
    7000\,{\rm km\,s^{-1}}$, and $cz > 7000\,{\rm km\,s^{-1}}$
    galaxies are red. The color boundary at $cz = 7000\,{\rm km\,s^{-1}}$
    corresponds to a distance $D \sim 100$~Mpc. \label{fig:figure16}}
\end{figure}

To estimate the amplitude of the cosmic variance, we divided our
sample covering $7.016$~sr of the sky into two equal-area hemispheres
split by the vertical plane passing through J2000 $\alpha =
12^\mathrm{h}51^\mathrm{m}26^\mathrm{s}$, the right ascension of the
north galactic pole (Figure~\ref{fig:figure16}). We call the hemisphere
covering J2000 $ \alpha = 00^\mathrm{h}51^\mathrm{m}26^\mathrm{s}$
through $\alpha = 12^\mathrm{h}51^\mathrm{m}26^\mathrm{s}$ ``RA1'' and
the other hemisphere ``RA2.''

The 3603 star-forming galaxies in RA1 produce $U_{\rm SF,1} = (1.75
\pm 0.06) \times10^{19} \mathrm{~W~Hz}^{-1} \mathrm{~Mpc}^{-3}$ and
the 3103 star-forming sources in RA2 produce $U_{\rm SF,2} = (1.35 \pm
0.04) \times10^{19} \mathrm{~W~Hz}^{-1} \mathrm{~Mpc}^{-3}$, where
these errors do not include cosmic variance.  The fractional
difference in $U_\mathrm{SF}$ between the two hemispheres is actually
$\sim 0.26$, so if the two halves of the sky are nearly independent,
the rms fractional uncertainty in their mean is $\sim 0.13$.  Thus our
spectroscopic subsample is large enough that cosmic variance exceeds
its Poisson and calibration variances.  Our estimate of the recent
``universal'' $U_\mathrm{SF}$ based on local measurements must include
the cosmic variance; it is
\begin{equation}\label{eqn:cosmicvarsf}
  U_\mathrm{SF} =  (1.54 \pm 0.20) \times 10^{19}
  \mathrm{~W~Hz}^{-1} \mathrm{~Mpc}^{-3}~.
\end{equation}
The corresponding numbers for radio sources primarily powered by AGNs
are $U_\mathrm{AGN,1} = (4.72 \pm 0.55) \times 10^{19} \mathrm{~W~Hz}^{-1}
\mathrm{~Mpc}^{-3}$ and $U_\mathrm{AGN,2} = (3.74 \pm 0.53) \times
10^{19} \mathrm{~W~Hz}^{-1} \mathrm{~Mpc}^{-3}$, so adding the cosmic
variance implies the recent universal AGN spectral energy density is
\begin{equation}\label{eqn:cosmicvaragn}
  U_\mathrm{AGN} =  (4.23 \pm 0.78) \times 10^{19}
  \mathrm{~W~Hz}^{-1} \mathrm{~Mpc}^{-3}~.
\end{equation}
Figure \ref{fig:figure16} suggests that bisecting the sky at the
chosen meridian gives a larger difference than most other choices
would have, so we believe the overall error estimates in
Equations~\ref{eqn:cosmicvarsf} and \ref{eqn:cosmicvaragn} are
conservative.

Figures~\ref{fig:figure13} and \ref{fig:figure14} show our luminosity
and power-density function data points with error bars that do not
include cosmic variance.  We note that the data still match, within those small
error bars, the smooth parametric fits shown as dotted curves.  We conclude
that cosmic variance affects the overall space density of galaxies but not their
detailed luminosity distributions.

We can also use our local sample to
estimate how the expansion dynamics of a $\Lambda$CDM universe might
be affected by density fluctuations on small scales.  The 
$\lambda = 2.16~\mu$m spectral luminosity densities of our sample
galaxies in RA1 and RA2 are
$1.734\times10^{20}\,{\rm W\,Hz^{-1}\,Mpc^{-3}}$ and
$1.065\times10^{20}\,{\rm W\,Hz^{-1}\,Mpc^{-3}}$, respectively.  Using
the $\lambda = 2.16~\mu$m luminosity as a proxy for baryonic mass
and assuming dark matter has a similar large-scale distribution, relative to the
mean matter density, RA1 and RA2 have densities $1.239$ and
$0.761$. For the global cosmological parameters
$\Omega_m=0.3$, $\Omega_{\Lambda} = 0.7$, and $\Omega_r =
8.5\times10^{-5}$,
\begin{align*}
  \Omega_{\rm RA1} &= (0.3\times1.239 + 0.7 + 8.6\times10^{-5}) = 1.0718 \\
  \Omega_{\rm RA2} &= (0.3\times0.761 + 0.7 + 8.6\times10^{-5}) = 0.9284.
\end{align*}
The ``local'' Hubble constant is proportional to $\Omega^{1/2}$, so in
regions RA1 and RA2 the local Hubble constant could be $H_{0,1}
\approx 72.5 \mathrm{~km~s}^{-1} \mathrm{~Mpc}^{-1}$ and $H_{0,2} =
67.5\,{\rm km\,s^{-1}\,Mpc^{-1}}$. This scatter is comparable to that
of published values of $H_0$, with the difference being that the lower
published value is a global measurement rather than a small-scale
measurement such as this. Regardless, differing densities on $\sim 100
\mathrm{~Mpc}$ scales may prevent local measurements
of the global $H_0$ to better than $\sim\pm2.5\,{\rm km\,s^{-1}}$.

\section{Recent Star Formation Rate Density}\label{sec:sfrd}

Radio continuum emission is a tight, nearly linear, and dust-unbiased
independent tracer of the SFRD $\psi$.  Steep-spectrum ($\langle
\alpha \rangle \approx -0.8$) synchrotron radiation from relativistic
electrons accelerated in the core-collapse SNRs of short-lived massive
($M > 8 ~M_\odot$) stars dominates the radio emission of SFGs at all
frequencies below $\nu \sim 30 \mathrm{~GHz}$, and flat-spectrum
($\alpha \approx -0.1$) free-free radiation from thermal electrons in
H\,\textsc{ii} regions ionized by by even more massive short-lived
stars emerges above 30\,GHz \citep{con92}. At $\nu \sim 1 \,\rm GHz$,
$\sim 90 \%$ of the radio emission from SFGs can be attributed to
synchrotron radiation and the remaining $\sim 10\%$ to free-free
emission. The FIR/radio correlation shows that SFR is
proportional to radio luminosity in all but the least-luminous SFGs
\citep{con91}, indicating that the constant of proportionality between
radio luminosity and SFR is remarkably insensitive to potentially
confounding variables such as interstellar magnetic field strength.

Thus SFR can be related to 1.4\,GHz luminosity by an equation of the form
\begin{equation}
  \frac{{\rm SFR}(M>5\,M_{\odot})}{M_{\odot}\,\rm yr^{-1}}
  = \frac {1} {x} \Biggl( \frac {L_{1.4\,\rm GHz}} {\mathrm{W~Hz}^{-1}} \Biggr)~,
\end{equation}
where $x$ is a dimensionless constant whose value has been found to range from $\sim
1.8 \times 10^{21}$ to $\sim
8.9 \times 10^{21}$.
For example, \citep{con02} reported
\begin{equation}\label{eqn:massivesfr}
  \frac{{\rm SFR}(M>5\,M_{\odot})}{M_{\odot}\,\rm yr^{-1}} =
  \frac{1}{4.6\times 10^{21}}\left(\frac{L_{1.4\,\rm GHz}}{\rm W\,Hz^{-1}}\right)~.
\end{equation}
Radio emission is insensitive to lower-mass stars. To account for their contribution to 
the total star-formation rate, we followed \cite{mad14} and
assumed a Salpeter initial mass function $\Psi(M) \propto M^{-2.35}$ over the
mass range $0.1\,M_{\odot}<M<100\,M_{\odot}$.  Then
the total star-formation rate is
\begin{equation}
{\rm SFR}(M>0.1\,M_{\odot}) \approx 5.5\,{\rm SFR}(M>5\,M_{\odot})~.
\end{equation}
Because the conversion factor between 1.4\,GHz luminosity and total
SFR is still uncertain, with values ranging from $5.5 x \sim 0.8
\times10^{21}$ to $1.7\times10^{21}$, we adopted the easily rescalable
midrange number $1.0\times10^{-21}$.  Dividing SFR and 1.4\,GHz
luminosity by volume gives the SFRD $\psi$ in terms of
$U_\mathrm{SF}$:
\begin{equation}
  \frac{\psi(M > 0.1~M_{\odot})}{M_{\odot}\,{\rm yr^{-1}\,Mpc^{-3}}}
  \approx 1.0\times10^{-21} \left(\frac{U_{\rm SF}}{{\rm
      W\,Hz^{-1}\,Mpc^{-3}}}\right)~.
\end{equation}
Then our measured $U_{\rm SF} = (1.54 \pm 0.20) \times10^{19}\,{\rm
  W\,Hz^{-1}}$ with the quoted error including cosmic variance
implies that the ``universal'' recent SFRD is
\begin{equation}
  \psi = (0.0154 \pm 0.0020) \, M_{\odot} \mathrm{~yr}^{-1} \mathrm{~Mpc^{-3}}~.
\end{equation}
This value of $\psi$ is lower than the $\psi = (0.022 \pm 0.001)
\allowbreak ~M_\odot \mathrm{~yr}^{-1} \allowbreak \mathrm{~Mpc^{-3}}$
(Poisson errors only) \citet{mau07} calculated using the higher
conversion factor $\psi = 1.13 \times 10^{-21} U_\mathrm{SF}$.
However, rescaling their conversion factor to $\psi = 1.0 \times
10^{-21} U_\mathrm{SF}$ and adding cosmic variance to their rms
uncertainty yields $\psi = 0.0195 \pm 0.0036 ~M_\odot
\mathrm{~yr}^{-1} \mathrm{~Mpc^{-3}}$.  Thus these two measurements of
$\psi$ agree within their uncertainties.

\begin{figure}[!h]
  \centering
  \includegraphics[trim={0cm 0.cm 2cm 6cm},clip,width=0.49\textwidth]
                  {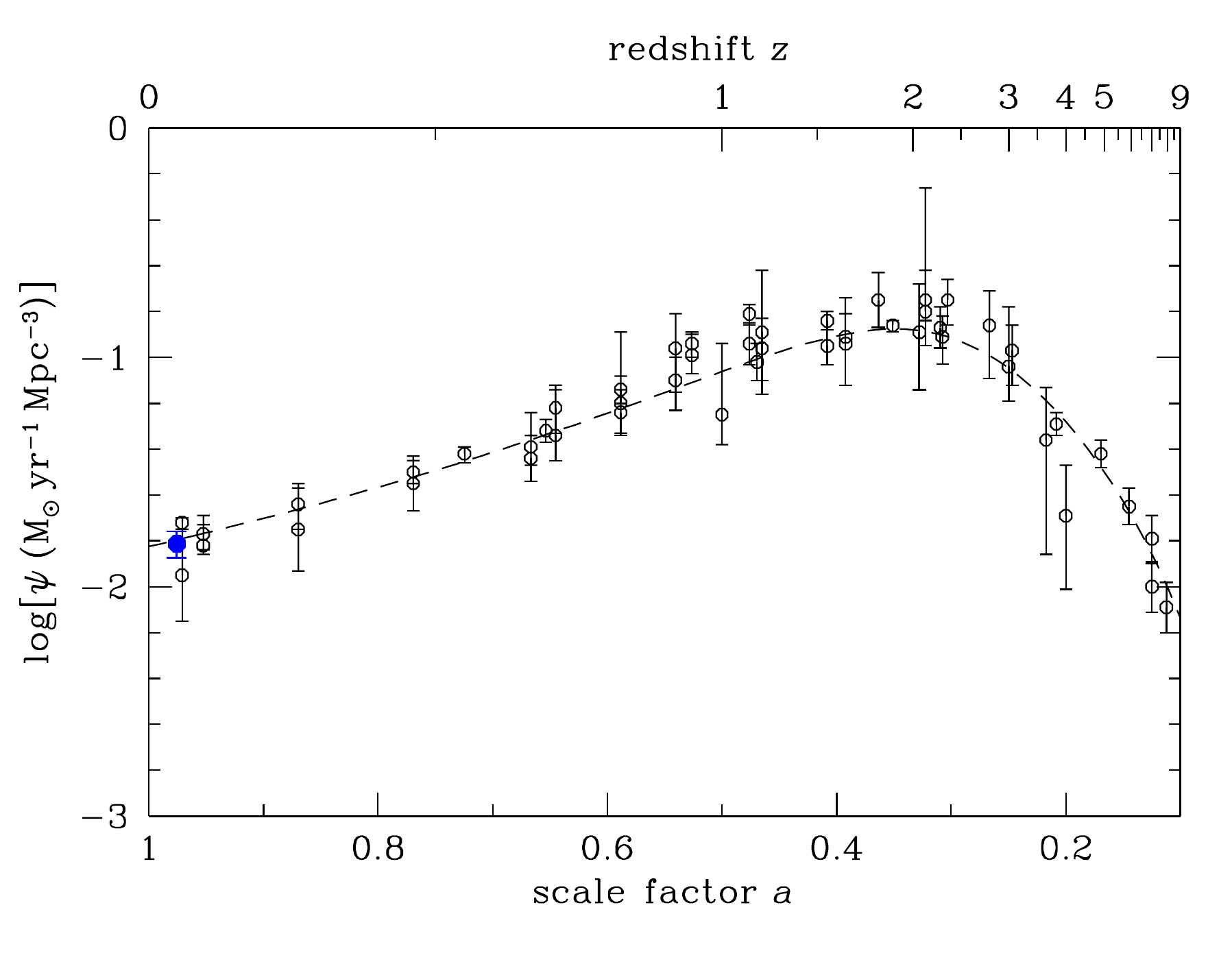}
 \caption{The \citet{mad14} SFRD model (Equation~\ref{eqn:madau}) is
   indicated by the dashed curve fitted to their black FIR and UV data
   points. The blue data point at expansion scale factor $a =
   (1+z)^{-1}\sim 0.98$ ($z = 0.026$) represents our estimate of the
   recent SFRD from based on $U_{\rm SF}$ measured at 1.4\,GHz. Lower
   abscissa: expansion scale factor $a = (1+z)^{-1}$.  Upper abscissa:
   redshift $z$.  Ordinate: star formation rate density ($M_\odot
   \mathrm{~yr}^{-1} \mathrm{~Mpc}^{-3}$). \label{fig:figure17}}
\end{figure}

Multiwavelength compilations of SFRD estimates can be found in
\cite{hop06} and \cite{mad14}. After scaling to the Salpeter IMF,
\cite{hop06} adopted the \citet{col01} parametric fit to describe
the evolution of the SFRD over the redshift range $0<z<7$:
\begin{equation}\label{eqn:hopkins}
\psi(z) = \frac{(a + b\,z)\,h}{1+(z/c)^d} \ {\rm M_{\odot}\,yr^{-1}\,Mpc^{-3}}~.
\end{equation}
For $h = 0.7$ they found $a = 0.0170$, $b = 0.13$, $c = 3.3$, and
$d=5.3$. At the weighted average redshift $\langle z \rangle \sim
0.026$ of our SFG sample, Equation~\ref{eqn:hopkins}
yields $\psi = 0.015~M_{\odot}~{\rm yr}^{-1}~{\rm Mpc}^{-3}$. From
a compilation of FUV and IR rest-frame measurements of $\psi$
spanning $0<z<8$, \cite{mad14} found the best-fit
function
\begin{equation}\label{eqn:madau}
  \psi(z) = 0.015 \frac {(1+z)^{2.7}} {1+[(1+z)/2.9]^{5.6}}
  ~M_\odot \mathrm{~yr}^{-1} \mathrm{~Mpc}^{-3}~.
\end{equation}
Equation~\ref{eqn:madau} gives $\psi(0.026) = 0.016~M_{\odot}
\mathrm{~yr}^{-1} \mathrm{~Mpc}^{-3}$.  Our $\psi =\allowbreak 0.0154
\pm 0.020\allowbreak M_{\odot}\,{\rm yr}^{-1}\,{\rm Mpc}^{-3}$
centered on $\langle z \rangle \approx 0.026$ agrees with both of
these independent SFRD evolutionary models. The blue point in Figure
\ref{fig:figure17} compares our measurement with the FIR and UV data
points and the dashed curve showing the \citet{mad14} model.

\acknowledgments This publication made use of data products from the
Two Micron All Sky Survey, which is a joint project of the University
of Massachusetts and the Infrared Processing and Analysis
Center/California Institute of Technology, funded by the National
Aeronautics and Space Administration and the National Science
Foundation.  This research made use of the NASA/IPAC Extragalactic
Database (NED), which is operated by the Jet Propulsion Laboratory,
California Institute of Technology, under contract with the National
Aeronautics and Space Administration. This publication made use of
data products from the Wide-field Infrared Survey Explorer, which is a
joint project of the University of California, Los Angeles, and the
Jet Propulsion Laboratory/California Institute of Technology, funded
by the National Aeronautics and Space Administration.  This material
is based upon work supported by the National Science Foundation
Graduate Research Fellowship under Grant No. DDGE-1315231.

We thank our anonymous referee for a careful reading and exceptionally
valuable suggestions for improving this paper.

\appendix

\section{A. 2MASX/NVSS Sky Coverage}\label{sec:skyareaappendix}

The 2MASX/NVSS sample covers the sky with J2000 $\delta > \delta_0 =
-40^\circ$ except for absolute galactic latitudes $\vert b \vert < b_0
= 20^\circ$.  The sample solid angle is the solid angle with $\delta >
\delta_0$ minus the solid angle with $\vert b \vert < b_0$, except for
(therefore plus) the solid angle with $\vert b \vert < b_0$ and
$\delta_0$:
\begin{equation}\label{eqn:total}
\Omega = \Omega(\delta > \delta_0) \,-\, \Omega(\vert b \vert < b_0)
\,+ \, \Omega(\vert b \vert < b_0, \delta < \delta_0)~.
\end{equation}

On a unit sphere the Cartesian coordinates corresponding to the J2000
equatorial coordinates $\alpha, \delta$ are
\begin{eqnarray}
x = & \sin \alpha \cos \delta \\
y = & \cos \alpha \cos \delta \nonumber\\
z = & \sin \delta ~~~~~~~\nonumber
\end{eqnarray}
The circle of constant declination $\delta$ has radius $r = (x^2 +
y^2)^{1/2} = \cos\delta$, so the solid angle covering all right
ascensions $\alpha$ and declinations north of $\delta_0 = -40^\circ$
is
\begin{equation}\label{eqn:omeganorth}
\Omega(\delta > \delta_0) =  2 \pi \int_{\delta_0}^{\pi/2} 
\cos \delta  \,d \delta 
 = 2 \pi (1 - \sin\delta_0)
\approx 10.3219 \mathrm{~sr}
\end{equation}
Likewise, the band covering all galactic longitudes $l$ and absolute
galactic latitudes $\vert b \vert < b_0 = 20^\circ$ covers solid angle
\begin{equation}\label{eqn:omegaplane}
\Omega(\vert b \vert < b_0) =  2 \pi \int_{-b_0}^{b_0} \cos b \, db =
 4 \pi \sin b_0 \approx 4.2980 \mathrm{~sr}
\end{equation}
The third term of Equation~\ref{eqn:total} is the solid angle with
$\vert b \vert < b_0$ and $\delta < \delta_0$. It can be written in
the form
\begin{equation}\label{eqn:thirdterm}
\Omega(\delta < \delta_0, \vert b \vert < b_0)
=  \int_{-b_0}^{b_0} \cos b \int_{l_\mathrm{min}(b)}^{l_\mathrm{max}(b)}\, dl \,db\, 
\end{equation}
where $l_\mathrm{max}(b) - l_\mathrm{min}(b)$ is the range of galactic
longitudes at galactic latitude $b$ and declination $\delta$.
Calculating that range requires converting between equatorial and
galactic coordinates.

The J2000 equatorial coordinates of the North Galactic Pole (NGP) are
$\alpha_\mathrm{p} = 12^\mathrm{h} 51^\mathrm{m} 26^\mathrm{s}$ and
$\delta_\mathrm{p} = +27^\circ 7' 42'' \approx 27.1283$~deg.\footnote{
  http://astronomy.swin.edu.au/cosmos/N/North+Galactic+Pole} The
2MASX/NVSS region spans all $\alpha$, so only $\delta_\mathrm{p}$
matters.  We can define ``shifted'' galactic coordinates $(\lambda,
b)$ with $\alpha_\mathrm{p} = 0$ so converting from $(\alpha, \delta)$
to $(\lambda, b)$ needs only a single rotation about the $x$ axis and
\begin{eqnarray}
x = & \sin \lambda \cos b \\
y = & \cos \lambda \cos b \nonumber\\
z = & \sin b~. ~~~~~\nonumber
\end{eqnarray}
Counterclockwise rotation through any angle $\psi$ about the $x$ axis
yields new coordinates
\begin{eqnarray}
x' = & x ~~~~~~~~~~~~~~~~~~\\
y' = & y \cos \psi + z \sin \psi \nonumber\\
z' = & z \cos \psi - y \sin \psi \nonumber
\end{eqnarray}
Rotating these coordinates \emph{clockwise}
by the codeclination of the galactic pole ($\pi / 2 - \delta_\mathrm{p}$)
corresponds to  $\psi = (\delta_\mathrm{p} - \pi / 2)$.  
Thus the three equations for $\lambda, b$ as functions of $\alpha,
\delta$, and $\psi$ are:
\begin{eqnarray}
x' = & \sin \lambda \cos b & = \sin \alpha \cos \delta \\
y' = & \cos \lambda \cos b & = (\cos \alpha \cos \delta) \cos \psi + \sin \delta \sin \psi \nonumber\\
z' = & \sin b  & =  \sin \delta \cos \psi - (\cos \alpha \cos \delta) \sin \psi  \nonumber
\end{eqnarray}

Solving the $z'$ equation for 
\begin{equation}
(\cos \alpha \cos \delta) = \frac{\sin \delta \cos \psi - \sin b}{\sin \psi}
\end{equation} and substituting this into the $y'$ equation gives
\begin{equation}
\cos \lambda \cos b = \Biggl( \frac{\sin \delta \cos \psi - \sin
  b}{\sin \psi} \Biggr) \cos \psi + \sin \delta \sin \psi
\end{equation}
The longitude $\lambda(b,\delta_0)$ at which galactic latitude $b$
crosses J2000 declination $\delta_0$ is
\begin{equation}\label{eqn:lambda}
\lambda(b, \delta_0) = \arccos \Biggl[
 \Biggl( \frac{\sin \delta_0 \cos \psi - \sin b}{\sin \psi \cos b} \Biggr) 
\cos \psi
+ \frac{\sin \delta_0 \sin \psi}{\cos b} \Biggr]
\end{equation}
Thus Equation~\ref{eqn:thirdterm} becomes
\begin{equation}\label{eqn:area3}
\Omega(\delta < \delta_0, \vert b \vert < b_0)
= \int_{-b_0}^{b_0} 2 \lambda(b, \delta_0) \cos b 
 \, db
\end{equation}
in $(\lambda, b)$ coordinates.  Integrating Equation~\ref{eqn:area3}
numerically for $\delta_0 = -40^\circ$ and $b_0 = 20^\circ$ gives
$\Omega(\delta < \delta_0, \vert b \vert < b_0) \approx 0.9920
\mathrm{~sr}$. Inserting Equations~\ref{eqn:omeganorth},
\ref{eqn:omegaplane}, and this result into Equation~\ref{eqn:total}
gives the total 2MASX/NVSS solid angle $\Omega \approx 7.0160
\mathrm{~sr}$.

\section{B. New Spectroscopic Redshifts}\label{sec:redshiftappendix}

We obtained spectra for 12 of the 19 galaxies lacking published
spectroscopic redshifts with the Dual Imaging Spectrograph (DIS) on
the Apache Point Observatory (APO) 3.5\,m telescope. Observations were
carried out over three half-nights occurring in October through
December of 2017. DIS is a medium dispersion double spectrograph that
has separated red and blue channels. The standard ``high'' resolution
DIS III grating setup B1200/R1200 was used. The wavelength ranges were
centered on the H$\beta$ and H$\alpha$ lines at the median redshift of
the 2MASX/NVSS sample, 5021\,\AA \ and 6780\,\AA \ for the blue and
red cameras, respectively. This resulted in a wavelength coverage of
$4401 - 5641$\,\AA \ and $6200 - 7360$\,\AA \ for the blue and red
channels, respectively.

Total exposure times ranged from 1620\,s to 3360\,s, taken in
intervals of 120\,s to 420\,s so as to mitigate cosmic-ray
contamination. The two galaxies with the weakest spectral lines were
observed on multiple nights to increase exposure time and improve the
quality of the redshift measurement. Bias and flat frames were
obtained before each observing run. Comparison spectra were obtained
before and after each observation run using a He, Ne, and Ar lamp.

The spectra were reduced and analyzed in a uniform manner with {\sc
  IRAF}. Initial 2D frames were bias-subtracted and flat-fielded using
subroutines in the {\sc CCDRED} package. Apertures were extracted with
the {\sc APEXTRACT} package. Dispersion functions were derived from
the HeNeAr lamp spectra and fit to the object frames using routines in
the {\sc ONEDSPEC} package. Multiple sub-exposures of each target were
combined for the blue and red spectra.

\begin{figure*}[ht]
\centering
\includegraphics[scale=0.6]{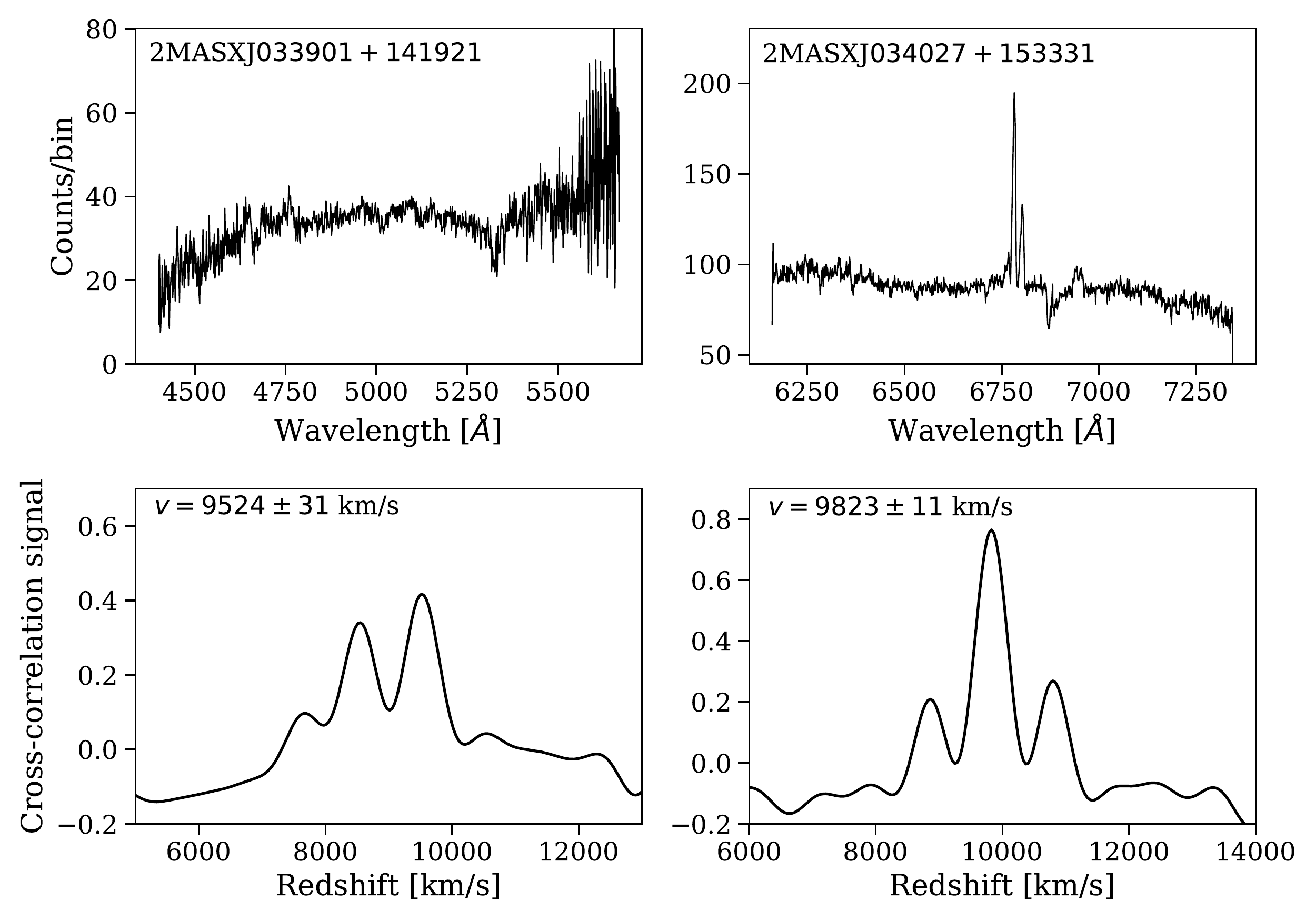}
\caption{Top panels: Examples of typical spectra obtained for the 12
  galaxies with the APO 3.5\,m telescope. Left: Absorption line
  spectrum with the DIS blue camera. Right: Emission line spectrum
  with the DIS red camera. Bottom panels: Results of the
  cross-correlation technique used to measure the redshifts for the
  corresponding top panel galaxies.\label{fig:figure18}}
\end{figure*}

The blue and red portions of the spectrum were combined and processed
using the {\sc XCSAO} procedure in the {\sc RVSAO} package to
determine barycentric radial velocities. Sample spectra are shown in
the top portion of Figure \ref{fig:figure18}. The {\sc XCSAO} routine
follows the cross-correlation technique developed by \cite{ton79}. We
used the SDSS galaxy
templates\footnote{http://classic.sdss.org/dr7/algorithms/spectemplates/},
specifically 23--28, in the cross-correlation. Hot pixels and the
unobserved wavelength range between the blue and red cameras were
ignored by the cross-correlation routine. Typical results from the
cross-correlation technique are shown in the bottom panel of Figure
\ref{fig:figure18} for the two galaxies shown in the top panel. The
resulting barycentric radial velocities are given in Table
\ref{tab:table5}.

The reliability of the velocities can be estimated by the $r$
statistic, a confidence measure. Calibration done in the development
of the {\sc XCSAO} cross-correlation routine \citep{kur98} suggests
that cross-correlations with $r > 3$ can be deemed reliable, but note
that many of the spectra in their test study with $2 < r < 3$ also
yield correct velocities. Of the twelve galaxies observed, all but one
of the spectra has $r > 3$. The exception, 2MASX J21352090+8906537,
has $r=2.55$ for the template with the highest cross-correlation
signal.

\begin{deluxetable*}{ccccccc}
\centering
\tablewidth{0pt}
\tablecaption{2MASS Supplemental Velocities}
\tablehead{
  \colhead{2MASX} & \colhead{R.A.} & \colhead{Dec.} & \colhead{$l$} & \colhead{$b$} &
    \colhead{$v = cz$} & \colhead{$\sigma_v$}  \\
  \colhead{J2000 name} & \colhead{hh:mm:ss.ss} & \colhead{dd:mm:ss.s} & \colhead{(deg)} & \colhead{(deg)} &
  \colhead{(km s$^{-1}$)} & \colhead{(km s$^{-1}$)} 
  }
\startdata
02570403$+$2000446  &  02:57:04.03  &  $+$20:00:44.6  &  159.08  &  $-33.90$ &  9500.6  &  14.4  \\  
03215557$+$2149375  &  03:21:55.57  &  $+$21:49:37.5  &  163.32  &  $-29.00$  &  \llap{1}4302.0  &  12.2  \\
03390103$+$1419217  &  03:39:01.03  &  $+$14:19:21.7  &  172.61  &  $-31.94$  &  9523.9  &  31.0  \\
03402770$+$1533113  &  03:40:27.70  &  $+$15:33:11.3  &  171.90  &  $-30.82$  &  9823.3  &  11.1  \\
04141963$+$2025240  &  04:14:19.63  &  $+$20:25:24.0  &  174.24  &  $-21.65$  &  6316.2  &  59.9  \\
07134975$+$8729044  &  07:13:49.75  &  $+$87:29:04.4  &  125.75  &  $+27.35$  &  \llap{1}5157.5  &  59.9  \\
17272375$+$1521110  &  17:27:23.75  &  $+$15:21:11.0  &   \hphantom{1}37.88  &   $+25.37$  &  9045.2  &  37.3  \\
17494097$+$5333541  &  17:49:40.97  &  $+$53:33:54.1  &   \hphantom{1}81.29  &   $+30.50$  &  \llap{2}8187.3  &  \llap{1}61.8 \\
17543888$+$6803287  &  17:54:38.88  &  $+$68:03:28.7  &   \hphantom{1}98.12  &   $+30.30$  &  \llap{2}3869.6  &  83.9  \\
20510128$-$1710242  &  20:51:01.28  &  $-$17:10:24.2  &   \hphantom{1}29.78  &  $-33.95$  &  \llap{1}9387.9  &  82.2 \\
21352090$+$8906537  &  21:35:20.90  &  $+$89:06:53.7  &  122.18  &   $+26.55$  &  21094.2\tablenotemark{a}  &  \llap{1}25.0 \\
23074944$-$1236479  &  23:07:49.44  &  $-$12:36:47.9  &   \hphantom{1}58.71   &  $-61.74$  &  \llap{2}0378.2  &  37.0  \\
\enddata
\tablecomments{All velocities in reference to the Solar System barycenter.
\tablenotetext{a}{Cross-correlation with the SDSS templates resulted in a $R$ value
less than $3$ ($R = 2.55$).}\label{tab:table5}}
\end{deluxetable*}

\clearpage

\end{document}